\documentclass[12pt]{article}
\usepackage{mathtools,amssymb,mathrsfs,microtype,tikz}
\usepackage[linktoc=all, colorlinks=true, linkcolor=blue, urlcolor=blue, citecolor=red]{hyperref}
\usepackage[backend=bibtex,style=numeric-comp,sorting=none,maxbibnames=99, minbibnames=99]{biblatex}
\makeatletter\@addtoreset{equation}{section}\makeatother

\DeclareMathOperator{\tr}{tr}
\DeclareMathOperator{\diag}{diag}
\DeclareMathOperator{\Arf}{Arf}

\renewcommand{\title}[1]{\vbox{\center\LARGE{#1}}\vspace{5mm}}
\renewcommand{\author}[1]{\vbox{\center\large#1}\vspace{5mm}}
\newcommand{\address}[1]{\vbox{\center\em#1}}

\begin{document}

\begin{titlepage}
\begin{center}
\vspace{5mm}
%\hfill {\tt HU-EP-09/40}\\
\hfill {\tt }\\
\vspace{8mm}

\title{\makebox[\textwidth]{\huge{Global Anomalies on the Hilbert Space}}}
\vspace{10mm}
Diego Delmastro,${}^{ab}$ %\footnote{\href{mailto:ddelmastro@perimeterinstitute.ca}{\tt ddelmastro@perimeterinstitute.ca}}
Davide Gaiotto,${}^{a}$ %\footnote{\href{mailto:??}{\tt ??}}
Jaume Gomis${}^{a}$%\footnote{\href{mailto:jgomis@perimeterinstitute.ca}{\tt jgomis@perimeterinstitute.ca}}
\vskip 7mm
\address{
${}^a$Perimeter Institute for Theoretical Physics,\\
Waterloo, Ontario, N2L 2Y5, Canada}
\address{
${}^b$ Department of Physics, University of Waterloo,\\ Waterloo, ON N2L 3G1, Canada}
\end{center}

\vspace{5mm}
\abstract{
We show that certain global anomalies can be detected in an elementary fashion by analyzing the way the symmetry algebra is realized on the torus Hilbert space of the anomalous theory. Distinct anomalous behaviours imprinted in the Hilbert space are identified with the distinct cohomology ``layers"  that appear in the classification of anomalies in terms of cobordism groups. We illustrate the manifestation of the layers in the Hilbert for a variety of anomalous symmetries and spacetime dimensions, including time-reversal symmetry, and both in systems of fermions and in anomalous topological quantum field theories (TQFTs) in $2+1d$. We argue that  anomalies  can imply an exact bose-fermi degeneracy in the Hilbert space, thus revealing a supersymmetric spectrum of states;   we provide a sharp characterization of when this phenomenon occurs and give nontrivial examples  in various dimensions, including in strongly coupled QFTs. Unraveling the anomalies of TQFTs leads us to develop the construction of the Hilbert spaces, the action of operators and the modular data in spin TQFTs,  material that can be read on its own. 
 }
  
  \vfill\eject

\vspace{20pt}

{\hypersetup{linkcolor=black}
\tableofcontents
\thispagestyle{empty}
}

\end{titlepage}

\section{Introduction and Summary}\label{sec:intro}
\setcounter{footnote}{0} 

Consider a system with a classical global symmetry group $G$. Powerful constraints on the dynamics can be derived by coupling the system to a background connection $A$ for the symmetry $G$. The system has an 't Hooft anomaly~\cite{tHooft:1979rat} if the non-invariance of the partition function under background gauge transformations generated by $g\in G$
\begin{equation}
Z[A] \mapsto e^{i\alpha(g,A)}Z[A]
\end{equation}
cannot be cancelled by a local counterterm constructed out of the background fields. This is the physics that endows anomalies with a cohomological formulation~\cite{Zumino:1983ew,Stora:1983ct,weinberg_1996}.

The anomaly $\alpha(g,A)$ is a local functional of the background connection and of the transformation $g\in G$. An 't Hooft anomaly is captured via anomaly inflow~\cite{CALLAN1985427} from a topological term in one dimension higher. Each such topological term can be thought of as the effective action characterizing a symmetry protected topological (SPT) phase~\cite{Chen:2013qty,Chen:2011pg,Senthil:2014ooa} with symmetry $G$ in one higher dimension. The topological term is gauge invariant on a closed manifold and reproduces the anomaly on a manifold with a boundary. Being topological, an 't Hooft anomaly is robust under deformations that preserve the symmetry, including renormalization group transformations. 't Hooft anomalies give physicists some of the very few clues into the nonpertubative dynamics of a quantum system. 

A combination of insights from condensed matter physics, particle physics, quantum information and mathematics has culminated in a conjecturally complete answer to the problem of classifying the possible anomalies in various dimensions~\cite{Chen:2011pg,Gu:2012ib,Wen:2013oza,KitaevZ16,Chen_2014,Kapustin:2014tfa,Kapustin:2014dxa,Freed:2016rqq}. This includes anomalies in bosonic as well as fermionic systems, for discrete and continuous internal symmetry groups as well as discrete spacetime symmetries such as time-reversal and parity.\footnote{Coupling to a time-reversal background requires defining the system on unoriented manifolds~\cite{PhysRevB.90.165134,Kapustin:2014tfa,Witten:2016cio,Tachikawa:2016cha,Barkeshli:2016mew}.} This has led to the topological classification of anomalies in terms of cobordism theory and generalized cohomology theories~\cite{Gu:2012ib,KitaevZ16,Kapustin:2014tfa,Kapustin:2014dxa,Freed:2016rqq,Xiong:2016deb,Gaiotto:2015zta,Kapustin:2017jrc,Gaiotto:2017zba,Thorngren:2018bhj,Wang:2018pdc,Yonekura:2018ufj,Witten:2019bou}.

Consider first a bosonic system, one which can be defined without a choice of spin structure of the underlying manifold. By Wigner's theorem, symmetries come in two flavours: linear and unitary, or antilinear and antiunitary, with time-reversal being the prototypical example of an antiunitary symmetry. Thus, the symmetry data of a bosonic system is specified by the pair 
\begin{equation}
(G,w_1)\,,
\end{equation}
where $G$ is a group and $w_1\in H^1(G,\mathbb Z_2)$ a certain cohomology class $w_1\colon G\rightarrow \mathbb Z_2$ that encodes the unitarity/antiunitarity of the group elements in $G$. The anomalies of a bosonic system with symmetry data $(G,w_1)$ in $D$ spacetime dimensions are classified by the twisted cobordism group~\cite{Kapustin:2014tfa}
\begin{equation} 
\Omega^{D+1}_\text{so}(G; w_1)\,. 
\end{equation}
 In low spacetime dimensions, for $D\leq 2$, the anomaly classification reduces to group cohomology: $\Omega^{D+1}_\text{so}(G; w_1)=H^{D+1}(G,U(1))$, extending the classic result that anomalies in quantum mechanics (i.e.~$D=1$) are classified by $H^{2}(G,U(1))$, that is, by the projective representations of $G$.\footnote{When $w_1$ is nontrivial the cocycle condition defining $H^*(G,K)$ is twisted by the action of $w_1$, which acts as an involution on $K$. This action is nontrivial for $K=U(1)$ and $K=\mathbb Z$ but trivial for $K=\mathbb Z_2$. In order to avoid clutter we do not write the twisting by $w_1$ explicitly.}
 In higher dimensions, $\Omega^{D+1}_\text{so}(G; w_1)$ can be reconstructed (losing some information about the addition law) from the Atiyah-Hirzebruch spectral sequence~\cite{doi:10.1142/9789814401319_0008}, that combines $H^{D+1}(G,U(1))$ with other cohomology groups of lower degrees.

Now recall the characterization of symmetries and classification anomalies of a fermionic system, which requires the choice of a ($G$-twisted) spin structure to be defined. A fermionic system has a universal and unbreakable $\mathbb Z^{F}_2$ unitary symmetry generated by fermion parity, denoted by $(-1)^F$. This symmetry induces a $\mathbb Z_2$-grading in the Hilbert space $\mathcal H$ of fermionic systems, which become super-vector spaces. Since (classically) symmetries cannot change the fermion parity, that is $[g,(-1)^F]=0$, the symmetry group $G_f$ acting on the local operators of a fermionic system is necessarily a $\mathbb Z^F_2$ central extension of a group $G$, such that $G=G_f/\mathbb Z^F_2$. Also, by virtue of Wigner's theorem, a symmetry can be either unitary or antiunitary. Therefore, the symmetries of a fermionic system are characterized by a cocycle $w_2\in H^2(G,\mathbb Z^F_2)$ specifying the $\mathbb Z^F_2$ central extension and by a cocycle $w_1\in H^1(G,\mathbb Z_2)$ encoding the unitarity/antiunitarity of group elements. The anomalies of a fermionic system with symmetry data\footnote{For example, for $G=\mathbb Z_2$, and taking $w_1$ and $w_2$ to be the nontrivial $\mathbb Z_2=\{0,1\}$ element in $H^1(\mathbb Z_2,\mathbb Z_2)$ and $H^2(\mathbb Z_2,\mathbb Z^F_2)$ yields the symmetry group generated by time-reversal $\mathsf T$ obeying $\mathsf T^2=(-1)^F$, sometimes denoted by $\mathbb Z_4^\mathsf T$. This is the relevant symmetry group of the celebrated topological superconductors.} 
\begin{equation} 
(G;w_1,w_2)
\end{equation}
in $D$ spacetime dimensions are classified by the twisted cobordism group~\cite{Kapustin:2014dxa,Thorngren:2018bhj}
\begin{equation} 
\Omega^{D+1}_\text{spin}(G; w_1,w_2)\,. 
\label{spinanom}
\end{equation}
State-of-the-art mathematical techniques allow for the computation of these twisted cobordism groups; see~\cite{Garcia-Etxebarria:2018ajm,Guo:2018vij,Wan:2018bns,Wan:2019fxh} for many relevant examples together with reviews aimed at physicists. A particularly convenient computational tool is again the Atiyah-Hirzebruch spectral sequence. The different ingredients that go into the computation of~\eqref{spinanom} in this spectral sequence can be given a nice physical interpretation in terms of layers in various dimensions (see below).

While the topological classification of anomalies is rather well understood, \emph{detecting} whether a physical system is anomalous can be a difficult task. Intuitively, one has to keep track of all the arbitrary choices required for a sharp definition of the system on a general background and then quantify the topological obstruction to the trivialization of these choices. A concrete calculation may involve hard-to-determine data characterizing the system.\footnote{For example, detecting anomalies in bosonic topological quantum field theories (TQFTs) requires knowing the $F$-symbols~\cite{Bulmash:2020flp}.} While detecting the anomalies induced by transformations connected to the identity of a Lie group $G$ is textbook material, the detection of \emph{global} anomalies, which includes anomalies for all discrete symmetries, is more subtle~\cite{WITTEN1982324,Wang:2018qoy}.\footnote{In the case of antiunitary symmetries it requires, for example, learning how to define spin TQFTs on unoriented manifolds, which is an open problem. Numerous interesting partial results have been obtained, however~\cite{Chan_2016,Wang:2016qkb,Tachikawa:2016cha,Tachikawa:2016nmo,Bhardwaj:2016dtk,Barkeshli:2017rzd,Turzillo:2018ynq,Kobayashi:2019xxg,Inamura:2019hpu}.} The approach is often indirect, for example by embedding some global anomalies into a perturbative ones (see for example the recent work~\cite{Davighi:2020uab} and references therein).

In this paper we exhibit an elementary method for detecting \emph{some} anomalies, based on constructing the Hilbert space of the theory on a flat (spatial) torus $T^{D-1}$ as well as determining how the algebra of symmetries is realized on the Hilbert space. This can be given the following physical interpretation. The anomaly of a $D$-dimensional theory can be represented by the class 
\begin{equation}
 \alpha_{D+1}\in \Omega^{D+1}_\text{spin}(G; w_1,w_2)\,.
\end{equation}
Studying the Hilbert space of the $D$-dimensional anomalous theory on a spatial torus\footnote{One could also study the reduction of the anomaly class on more general manifolds, potentially detecting more anomalies.} produces upon integration a class 
\begin{equation}
\tilde \alpha_2=\int_{T^{D-1}} \alpha_{D+1}\,.
\end{equation}
The class $\tilde \alpha_2$ can be viewed as the effective anomaly class of a quantum mechanical theory in $0+1d$, which we recognize from the properties of the Hilbert space. As a result, we expect to be able to detect this way all anomalies whose cobordism class can be recognized from the values on manifolds of the form $T^{D-1} \times \Sigma_2$, equipped with generic flat connections, spin structures, etc. 

This perspective also shows that a torus compactification can provide useful anomaly information only if the relevant structures -- either the background $G$ connection or spin structure -- do not extend to one higher dimension (i.e.~if they are not the boundary of a manifold in one higher dimension). Indeed, if these structures were all bounding such that $T^{D-1}=\partial M^D$, then $ \int_{T^{D-1}} \alpha_{D+1}=\int_{M^D}d\alpha_{D+1}=0$, and the effective anomaly in $0+1d$ vanishes. This means that in order to detect the anomaly in the torus Hilbert space we must either turn on non-trivial holonomies for the symmetry $G$ or we must consider periodic boundary conditions on the torus for fermionic theories -- or both.\footnote{Turning on non-trivial holonomies for the symmetry $G$ defines a $G$-twisted Hilbert spaces, which are the Hilbert spaces where to detect anomalies if the spin structures are bounding.} 

In practice, we find that this method captures a surprisingly large amount of anomaly information. This is especially true for fermionic systems.

In order to illustrate how various anomalies are manifested in the Hilbert space, it is useful to recall some ingredients of the (partial) reconstruction of $\Omega^{D+1}_\text{spin}(G; w_1,w_2)$ via the Atiyah-Hirzebruch spectral sequence. The starting point is a collection of layers in various degrees\footnote{Recall that the action of $w_1$ is trivial on $\mathbb Z_2$ coefficients.} (see e.g. \cite{Kapustin:2017jrc,Gaiotto:2017zba,Thorngren:2018bhj})
\begin{equation}\label{eq:layers}
\begin{alignedat}{2}
&\vdotswithin{\in}&\\
\nu_{D-2}&\in H^{D-2}(G,\mathbb Z)&&\text{$p_x+ip_y$ layer} \\
\nu_{D-1}&\in H^{D-1}(G,\mathbb Z_2)&&\text{Arf layer}\\
\nu_{D}&\in H^{D}(G,\mathbb Z_2)&&\text{$\psi$ layer}\\
\nu_{D+1}&\in H^{D+1}(G,U(1))&\qquad&\text{Bosonic layer}
\end{alignedat}
\end{equation}
with nontrivial differentials connecting the various classes. Each layer has a physical and geometric interpretation (see section~\ref{sec:layer} for more details). In particular, the groups which appear in the second slot of $H^{D-k}(G,\,\cdot\,)$ are the groups of $k$-dimensional SPT phases with no symmetries% (recall that $\mathbb Z^{F}_2$ is unavoidably present in fermonic systems)
. We summarize them in table~\ref{tablespt}.
\begin{table}[!h]
\arraycolsep=10pt\def\arraystretch{1.3}
\begin{equation*}
\begin{array}{|c|c|c|c|}\hline
&0+1d&1+1d&2+1d \\\hline
\text{SPTs}&\mathbb Z_2&\mathbb Z_2&\mathbb Z \\\hline
\text{generator}&\psi&\Arf&p_x+ip_y\ \text{aka}\ SO(1)_1\\\hline
Z&S_\pm\mapsto \pm1&\Sigma\mapsto (-1)^{\Arf(\Sigma)}&M_3\mapsto e^{i\mathrm{CS}_\text{grav}[M_3]} \\\hline
\end{array}
\label{tablespt}
\end{equation*}
\caption{The first row gives the classification of SPT phases with no symmetries, the second the generators of the SPT classes, and the last the partition functions of the generators. $S_\pm$ denotes a circle with periodic/antiperiodic ($\text{R}/\text{NS}$) boundary conditions; $\Sigma$ is a compact Riemann surface, and $\Arf(\Sigma)$ is the Arf-invariant of the surface with spin structure, which evaluates to $0$ on even and to $1$ on odd spin structures; and $M_3$ is a three-manifold, with $\mathrm{CS}_\text{grav}=\frac{1}{4\pi}\int_{M_3}\text{tr}(\omega\,\mathrm d\omega+\frac23\omega^3)$.}
\end{table}

The endpoint of the Atiyah-Hirzebruch spectral sequence calculation is the \emph{associated graded} of a filtration of $\Omega^{D+1}_\text{spin}(G; w_1,w_2)$: the addition law on the $k$-th layer is modified by unknown carry-overs from higher layers, which are somewhat tricky to compute. Physically, that means that even if the non-trivial differentials vanish, we can only really assign a specific value to $\nu_{D-k}$ if all $\nu_{D-k'}$ with $k'>k$ vanish, or we can only discuss the difference in the $\nu_{D-k}$ anomaly of two theories for which all $\nu_{D-k'}$ with $k'>k$ are the same. 

We now demonstrate the Hilbert space manifestation of the layers in the anomalies of $0+1d$ fermionic systems with an antiunitary time-reversal symmetry $\mathsf T$ with $\mathsf T^2=1$, so that $G_f=\mathbb Z^\mathsf T_2\times \mathbb Z_2^F$. The anomalies of such a system are classified by $\Omega_\text{spin}^2(\mathbb Z_2;1,0)=\Omega^2_{\text{pin}^-}=\mathbb Z_8$ \cite{Fidkowski:2009dba}. The anomaly arises from three layers
\begin{equation}\label{eq:intro_system1d}
\begin{alignedat}{2}
\nu_{0}&\in H^{0}(\mathbb Z^\mathsf T_2,\mathbb Z_2)\simeq \mathbb Z_2&&\text{Arf layer}\\
\nu_{1}&\in H^{1}(\mathbb Z^\mathsf T_2,\mathbb Z_2)\simeq \mathbb Z_2&&\text{$\psi$ layer}\\
\nu_{2}&\in H^{2}(\mathbb Z^\mathsf T_2, U(1))\simeq \mathbb Z_2&\qquad&\text{Bosonic layer}
\end{alignedat}
\end{equation}
We can think about these three groups as compiling into $\mathbb Z_8$, corresponding to the binary expansion
\begin{equation}
\nu=\nu_0+2\nu_1+4\nu_2\quad \mod8\,.
\end{equation}
 The simplest $0+1d$ system with anomaly $\nu\in\mathbb Z_8$ is a set of $\nu$ free massless Majorana fermions, with time-reversal acting as $\mathsf T(\psi(t))=\psi(-t)$ on all $\nu$ fermions. At the level of operators, this system has symmetries generated by $\mathsf T$ and $(-1)^F$, these two operations commuting and being both of order two, i.e., $G_f=\mathbb Z_2^{\mathsf T}\times\mathbb Z_2^F$. At the level of the Hilbert space, the anomaly $\nu$ is manifested through the following anomalous pattern:
\begin{itemize}
\item \underline{$\nu=$ odd:} There is no graded Hilbert space $\mathcal H$. This arises from the Arf layer $\nu_0$.

\item \underline{$\nu=2\mod4$:} There is a graded Hilbert space $\mathcal H$ but the symmetry generators on $\mathcal H$ do not commute. Instead, they anti-commute:
\begin{equation}
\{\mathsf T,(-1)^F\}=0\,.
\end{equation}
This arises from the fermion $\psi$ layer $\nu_1$.

\item \underline{$\nu=4\mod8$:} There is a graded Hilbert space $\mathcal H$ with $[ \mathsf T,(-1)^F]=0$ on it, but the symmetry algebra $\mathsf T^2=1$ is realized projectively on $\mathcal H$, that is 
\begin{equation}
 \mathsf T^2=-1\qquad\text{on}\qquad\mathcal H\,.
\end{equation}
This arises from the bosonic layer $\nu_2$. 
\end{itemize}

%\note{A more precise statement concerning the $\nu$ odd case. Any $\mathbb Z_2$-graded algebra is either isomorphic to $\mathcal H$ of to $\mathcal H\otimes\mathrm{Cl}(1)$, where $\mathcal H$ is a $\mathbb Z_2$-graded vector space (see e.g.~\url{https://doi.org/10.1103/PhysRevB.98.125101}). The ``lack of Hilbert space'' really means that the space of states is $\mathcal H\otimes\mathrm{Cl}(1)$ instead of just $\mathcal H$.}

As we compactify higher-dimensional systems on tori, we will use this characterization to recognize the image of various anomalies.\footnote{An elegant instance of this general idea is Witten's $SU(2)$ anomaly \cite{WITTEN1982324}, which is described by a cobordism class $\eta\wedge c_2(F)$ \cite{Wan:2019fxh}, that when integrated over a four sphere with background gauge fields with minimal instanton number, yields the SPT class $\eta$ in $0+1d$ with no symmetries, which describes the $\psi$-phase (see table~\ref{tablespt}). Therefore the $SU(2)$ global anomaly is detected as an anomaly in $(-1)^F$ due to a fermion zero mode in the instanton background and arises from the $\psi$-layer.}

Let us explain our approach to detecting anomalies in the celebrated example of topological superconductors. Consider the anomalies of $2+1d$ fermionic systems with time-reversal symmetry $\mathsf T$ obeying $\mathsf T^2=(-1)^F$. The symmetry group is $G_f=\mathbb Z_4^\mathsf T$, with $G=G_f/\mathbb Z_2^F=\mathbb Z_2^\mathsf T$ and the symmetry is twisted by the nontrivial $\mathbb Z_2$ classes $w_1$ and $w_2$ in $H^1(\mathbb Z_2^\mathsf T,\mathbb Z_2)$ and $H^2(\mathbb Z_2^\mathsf T,\mathbb Z^F_2)$. The anomalies are classified by $\Omega_\text{spin}^4(\mathbb Z_2,1,1)=\Omega_{\text{pin}^+}^4=\mathbb Z_{16}$ \cite{metlitski2014interaction,Wang_2014,Witten:2015aba,Kapustin:2014dxa}. By anomaly inflow, this is the same as the classification of topological superconductors in $3+1d$. The anomalies are constructed from the following layers
\begin{equation}\label{eq:supercond}
\begin{alignedat}{2}
\nu_{1}&\in H^{1}(\mathbb Z_2^\mathsf T,\mathbb Z)\simeq \mathbb Z_2&&\text{$p_x+ip_y$ layer} \\
\nu_{2}&\in H^{2}(\mathbb Z_2^\mathsf T,\mathbb Z_2)\simeq \mathbb Z_2&&\text{Arf layer}\\
\nu_{3}&\in H^{3}(\mathbb Z_2^\mathsf T,\mathbb Z_2))\simeq \mathbb Z_2&&\text{$\psi$ layer}\\
\nu_{4}&\in H^{4}(\mathbb Z_2^\mathsf T,U(1))\simeq \mathbb Z_2&\qquad&\text{Bosonic layer}
\end{alignedat}
\end{equation}
These four groups compile into $\mathbb Z_{16}$, corresponding to the binary expansion
\begin{equation}
\nu=\nu_1+2\nu_2+4\nu_3+8\nu_4\quad\mod16\,.
\end{equation}
 
Anomalies $\nu\in \mathbb Z_{16}$ can be detected by studying the Hilbert spaces $\mathcal H_{XY}$ of the theory on the two-torus $T^2$, which depend on the choice of spin structure on $T^2$, where $X,Y\in\{\text{NS},\text R \}$. This gives rise to the Hilbert spaces associated to even spin structures $\mathcal H_\text{NS-NS}, \mathcal H_\text{NS-R},\mathcal H_\text{R-NS}$, and to the odd spin structure $\mathcal H_\text{R-R}$. As explained above, anomalies can only appear in $\mathcal H_\text{R-R}$, as the other three spin structures are bounding.

The following anomalies can be detected on the Hilbert space, as we show in both the study of spin TQFTs and fermions in $2+1d$:
\begin{itemize}
 \item \underline{$\nu$ odd:} In $\mathcal H_\text{R-R}$ the classically $(-1)^F$-even time-reversal symmetry generator $\mathsf T$ becomes $(-1)^F$-odd, thus changing the parity of the states in $\mathcal H_\text{R-R}$. 
 This corresponds to $\mathsf T$ anticommuting with $(-1)^F$ instead of commuting in $\mathcal H_\text{R-R}$:
\begin{equation}
\{\mathsf T,(-1)^F\}=0\,.
\end{equation}
This anomalous behaviour is associated with the $p_x+ip_y$ layer in~\eqref{eq:supercond}. For $\nu$ even $[\mathsf T,(-1)^F]=0$ on $\mathcal H_{XY}$.

\item \underline{$\nu=2\mod4$:} The $\mathbb Z_4^\mathsf T$ symmetry algebra on $\mathcal H_{XY}$ is 
\begin{equation}
\mathsf T^2=(-1)^F\times(-1)^{\Arf(T^2)}\qquad\text{on}\qquad\mathcal H_{XY}\,.
\end{equation}
The symmetry algebra is undeformed on the Hilbert spaces with even spin structure and deformed in the Hilbert space with odd spin structure. This anomalous behaviour is associated with the Arf layer in~\eqref{eq:supercond}. For $\nu=0\mod4$, $\mathsf T^2=(-1)^F$ on $\mathcal H_{XY}$.

\item The next two layers $\nu_3$ and $\nu_4$, corresponding to $\nu=4\mod8$ and $\nu=8\mod16$, are not visible on the torus Hilbert space and require other observables to detect them.
\end{itemize}

The analysis of anomalies for time-reversal symmetry $\mathsf T^2=(-1)^F$ in the Hilbert space of spin TQFTs~\cite{Dijkgraaf:1989pz} requires constructing $\mathcal H_{XY}$ in the first place, and also learning how to compute the action of the operators (Wilson lines) on $\mathcal H_{XY}$. We explain how to do this for arbitrary spin TQFTs. We show that the Hilbert spaces $\mathcal H_{XY}$ and the action of $(-1)^F$, the matrix elements of operators and the spin modular data can be unambiguously constructed from the data of a suitable bosonic shadow/parent TQFT (encapsulated in a unitary modular tensor category). This involves some novel ingredients which require, for example, the use of some $F$-symbols of the bosonic TQFT. This construction not only allows us to study the anomalies of time-reversal symmetry, but it is interesting in its own right and can be read on its own.

Another interesting example where the anomaly layers can be detected on the Hilbert space is in $1+1d$ fermionic systems with a unitary $\mathbb Z_2$ symmetry. The overall symmetry is $G_f=\mathbb Z_2\times \mathbb Z^F_2$ and the anomalies are classified by $\Omega_\text{spin}^3(\mathbb Z_2,0,0)=\mathbb Z_8$ \cite{Gu:2013azn,Kapustin:2014dxa,Wang:2016lix,Kapustin:2017jrc}, constructed from the layers 
\begin{equation}\label{eq:boso2}
\begin{alignedat}{2}
\nu_{1}&\in H^{1}(\mathbb Z_2,\mathbb Z_2)\simeq \mathbb Z_2&&\text{Arf layer}\\
\nu_{2}&\in H^{2}(\mathbb Z_2,\mathbb Z_2)\simeq \mathbb Z_2&&\text{$\psi$ layer}\\
\nu_{3}&\in H^{3}(\mathbb Z_2,U(1))\simeq \mathbb Z_2&\qquad&\text{Bosonic layer}
\end{alignedat}
\end{equation}
These three groups compile into $\mathbb Z_{8}$, corresponding to the binary expansion
\begin{equation}
\nu=\nu_1+2\nu_2+4\nu_3\quad\mod8\,.
\end{equation}

The simplest example of a $1+1d$ theory with symmetry $\mathbb Z_2\times \mathbb Z^F_2$ that realizes the $\nu\in \mathbb Z_{8}$ anomaly is a system of $\nu$ Majorana fermions. The generator of $\mathbb Z_2$ is the chiral symmetry $g=(-1)^{F_{L}}$, which acts trivially on the right-moving fermions and negates the left-moving fermions.

The anomaly $\nu\in\mathbb Z_8$ can be detected by studying the untwisted Hilbert space $\mathcal H_{X}$ and the $\mathbb Z_2$-twisted Hilbert space $\mathcal H^g_{X}$ of the $\nu$ Majorana fermions, where $X\in\{\text{NS},\text R\}$ labels the spin structure on the (spatial) circle. We observe the following pattern:
\begin{itemize}
\item \underline{$\nu$ odd:} The theory does not have proper graded twisted Hilbert spaces $\mathcal H^g_\text{NS}$ and $\mathcal H^g_\text R$. %While the $\mathbb Z_2$-twisted partition function is well defined, it does not lead to a Hilbert space interpretation, with an integral number of states at each level.

Also, while the untwisted Hilbert spaces $\mathcal H_{X}$ are well-defined, $(-1)^{F_L}$ and $(-1)^F$ do not commute on $\mathcal H_\text R$
\begin{equation}
\{(-1)^{F_L},(-1)^F\}=0\qquad \hbox{on}~ \mathcal H_R\,.
\end{equation}
For $\nu$ even $\mathcal H_{X}$ and $\mathcal H^g_{X}$ are properly graded and $[(-1)^{F_L},(-1)^F]=0$.

\item The $\nu=2\mod 4$ and $\nu=4\mod 8$ layers are not visible on the Hilbert space as an anomalous realization of symmetry or a projective representation. Indeed reducing the $H^3$ class in~\eqref{eq:boso2} on the circle produces a trivial class in $H^2(\mathbb Z_2,U(1))$, signaling that there are no nontrivial projective representations of $\mathbb Z_2$ in $\mathcal H_{X}$ or $\mathcal H^g_{X}$. We note, however, that the anomaly can be detected by measuring the spin of states in the twisted Hilbert spaces $\mathcal H^g_{X}$ (see e.g.~\cite{Lin:2019kpn} for a similar discussion for bosonic systems). In an anomalous theory this spin has a fractional part, which means that the rotation symmetry is realized projectively.
 
\end{itemize}

An interesting application of these results is the following. As explained above, some anomalies imply that the symmetry generator is fermion-odd in the Hilbert space $\mathcal H$ with the appropriate (non-bounding) structure: the operator that implements the symmetry anticommutes with $(-1)^F$ instead of commuting. This immediately implies that the spectrum of the theory is supersymmetric, namely for any state in $\mathcal H$ there is a partner with the same energy and with opposite fermion parity. This property of the theory is rather surprising: the bose-fermi degeneracy is a consequence of an anomaly instead of a conventional supersymmetry. This provides a unified perspective on several observations in the literature, and it leads to generalizations and new predictions:
\begin{itemize}
\item Any $0+1d$ theory with an antiunitary $\mathbb Z_2^{\mathsf T}$ symmetry and an odd number of Dirac fermions has exact bose-fermi degeneracy. The supersymmetric spectrum of an odd number of free Dirac fermions was described in~\cite{PhysRevLett.102.187001}, and has been studied more recently in~\cite{Prakash:2020wyl,Turzillo:2020mbr,Prakash:2020krs}. From our perspective, any theory with a $\mathbb Z_2^{\mathsf T}$ anomaly $\nu=2\mod 4$ has a supersymmetric spectrum.

\item Any $1+1d$ theory with a unitary chiral $\mathbb Z_2$ symmetry and an odd number of Majorana fermions has exact bose-fermi degeneracy in the untwisted Ramond Hilbert space $\mathcal H_\text R$. This includes the supersymmetric spectrum of $SU(N)$ adjoint QCD with $N$ even in $1+1d$ recently discussed in~\cite{Cherman:2019hbq} (the spectrum is supersymmetric in spite of the fact that Lagrangian of adjoint QCD is not supersymmetric). The fact that that a $\mathbb Z_2$-symmetric theory with an odd number of Majorana fermions has $\{(-1)^{F_L},(-1)^F\}=0$ in $\mathcal H_\text R$ implies that any such a theory will have a supersymmetric spectrum. This includes 
examples in Yang-Mills with $Spin(N)$ gauge group, e.g.~in the fundamental representation for $N$ odd and in the traceless symmetric representation for $N=0,3\mod4$. It would be interesting to exhibit this bose-fermi degeneracy explicitly.

\item Any $2+1d$ theory with antiunitary $\mathbb Z_4^{\mathsf T}$ symmetry and an odd number of Majorana fermions has exact bose-fermi degeneracy in the odd spin structure Hilbert space $\mathcal H_\text{R-R}$. This is a nontrivial prediction for the spectrum of gauge theories in $2+1d$ theories, which are strongly coupled in the infrared. An instance of a theory that should have a supersymmetric spectrum is $SO(N)$ gauge theory (with vanishing Chern-Simons coupling) with a fermion in the traceless symmetric representation. Time-reversal invariance requires that $N$ is even and $\nu$ odd further requires that $N=0\mod 4$. While the Lagrangian of this theory is not supersymmetric, the anomaly implies that the spectrum is nonetheless supersymmetric. We can provide nontrivial evidence for this claim. In~\cite{Gomis:2017ixy} the infrared dynamics of this theory was proposed to be captured by $SO(\frac{N+2}{2})_\frac{N+2}{2}$ Chern-Simons theory. Using the formulae in~\cite{Delmastro:2020dkz} for the number\footnote{In~\cite{Delmastro:2020dkz} it was shown that $SO(2n+1)_{2k+1}$ Chern-Simons theory has $\binom{n+k-1}{k-1}$ bosons and $\binom{n+k-1}{k}$ fermions in the   Hilbert space $\mathcal H_\text{R-R}$. The spectrum is supersymmetric for $n=k$, when the theory is time-reversal invariant.} of bosonic and fermionic states in $\mathcal H_\text{R-R}$ of this Chern-Simons theory we find that the spectrum is indeed supersymmetric! Our argument applies to other gauge theories with higher rank real representations and is a nontrivial prediction of their spectrum.

\end{itemize}

The next layer, measuring the projectivity of the symmetry algebra on $\mathcal H$, also has nontrivial implications, the most famous being Kramers theorem. From our analysis one can conclude that any theory in $2+1d$ with $\mathbb Z_4^{\mathsf T}$ symmetry and anomaly $\nu=2\mod4$ has (at least) two-fold degeneracy in the fermionic part of the even-spin-structure Hilbert spaces, and in the bosonic part of the odd-spin-structure Hilbert space. When $\nu=0\mod4$ there is (at least) two-fold degeneracy for all the fermionic states, in any of the spin structures.

Finally, we should stress that our analysis may not yet capture all the information about anomalies which is encoded in the torus Hilbert spaces. Isometries of the internal space will act on the Hilbert space of a compactified theory. As a result, one could study anomalies for the combination of the original symmetries and the new internal symmetries of the compactified system. We leave this to future work. 

The plan for the rest of the paper is as follows. In section~\ref{sec:free_fermion_hilbert} we study free fermions in various dimensions and illustrate how anomalies manifest themselves at the level of their Hilbert space. We consider antiunitary time-reversal symmetry in $0+1$ and $2+1$ dimensions, where the algebra is $\mathsf T^2=1$ and $\mathsf T^2=(-1)^F$, respectively; and we also consider unitary chiral symmetry in $1+1$ dimensions with algebra $g^2=1$. After that, in section~\ref{sec:t_reversal} we consider the same problem in $2+1d$ spin TQFTs. We study how their anomalies are seen by constructing their Hilbert spaces. Here we revisit the algebra $\mathsf T^2=(-1)^F$ and find the same behaviour as in the case of free fermions. In appendix~\ref{sec:condensation} we describe how to construct the Hilbert spaces of fermionic TQFTs, that is TQFTs that depend on the spin structure. We demonstrate this framework through examples in appendix~\ref{sec:examples_spin_TQFT}, where we work out in some detail the construction of the Hilbert space for several interesting TQFTs. This last appendix also includes a few remarks about a more exotic time-reversal symmetry with algebra $\mathsf T^2=\mathsf C$, where $\mathsf C$ denotes charge conjugation, a unitary $\mathbb Z_2$ symmetry.

\section{Anomalies from Layers}\label{sec:layer}
The classification of SPT phases in terms of generalized cohomology/cobordism and associated ``layers'' is somewhat forbidding, but has a rather transparent physical meaning. Ultimately, we want to have a procedure to associate a partition function to a manifold equipped with appropriate structures. First, we can triangulate the manifold, equipping it with a discretization of the various structures we want to endow it with: a flat connection along the edges of the triangulation, some discrete version of the spin structure and orientation, etc. Next, we 
can take the cell decomposition $C$ dual to the triangulation, and place on the facets of $C$ some collection of invertible TFTs (meaning here SPTs with no symmetries) of appropriate dimension, following some rules which take into account the discrete data we put on the manifold. The partition function is then defined as the partition function of the collection of 
invertible TFTs. 

The ``layers'' of the cohomology theory are simply a way to encode which rules we use to place invertible TFTs on facets. The differential in the generalized cohomology theory imposes the constraint that the final answer should be independent of the choice of triangulation as well as any other choices made at intermediate steps of the construction. It also identifies pairs of rules which give the same final answer.

As an example, consider orientable, spin SPTs for unitary symmetries. A discretized $G$ flat connection is given as a collection of $G$ elements on the edges of the triangulation.
\begin{enumerate}
\item If we were to include only the bottom layer, we would leave all facets bare and only focus on vertices of $C$. At each vertex we place some complex phase (aka elements of $U(1)$) determined by the group elements along the edges of the dual simplex. This is literally the cochain $\nu_{D+1}$ representing an element in the group cohomology $H^{D+1}(G,U(1))$. The cocycle condition ensures that the partition function defined as the product of all the phases is independent of the choice of triangulation and gauge. Coboundaries give partition functions which evaluate to $1$ in a trivial manner.
\item Following \cite{Gu:2012ib}, the next refinement of the story involves placing a fermionic one-dimensional Hilbert space along some of the edges of $C$. 
The choice is the cochain $\nu_{D}$ representing an element in the group cohomology $H^{D}(G,\mathbb Z_2)$. 
The cocycle condition ensures that each vertex is connected to an even number of fermionic edges. At each vertex we now get to pick a vector in the (one-dimensional, Grassmann even) tensor product of these vector spaces. This is roughly the same as a choice of $\nu_{D+1}$, but not canonically, because of sign ambiguities in the tensor product. The Grassmann combinatorics needed to rearrange the tensor products when contracting states at the endpoints of 
fermionic edges, as well as the (spin structure dependent) signs arising from fermion loops contribute to the overall sign of the partition function.
\item At the next level of refinement, we can place Arf theories on some two-dimensional facets according to some $\nu_{D-1}$. The cocycle condition ensures that we have an even number of Arf facets impinging on an edge, but the edge must now carry a specific choice of how to gap the corresponding Majorana modes. The two possible choices have opposite Grassmann parity, so this choice is similar but not canonically equivalent to a choice of $\nu_{D}$, etc. The evaluation of the partition function will now require a careful manipulation of the Majorana modes. 
\item Next, we can place $SO(n)_{\pm 1}$ Chern-Simon theories on three-dimensional facets according to some $\nu_{D-2}$. The cocycle condition insures that we have the same number of chiral and anti-chiral fermions at two-dimensional facets, but facets must now carry a specific choice of how to gap these $2d$ fermions. Two inequivalent choices differ by a factor of Arf. The evaluation of the partition function must now cope with this extra level of complication. 
\item In principle, we could continue, selecting some invertible fermionic theories to place on four-dimensional facets, etc. In practice, no 
non-trivial invertible theories are expected to exist up to dimension $7$, so we can safely stop here for most physical systems. 
\end{enumerate} 

On general grounds, the differential in the generalized cohomology theory takes a triangular form, with the diagonal being the standard differential for $H^{D+1-k}(G,T_k)$, where $T_k$ is the group of invertible theories we can place on the $k$-th dimensional facets. The off-diagonal components of the differential are non-trivial and somewhat tricky to compute far from the diagonal. Furthermore, the ``stacking'' operation on generalized cohomology classes, i.e.~the sum of anomalies, is also defined in a triangular manner, with the diagonal being the usual operation of stacking invertible theories. 

As one compactifies an SPT on, say, a circle, one can take a triangulation of the $D$-dimensional manifold $M$ and refine it to 
a triangulation of $M \times S^1$ in a systematic way. Applying the rules above to $M \times S^1$ and reducing them to some evaluation on the triangulation of $M$ one can figure out the resulting SPT theory in one dimension lower. This was done for the Gu-Wen layer in \cite{Tantivasadakarn:2017xbg}, but 
has not been done in full generality. 

\section{Anomalies in free fermion Hilbert space}\label{sec:free_fermion_hilbert}

In this section we demonstrate how the Hilbert space on the torus detects a variety of anomalies in systems of free fermions in various dimensions. 

\subsection{Anomalous $\mathbb Z^{\mathsf T}_2$ in 0+1 dimensions}\label{sec:timeone}

The anomalies of a fermionic system in 0+1 dimensions with an antiunitary time-reversal symmetry $\mathsf T$ with $\mathsf T^2=1$, so that $G_f=\mathbb Z^\mathsf T_2\times \mathbb Z_2^F$, are classified by $\Omega_\text{spin}^2(\mathbb Z_2;1,0)=\Omega^2_{\text{pin}^-}=\mathbb Z_8$. These anomalies arise from three layers 
\begin{equation}
\begin{aligned}
\nu_{0}&\in H^{0}(\mathbb Z^\mathsf T_2,\mathbb Z_2)\simeq \mathbb Z_2\,\\
\nu_{1}&\in H^{1}(\mathbb Z^\mathsf T_2,\mathbb Z_2)\simeq \mathbb Z_2\, \\
\nu_{2}&\in H^{2}(\mathbb Z^\mathsf T_2,U(1))\simeq \mathbb Z_2\,,
\end{aligned}
\label{eq:system1d}
\end{equation}
which generate the $\mathbb Z_8$ anomaly. 
%\note{A representative of the class $\nu\in\mathbb Z_8$ is given by the system of $\nu$ free Majorana fermions.}

We shall study the $\mathbb Z_8$ anomaly in a system of free fermions. Related considerations can be found in~\cite{Kaidi:2019tyf,Stanford:2019vob,Turzillo:2017wjl}.
 
Consider $\nu$ Majorana fermions in $0+1$ dimensions
\begin{equation}
{\mathcal L}= \sum_{a=1}^\nu\ \frac{i}{4}\psi^a \partial_t\psi^a\,.
\label{eq:fermionsone}
\end{equation}
The theory has a $\mathbb Z^{\mathsf T}_2$ time-reversal symmetry which acts as\footnote{For the purposes of studying anomalies it suffices to take all fermions to transform with the same sign under $\mathsf T$. If a fermion is assigned the transformation $\mathsf T \psi_-(t)=- \psi_-(-t)\mathsf T$, we can then write a $\mathbb Z^{\mathsf T}_2$-invariant mass term $i\psi_+\psi_-$ that couples a pair of fermions which transform with opposite signs under $\mathsf T$. This
lifts both fermions and therefore without loss of generality we can focus on a collection of fermions that transform with the same sign under $\mathsf T$.}
\begin{equation}
\mathsf T \psi^a(t)= \psi^a(-t)\mathsf T
\label{eq:timeoned}
\end{equation}
and $\mathbb Z_2^F$ fermion parity
\begin{equation}
\{(-1)^F,\psi^a(t)\}=0\,.
\end{equation}
It is known that a $\mathbb Z^\mathsf T_2$-symmetric quartic interaction that gaps out the fermions can be written for $\nu=8$~\cite{Fidkowski:2009dba}. This realizes in the fermion system the $\mathbb Z_8$ anomaly expected from the cobordism classification.

Canonical quantization of~\eqref{eq:fermionsone} leads to a Clifford algebra of rank $\nu$
\begin{equation}
 \{\psi^a,\psi^b\}= 2\delta^{ab}\qquad a,b=1,2,\ldots,\nu\,.
\end{equation}
We now proceed to identifying the anomaly layers~\eqref{eq:system1d}. Each layer is implemented in a characteristic way in the fashion that symmetries are realized on the Hilbert space $\mathcal H$.

\begin{itemize}
\item $ \nu$ odd:

\noindent 
There is a rather severe anomaly for $\nu$ odd as the operator $(-1)^F$ generating the $\mathbb Z_2^F$ symmetry does not exist. The theory does not admit a proper graded Hilbert space of states. Equivalently stated, the Clifford algebra of odd rank has two irreducible representations, and $(-1)^F$ exchanges them, instead of acting within an irreducible representation. This anomaly is associated with the $H^0(\mathbb Z_2^\mathsf T,\mathbb Z_2)=\mathbb Z_2$ layer, the Arf layer. 
\smallbreak

This anomaly due to the lack of proper Hilbert space can also be detected by studying partition functions. Consider the partition function on the circle with antiperiodic (NS) and periodic (R) boundary conditions. The partition function with NS boundary conditions is\footnote{This partition function can be evaluated by taking the square root of partition function of $2\nu$ Majorana fermions, which has a $2^\nu$-dimensional Hilbert space. It can also be computed by zeta-regularizing $Z\equiv\text{Pf}(i\partial_t)^\nu=\prod_{n\in\mathbb Z} \lambda^\nu_n$, where the eigenvalues of the $0+1d$ Dirac operator are $\lambda_n=n+1/2$ in the NS sector, and $\lambda_n=n$ in the R sector.}
\begin{equation}
Z_\text{NS}=2^{\,\nu/2}\,.
\label{eq:Z_NS_1d}
\end{equation}
Nominally, this partition function should count the number of states in $\mathcal H$, that is $Z_\text{NS}=\tr_\mathcal H(\boldsymbol1)=\dim(\mathcal H)$. The answer~\eqref{eq:Z_NS_1d} mirrors the statement that there is no proper Hilbert space for $\nu$ odd, as $2^{\nu/2}$ is not an integer. Likewise, while the partition function with R boundary conditions vanishes due to the presence of zero-modes, i.e.~$Z_\text{R}=\tr_\mathcal H(-1)^F=0$, the correlator
\begin{equation}
\langle \psi^1\psi^2\cdots\psi^\nu\rangle_\text{R}
\label{eq:corr_R_1d}
\end{equation}
is non-vanishing, as the insertions compensate the zero-modes. This observable~\eqref{eq:corr_R_1d} changes sign under the action of $(-1)^F$, signaling that $(-1)^F$ is anomalous as the $\mathbb Z_2^F$ Ward identities are violated.

\item $ \nu=2\mod 4$. For $\nu$ even the theory has a well-defined Hilbert space and operator $(-1)^F$ acting on it. The Clifford algebra of even rank $\nu$ has a unique irreducible representation of dimension $2^{\nu/2}$, thus all representations are unitarily equivalent, and we can study the implementation of symmetries in any choice of basis. We can construct $\mathcal H$ by defining the creation and annihilation operators 
\begin{equation}
\psi_{\pm}^A=\frac{1}{2}\big(\psi^{2A-1}\pm i\psi^{2A}\big) \qquad A=1,\ldots, \nu/2\,,
\end{equation}
 which obey
\begin{equation}
 \{\psi_+^A, \psi_-^B\}=\delta^{AB}\,,~ \{\psi_+^A, \psi_+^B\}= \{\psi_-^A, \psi_-^B\}=0\qquad A,B=1,\ldots, \nu/2\,,
\label{eq:cliffone}
\end{equation}
We define the vacuum by 
\begin{equation}
\psi_-^A|0\rangle=0\qquad A=1,\ldots, \nu/2\,,
\label{eq:vacone}
\end{equation}
and create the whole module by acting with the different $\psi_+^A$ on it. Time-reversal acts by exchanging the creation and annihilation operators (see~\eqref{eq:timeoned} and recall that $\mathsf T$ is antilinear)
\begin{equation}
 \mathsf T\psi_\pm^A=\psi_\mp^A\mathsf T\,.
\label{eq:antic}
\end{equation}
 This allows us to determine the action of $\mathsf T$ on the vacuum $|0\rangle$ by considering the most general state 
\begin{equation}
 \mathsf T |0\rangle= \alpha |0\rangle +\alpha_A\psi_+^A |0\rangle+\cdots +\alpha_{12\ldots \nu/2} \psi_+^1\psi_+^2\cdots\psi_+^{\nu/2} |0\rangle\,,
\label{eq:actwithTone}
\end{equation}
for some yet-to-be-fixed coefficients $\{\alpha\}$. Acting on both sides with $\psi^A_-$ and using~\eqref{eq:cliffone},~\eqref{eq:vacone} and~\eqref{eq:antic} we conclude that all but the last coefficient vanish, namely\footnote{This corresponds to what is usually referred to as \emph{particle-hole symmetry}: time-reversal exchanges $\psi_+$ and $\psi_-$, so a state full of $\psi_+$ is mapped to a state full of $\psi_-$, and vice-versa.}
\begin{equation}
\mathsf T |0\rangle= \alpha_{12\ldots \nu/2} \psi_+^1\psi_+^2\cdots\psi_+^{\nu/2} |0\rangle\,,
\end{equation}
with $|\alpha_{12\ldots \nu/2}|=1$ so that state is normalized. Note that $\mathsf T$ adds $\nu/2$ fermionic modes, so it changes the fermion parity of the state if $\nu/2$ is odd. This implies that the $\mathbb Z^\mathsf T_2\times \mathbb Z_2^F$ symmetry generators on $\mathcal H$ obey
\begin{equation}
\{\mathsf T,(-1)^F\}=0\qquad\text{for $\nu=2 \mod 4$}\,.
\end{equation}
 and 
\begin{equation}
[\mathsf T,(-1)^F]=0\qquad\text{for $\nu=0 \mod 4$}\,,
\end{equation}
Therefore, the anomaly corresponding to the $\nu= 2 \mod 4$ layer is detected by virtue of the operators $\mathsf T$ and $(-1)^F$ anticommuting in $\mathcal H$. This anomaly is associated with the $H^1(\mathbb Z_2^\mathsf T ,\mathbb Z_2)=\mathbb Z_2$ layer, the $\psi$-layer.

\item $\nu=4\mod8$. For $\nu=4\mod8$ the theory has a proper Hilbert space and $[\mathsf T,(-1)^F]=0$. We now proceed to study how the $\mathbb Z_2^{\mathsf T}$ symmetry is realized on the Hilbert space. Acting with $\mathsf T$ again in~\eqref{eq:actwithTone} yields 
\begin{equation}
\mathsf T^2|0\rangle= |\alpha_{12\ldots \nu/2}|^2\psi_-^1\psi_-^2\cdots\psi_-^{\nu/2} \psi_+^1\psi_+^2\cdots\psi_+^{\nu/2}|0\rangle=(-1)^{\nu/4(\nu/2-1)}|0\rangle\,.
\end{equation}
Therefore, for $\nu=4\mod8$ the $\mathbb Z_2^{\mathsf T}$ symmetry is realized projectively on the Hilbert space, that is
\begin{equation}
\mathsf T^2=-1\,.
\end{equation}
Therefore, the anomaly corresponding to the $\nu= 4 \mod 8$ layer is detected by virtue of the $\mathbb Z_2^{\mathsf T}$ symmetry being realized projectively on the Hilbert space.\footnote{We note that $\mathsf T^2=-1$
for $\nu=4, 6\mod 8$ and $\mathsf T^2=1$ for $\nu=0,2 \mod 8$. This follows from the properties of the antilinear involution acting on the Clifford algebra as $\mathsf T^{-1} \gamma^a \mathsf T=\gamma^a$, or equivalently $U\gamma^a U^{-1}=(\gamma^a)^*$, where we have written $\mathsf T=UK$, with $K$ denoting complex conjugation and $U$ a unitary. Therefore, $\mathsf T^2=U^*U=\pm 1$ with $\mathsf T^2=1$ corresponding to a real involution and $\mathsf T^2=-1$ to a pseudoreal/quaternionic involution. Using the explicit form of the gamma matrices we constructed we find the signs as discussed, implying that for $\nu=0,2\mod 8$ and $\nu=4,6\mod 8$ the Clifford algebra admits a real and quaternionic involution respectively.} 
This anomaly is associated with the $H^2(\mathbb Z_2^\mathsf T,U(1))=\mathbb Z_2$ layer, the bosonic layer.

\end{itemize}

\paragraph{SYK model.} To close this section we can consider a very simple application of these results. One of the most well-known systems in $0+1d$ is the celebrated SYK model~\cite{KitaevSYK,PhysRevLett.70.3339}, which consists of a system of $N$ Majorana fermions interacting via four-fermi terms
\begin{equation}
\mathcal L=\sum_a \frac i4\psi^a\partial_t\psi^a-\sum_{abcd}J_{abcd}\psi^a\psi^b\psi^c\psi^d\,,
\end{equation}
where the coupling constants $J$ are real. This Lagrangian is invariant under $\mathsf T(\psi^i(t))=\psi^i(-t)$, and therefore all the conclusions from our previous discussion hold. The time-reversal anomaly of the system is $\nu=N\mod8$. We immediately conclude that,
\begin{itemize}
\item If $N$ is odd, the SYK model does not admit a satisfactory ($\mathbb Z_2$-graded) Hilbert space.
\item If $N$ is even, $N=2\mod4$, then $\mathsf T$ is fermion-odd, and therefore the spectrum of the Hamiltonian is (at least) two-fold degenerate, with energy doublets having opposite fermion parity (the Hilbert space is supersymmetric).
\item If $N$ is even, $N=4\mod 8$, then $\mathsf T$ squares to $-1$, and therefore the spectrum of the Hamiltonian is (at least) two-fold degenerate, with energy doublets having the same fermion parity (they are Kramers doublets).
\item If $N$ is even, $N=0\mod8$, the symmetry is non-anomalous and we cannot conclude anything about the spectrum of the Hamiltonian. Unless we tune the coefficients $J$ to have some special symmetry, we do not expect any degeneracy in the Hilbert space.
\end{itemize}

It is a rather entertaining exercise to explicitly check these claims by numerically diagonalizing the SYK Hamiltonian for small values of $N$. We also note that similar ideas can be found in e.g.~\cite{You_2017,PhysRevLett.124.236804}.

%PhysRevLett.124.236804

%\note{Perhaps we should stress that $\mathsf T^2=-1$ and $\mathsf T^2=(-1)^F$ are not the same thing. Sometimes people use the notation $\mathsf T^2=-1$ when they actually mean $\mathsf T^2=(-1)^F$.}
%
%{\bf Feel free to add a comment you like}

\subsection{Anomalous $\mathbb Z^{\mathsf T}_4$ in 2+1 dimensions}\label{sec:T2_F_3d}

The anomalies of a fermionic system in $2+1d$ with an antiunitary time-reversal symmetry $\mathsf T^2=(-1)^F$ are classified by $\Omega^4(\mathbb Z_2;1,1)=\Omega^4_{\text{pin}^+}=\mathbb Z_{16}$. These anomalies arise from four layers~\eqref{eq:supercond}
\begin{equation}
\begin{aligned}
\nu_{1}&\in H^{1}(\mathbb Z^\mathsf T_2,\mathbb Z)\simeq \mathbb Z_2\\
\nu_{2}&\in H^{2}(\mathbb Z^\mathsf T_2,\mathbb Z_2)\simeq \mathbb Z_2 \\
\nu_{3}&\in H^{3}(\mathbb Z^\mathsf T_2,\mathbb Z_2))\simeq \mathbb Z_2\\
\nu_{4}&\in H^{4}(\mathbb Z^\mathsf T_2,U(1))\simeq \mathbb Z_2\,,
\end{aligned}
\end{equation}
which compile into $\mathbb Z_{16}$. 

In this section we study this anomaly in a system of free Majorana fermions, and in section~\ref{sec:t_reversal} we will do the same in time-reversal symmetric spin TQFTs.

Consider a system of $\nu$ Majorana fermions $\psi$. We shall work in the Majorana basis where the gamma matrices are real,
\begin{equation}
\gamma^0=i\sigma^2,\qquad \gamma^1=\sigma^1,\qquad\gamma^2=\sigma^3\,.
\end{equation}
In this basis the Majorana condition is simply $\psi^*=\psi$ so $\psi$ is a real two-component Grassmann-odd spinor. We can without loss of generality take time-reversal to act as
\begin{equation}
\mathsf T(\psi(t))=\pm\gamma^0\psi(-t)\,.
\end{equation}
Given a pair of fermions transforming with opposite signs, we can write down a $\mathsf T$-invariant mass term, which means that such a pair does not contribute to anomalies. Therefore, as far as anomalies is concerned, we can take all fermions to transform with the same sign, say $+1$. It is known that a $\mathsf T$-invariant interaction exists with $16$ fermions that lifts all of them~\cite{fidkowski2013nonabelian,Wang_2014,metlitski2014interaction,You_2014,KitaevZ16,Witten:2015aba,Tachikawa:2016xvs}.

We now construct the torus Hilbert space of the system and study how the time-reversal anomaly manifests itself on it. A subtle but important difference in $2+1d$ as opposed to the examples in $0+1d$ and $1+1d$ is that $\Omega^\text{spin}_4=\mathbb Z$ contains a free part: in $2+1d$ there exists a purely gravitational SPT. This invertible theory is intertwined with time-reversal in an interesting way, which we review next. The generator of SPTs with no symmetry in $2+1d$ is given by the spin TQFT denoted by $SO(1)_1$, corresponding to the super Ising category~\cite{Tachikawa:2016cha,Seiberg:2016rsg,Aharony:2016jvv,Cordova:2017vab}. The partition function of this theory is $e^{-i\mathrm{CS}_\text{grav}}$, where locally
\begin{equation}
\mathrm{CS}_\text{grav}=\frac{1}{4\pi}\int_{M_3}\text{tr}\left(\omega\,\mathrm d\omega+\frac23\omega^3\right)\,,
\end{equation}
where $\omega$ is the spin connection for the gravitational background of $M_3$. An arbitrary SPT with no symmetry is given by a number $n\in\mathbb Z$ of copies of the generator, namely $SO(n)_1:=SO(1)_1^n$, whose partition function is $e^{-in\mathrm{CS}_\text{grav}}$. As a spin TQFT, $SO(n)_1$ can be obtained by condensing a certain fermion in the bosonic TQFT $Spin(n)_1$, that is by gauging a certain $\mathbb Z_2$ one-form symmetry (see section~\ref{sec:t_reversal}). Note that the Chern-Simons form $\mathrm{CS}_\text{grav}$ is a volume form, so it is odd under time-reversal.

If the manifold is non-trivial, the fermions automatically couple to the Chern-Simons term for the background gravitational field, because the Dirac operator contains a piece proportional to the spin connection. In $2+1d$ time-reversal acts both on the fermions and on the Chern-Simons interactions, and the combined system is only time-reversal invariant if the coefficient of the latter is adjusted appropriately. This behaviour should be thought of as a mixed time-reversal-gravitational anomaly, and it can be ascribed to a controlled non-invariance of the fermion path-integral measure $[D\psi]$. This non-invariance is a topological phase, the eta invariant $\eta$, and we can summarize the anomaly as the statement that each massless Majorana fermion $\psi$ transforms as
\begin{equation}
\mathsf T\colon [D\psi]\mapsto e^{-i\pi\eta/2}[D\psi]\,.
\end{equation}
In absence of other background fields, the eta invariant is precisely the gravitational Chern-Simons term,
\begin{equation}
\frac12\pi\eta=\mathrm{CS}_\text{grav}\mod 2\pi \mathbb Z\,.
\end{equation}
In this sense, time-reversal does not map the QFT of a single massless fermion into itself, but rather into itself tensored with a copy of the SPT $SO(1)_1$; schematically
\begin{equation}\label{eq:T_psi_SPT}
\mathsf T\big( \text{massless }\psi\big)=\text{massless }\psi\times SO(1)_1\,.
\end{equation}

In order to compensate for the anomalous phase $e^{-i\pi\eta/2}$, we formally need to attach to each massless Majorana fermion a copy of $\frac12\mathrm{CS}_\text{grav}$, i.e., to a copy of a ``square root'' of $SO(1)_1$. The combined object $e^{\frac12\mathrm{CS}_\text{grav}}[D\psi]$ is now time-reversal invariant. In the notation of~\eqref{eq:T_psi_SPT}, we formally need to move ``half'' of $SO(1)_1$ to the left, so as to have $\mathsf T$ mapping a QFT into itself instead of into a second QFT.

The discussion above is equivalent to the statement that a massless Majorana fermion carries chiral central charge $c=1/4$ (this is also known as the \emph{framing anomaly}~\cite{witten1989}; recall that $c$ measures the coupling of the theory to $\mathrm{CS}_\text{grav}$). As $c$ is odd under time-reversal, a system with $c\neq0$ is not invariant by itself, but must be coupled to a suitable SPT, whose central charge is $-c$, in order to make the total central charge zero. The generator of SPTs $SO(1)_1$ has $c=1/2$, so in order to compensate for the $c=1/4$ of the fermion we formally need to couple it to a square root of $SO(1)_1$. More generally, given an arbitrary number $\nu$ of massless Majorana fermions, the system $\psi^\nu$ is not actually time-reversal invariant, but the combined system $\psi^{\nu}\times SO(\nu/2)_{-1}$ is. Naturally, if the number of fermions $\nu$ is odd, the coefficient of $\mathrm{CS}_\text{grav}$ is not properly normalized, and the system does not make sense as a purely $2+1d$ object: we either give up time-reversal invariance and drop the gravitational counterterm, or we keep the symmetry and regard the system as the boundary of a $3+1d$ theory. For $\nu$ even, we can maintain time-reversal invariance and still have a conventional $2+1d$ theory, but only after coupling the fermions to $SO(\nu/2)_{-1}$. For now, we will consider the $\nu$ fermions alone, and later on we will study the effect of turning on $SO(\nu/2)_{-1}$ for $\nu$ even.

With this in mind, let us go back to studying the system of $2+1d$ $\nu$ massless Majorana fermions on the torus $T^2$. Anomalies, being renormalization-group invariant, always arise in the realization of the symmetry on the low energy states; therefore, in order to detect the anomalies, it suffices to look at the vacuum sector. For even spin structure on $T^2$ there are no zero modes and no anomalies; this agrees with the general discussion in section~\ref{sec:intro} where we argued that anomalies can only be detected on manifolds that do not bound.

For odd spin structure there are zero modes and potential anomalies. Roughly, the system with odd spin structure on $T^2$ behaves as $2\nu$ copies of the $0+1d$ system of Majoranas we analyzed earlier, the factor of $2$ being due to the fact that each $\psi$ has two real components instead of one. In this sense, the analogous to the first layer in $0+1d$ is never activated in $2+1d$, because the number of Majorana components is always even. In other words, the Hilbert space $\mathcal H_{XY}$ of $2+1d$ Majorana fermions is always well-defined, regardless of the parity of the number of fermions. But the other two layers, those measured by the fermion parity of $\mathsf T$ and the sign in $\mathsf T^2=\pm1$, are potentially activated. The first one is measured by the parity of $\nu$, and the second one by the parity of $\nu/2$. We will exhibit the following anomalous behavior in the Hilbert space:
\begin{itemize}
\item $\nu$ odd: In $\mathcal H_\text{R-R}$ time-reversal is fermion-odd, it anticommutes with $(-1)^F$
\begin{equation}
\{\mathsf T,(-1)^F\}=0\,.
\end{equation}
This anomaly is associated to the $p_x+i p_y$ layer $\nu_1$. For $\nu$ even $[\mathsf T,(-1)^F]=0$.

\item $\nu=2\mod 4$: In the even spin structure Hilbert spaces $\mathcal H_\text{NS-NS},\mathcal H_\text{NS-R},\mathcal H_\text{R-NS}$ time-reversal satisfies the standard algebra $\mathsf T=(-1)^F$, but in the odd spin structure Hilbert space $\mathcal H_\text{R-R}$ this algebra is realized projectively, namely $\mathsf T^2=-(-1)^F$. In other words, the time-reversal symmetry on $\mathcal H_{XY}$ satisfies 
\begin{equation}
\mathsf T^2=(-1)^F\times(-1)^{\mathrm{Arf}(T^2)}\,.
\end{equation}
This anomaly is associated to the Arf layer $\nu_2$.
\end{itemize}
The next two layers, $\nu_3,\nu_4$, which measure $\nu\mod 8$ and $\nu\mod16$, respectively, are invisible on the torus Hilbert spaces.

The discussion regarding the first two layers is essentially identical to the $0+1d$ case, so we only highlight the differences. The fermions now depend on both time $t$ and the spatial coordinate $\boldsymbol x$, which we take to coordinatize a torus $T^2$. The Hilbert space associated to this spatial slice is built by acting with the spatial modes on the vacuum sector. If $\mathrm{Arf}(T^2)=0$, then there are no zero-modes, and the vacuum Hilbert space is trivial: there is a unique vacuum state $|0\rangle$. Therefore, here time-reversal acts quite trivially: $\mathsf T$ is fermion-even and satisfies $\mathsf T^2=(-1)^F$ on the nose: neither layer is activated. In order to detect the anomaly we have to look at the non-bounding torus, i.e., where both boundary conditions are periodic such that $\Arf(T^2)=1$. Here there is a single zero-mode for each Majorana fermion, which is spatially constant. In what follows we shall study this vacuum module generated by these zero-modes in $\mathcal H_\text{R-R}$.

First of all, since $\mathsf T(\psi)=\gamma^0\psi$ where $\gamma^0=i\sigma^2$, time-reversal acts on the two components of the Majorana fermion as
\begin{equation}
\mathsf T(\psi^1)=+\psi^2,\qquad \mathsf T(\psi^2)=-\psi^1\,.
\end{equation}
In terms of the complex spinor $\Psi=\frac{1}{\sqrt{2}}(\psi^1+i\psi^2)$ this becomes
\begin{equation}
\mathsf T(\Psi)=i\Psi^*\,.
\end{equation}
The Hilbert space is built by declaring that $\Psi_i|0\rangle=0$ for all $i=1,2,\dots,\nu$, and by repeatedly acting with $\Psi_i^*$ on $|0\rangle$. The action of time-reversal on the whole vacuum Hilbert space is uniquely fixed in terms of its action on $|0\rangle$, which again reads
\begin{equation}
\mathsf T|0\rangle=\Psi_1^*\Psi^*_2\cdots\Psi^*_\nu|0\rangle
\end{equation}
up to an inconsequential phase. We thus see that, indeed, if $\nu$ is odd $\mathsf T$ anticommutes with $(-1)^F$. Now, if we act with $\mathsf T$ twice we get
\begin{equation}
\begin{aligned}
\mathsf T^2|0\rangle&=\mathsf T\Psi_1^*\Psi^*_2\cdots\Psi^*_\nu|0\rangle\\
&=(-i)^\nu\Psi_1\Psi_2\cdots\Psi_\nu\mathsf T|0\rangle\\
&=(-i)^\nu\Psi_1\Psi_2\cdots\Psi_\nu\Psi_1^*\Psi^*_2\cdots\Psi^*_\nu|0\rangle\\
%&=(-i)^\nu(-1)^{\nu(\nu-1)/2}|0\rangle
&=i^{-\nu^2}|0\rangle\,.
\end{aligned}
\end{equation}
%(For $\nu$ odd the sign of $\mathsf T^2$ is ambiguous because had we defined the Hilbert space interchanging $\Psi,\Psi^*$, i.e., by $\Psi^*|0\rangle=0$, then we would have found $\mathsf T^2=i^{+\nu^2}$ instead. So it really only makes sense to ask about the second layer if the first one is trivial, I'd say.)
%(The value of $\mathsf T^2$ was computed only on the state $|0\rangle$, but $\mathsf T^2=(-1)^F$ when acting on $\Psi_i$, which means that the phase above persists for all states in the module.)
When $\nu$ is even we get $\mathsf T^2|0\rangle=+|0\rangle$. More generally, as $\mathsf T^2=(-1)^F$ when acting on the creation operators, the relation $\mathsf T^2|0\rangle=+|0\rangle$ lifts to $\mathsf T^2=(-1)^F$ on the whole Hilbert space.

We now return to the effect of the gravitational SPT $SO(\nu/2)_{-1}$ for $\nu$ even that is needed in order to have a time-reversal symmetric theory. This SPT has a unique state on any spin structure Hilbert space $\mathcal H_{XY}$. The fermion parity of this state is known to be $(-1)^F=(-1)^{\Arf(T^2)\nu/2}$ (see~\cite{Gaiotto:2015zta,Delmastro:2020dkz} and appendix~\ref{app:sos}). Therefore, the $\mathsf T$-invariant combined system $\psi^\nu\times SO(\nu/2)_{-1}$ has a time-reversal algebra
\begin{equation}
\mathsf T^2=(-1)^F\times(-1)^{\Arf(T^2)\nu/2}\,.
\end{equation}
This means that the operator algebra $\mathsf T^2=(-1)^F$ is undeformed for $\nu=0\mod4$, while it gets deformed by the Arf theory for $\nu=2\mod4$ in $\mathcal H_\text{R-R}$, as claimed.

\subsection{Anomalous $\mathbb Z_2$ in 1+1 dimensions}\label{sec:chiral}

The anomalies of a fermionic system in $1+1d$ with a unitary $\mathbb Z_2$ symmetry such that $G_f=\mathbb Z_2\times \mathbb Z_2^F$ are classified by $\Omega_\text{spin}^2(\mathbb Z_2;0,0)=\mathbb Z_8$. These anomalies arise from three layers 
\begin{equation}
\begin{aligned}
\nu_{1}&\in H^{1}(\mathbb Z_2,\mathbb Z_2)\simeq \mathbb Z_2\,\\
\nu_{2}&\in H^{2}(\mathbb Z_2,\mathbb Z_2)\simeq \mathbb Z_2\, \\
\nu_{3}&\in H^{3}(\mathbb Z_2,U(1))\simeq \mathbb Z_2\,,
\end{aligned}
\label{eq:system2d}
\end{equation}
which generate the $\mathbb Z_8$ anomaly.

Consider $\nu$ Majorana fermions in $1+1d$\footnote{The anomalies in $1+1d$ are actually $\mathbb Z_8\times\mathbb Z$, the second factor being the gravitational anomaly. We take $\nu_L=\nu_R=\nu$ to cancel this gravitational anomaly and focus directly on the $\mathbb Z_8$ factor.}
\begin{equation}
{\mathcal L}= \sum_{a=1}^\nu\, i\psi^a_L \partial_+\psi^a_L+i\psi^a_R \partial_-\psi^a_R\,,
\end{equation}
where $\partial_\pm=\partial_t\pm\partial_x$. This system has a chiral $\mathbb Z_2$ unitary symmetry generated by $g=(-1)^{F_L}$ which combines with the nonchiral $\mathbb Z_2^F$ symmetry generated by $(-1)^{F}$ to yield the symmetry group $G_f=\mathbb Z_2\times \mathbb Z_2^F$. These symmetries act on the fermions as 
\begin{equation}
\begin{aligned}
\{(-1)^{F_L},\psi^a_L\}&= [(-1)^{F_L},\psi^a_R]=0\\
\{(-1)^{F},\psi^a_L\}&= \{(-1)^{F},\psi^a_R\}=0\,.
\end{aligned}
\label{eq:anticomm2d}
\end{equation}
It is known that a $\mathbb Z_2$-symmetric interaction that gaps out the fermions can be written for $\nu=8$~\cite{Ryu_2012,Qi_2013,Yao_2013,Gu:2013azn}. This realizes in the fermion system the $\mathbb Z_8$ anomaly expected from the cobordism classification.

We now analyze the anomaly layers that can be detected in the Hilbert space. We discuss in turn the Hilbert space $\mathcal H_X$ and the $\mathbb Z_2$-twisted Hilbert space $\mathcal H^g_X$, where $X\in\{\text{NS},\text R\}$ denotes the spin structure on the spatial circle. The twisted Hilbert space $\mathcal H^g_X$ is defined by quantizing in the presence of a nontrivial $\mathbb Z_2$ (flat) connection around the circle for the $\mathbb Z_2$ symmetry.

In order to detect the anomalies we proceed to study the implementation of symmetries on the zero-mode operators in $\mathcal H_X$ and $\mathcal H^g_X$ in turn.

{\begin{center}\underline{Anomalies in $\mathcal H_X$} \end{center}}

Since in the NS sector there are no fermion zero-modes, there is a unique, trivial vacuum and symmetries are realized on $\mathcal H_\text{NS}$ in a non-anomalous fashion. In the R sector there are fermion zero-modes which upon quantization furnish a Clifford algebra of rank $2\nu$
\begin{equation}
\{\psi_L^a,\psi_L^b\}= \{\psi_R^a,\psi_R^b\}=2\delta^{ab}\,,\qquad \{\psi_L^a,\psi_R^b\}=0\qquad a,b=1,2,\ldots,\nu\,.
\end{equation}
This Clifford algebra has a unique irreducible representation of dimension $2^\nu$, thus all representations are unitarily equivalent, and we can study the implementation of symmetries in any choice of basis. We can construct $\mathcal H_\text{R}$ by defining the creation and annihilation operators $\psi_+^a=\frac12(\psi_R^a+i\psi_L^a)$ and $\psi_-^a=\frac12(\psi_R^a-i\psi_L^a)$, such that $\psi_-^a|0\rangle=0$. It follows from~\eqref{eq:anticomm2d} that
\begin{equation}
(-1)^{F_L}\psi_+^a=\psi_-^a(-1)^{F_L}\,.
\end{equation}
The $\mathbb Z_2$ symmetry generator thus maps the empty vacuum to the completely filled state
\begin{equation}
(-1)^{F_L}|0\rangle=\alpha\psi_+^1\psi_+^2\cdots \psi_+^\nu|0\rangle\,.
\end{equation}
for some phase $\alpha$. This implies that the $\mathbb Z_2\times \mathbb Z_2^F$ symmetry generators on $\mathcal H_R$ obey
\begin{equation}
\{(-1)^{F_L},(-1)^F\}=0\qquad\text{for $\nu$ odd}\,,
\end{equation}
 and 
\begin{equation}
[(-1)^{F_L},(-1)^F]=0\qquad\text{for $\nu$ even}\,.
\end{equation}
Therefore, the anomaly corresponding to the $\nu$ odd layer is detected by virtue of the operators $(-1)^{F_L}$ and $(-1)^F$ anticommuting in $\mathcal H_R$. This has also been noticed in~\cite{Thorngren:2018bhj,Karch:2019lnn}.

The anomaly associated to the $\nu$ odd layer can also be detected in the torus partition function with periodic boundary conditions around both the spatial circle and temporal circle, that is with $(\text R,\text R)$ boundary conditions along the two cycles of the torus. The zero-modes in $\mathcal H_\text R$ imply that the partition function vanishes, but the partition function with fermion zero-modes saturated is nonvanishing:
\begin{equation}
\langle \psi_L^1\psi_L^2\cdots\psi_L^\nu \psi_R^1\psi_R^2\cdots\psi_R^\nu\rangle\neq 0\,.
\end{equation}
This implies that the $(-1)^{F_L}$ Ward identities are violated for $\nu$ odd, that is, there is an anomaly for the chiral $\mathbb Z_2$ symmetry.

{\begin{center}\underline{Anomalies in $\mathcal H^g_X$} \end{center}}

This Hilbert space is constructed by imposing boundary conditions twisted by $(-1)^{F_L}$ when fermions are transported around the spatial circle. This yields different boundary conditions for the left-moving and right-moving fermions, which we we will denote by $[X_L,X_R]$, where $X_{L/R}\in\{\text{NS},\text R\}$.\footnote{This is to be contrasted with the boundary condition in the untwisted Hilbert space, where $X_L=X_R$.}
 
Let us consider $\mathcal H_X^g$ in turn:
 
 \smallskip
 \noindent
$\boldsymbol{\mathcal H^g_\mathrm{NS}}$. This corresponds to $[\text R,\text{NS}]$ boundary conditions on the fermions. There are $\nu$ zero modes from the left movers and none from the right movers. The zero mode algebra is thus a Clifford algebra of rank $\nu$
\begin{equation}
 \{\psi_L^a,\psi_L^b\}=2\delta^{ab}\qquad a,b=1,2,\ldots,\nu\,.
\end{equation}

\begin{itemize}
\item $\nu$ odd. There is a rather severe anomaly for $\nu$ odd as the operator $(-1)^F$ generating $\mathbb Z_2^F$ obeying $\{(-1)^F, \psi_L^a \}=0$ does not exist. The theory does not admit a proper graded Hilbert space of states. Equivalently stated, the Clifford algebra of odd rank has two irreducible representations, and $(-1)^F$ exchanges them, instead of acting within an irreducible representation. This anomaly is associated with the $H^1(\mathbb Z_2,\mathbb Z_2)=\mathbb Z_2$ layer, the Arf layer. 
 
The $\mathbb Z_2$ anomaly associated to the Arf layer and the corresponding lack of a Hilbert space can also be detected in the torus partition function with antiperiodic boundary conditions around both the spatial circle and the temporal circle, that is, with $(\text{NS},\text{NS})$ boundary conditions along the two cycles of the torus. This partition function is given by
\begin{equation}
 Z^{(g,1)}_{(\text{NS},\text{NS})}=(\sqrt{2}\chi_\sigma)^\nu (\overline\chi_1+\overline\chi_\epsilon)^\nu\,,
\label{eq:parttwist}
\end{equation}
where $\chi_1,\chi_\sigma$ and $\chi_\epsilon$ are the Virasoro characters with weight $0,1/16$ and $1/2$. For $\nu$ odd, the partition function indeed does not have an integral expansion, and thus there is no suitable Hilbert space. The dimensions of the modules are integers times a factor of $\sqrt2$, consistent with the fact that the anomaly is due to a dangling $\mathrm{C}\ell(1)$, whose dimension is formally $\sqrt2$.
 
\item $\nu$ even. For $\nu$ even the theory has a well-defined Hilbert space $\mathcal H^g_\text{NS}$ and well defined $(-1)^F$ and $(-1)^{F_L}$ operators obeying $[(-1)^{F_L},(-1)^F]=0$ in the Hilbert space. The $\mathbb Z_2$ symmetry generator maps the empty vacuum in $\mathcal H^g_X$ to itself up to phase $(-1)^{F_L}|0\rangle=\alpha|0\rangle$. 
 
%It is worth mentioning that the theory with $\nu=4 \mod 8$ has a bosonic $\mathbb Z_2$ anomaly, measured by $H^3(\mathbb Z_2, U(1))=\mathbb Z_2$. While this anomaly is not visible in the way the symmetry is realized in the Hilbert space, it can be detected by the presence of states with anomalous spin in the Hilbert space (see e.g.~\cite{Lin:2019kpn} for a similar discussion for bosonic systems). This anomaly can be detected from the torus partition function with $(\text{NS},\text{NS})$ boundary conditions~\eqref{eq:parttwist}. Under the modular transformation $T_\text{Dehn}^2$ all spin structures in the torus are invariant, which implies that in a non-anomalous theory $T_\text{Dehn}^2=1$ and the spin of states in $\mathcal H^g_\text{NS}$ should be $\mathbb Z/2$. Instead, in an anomalous theory, $T_\text{Dehn}^2=-1$ and the spin of states in $\mathcal H^g_\text{NS}$ can be $1/4+\mathbb Z/2$. From~\eqref{eq:parttwist} we see that indeed for $\nu=4 \mod 8$ fermions, the states in $\mathcal H^g_\text{NS}$ have spins $1/4+\mathbb Z/2$. This diagnoses that the system with $\nu=4 \mod 8$ fermions has a $\mathbb Z_2$ anomaly arising from the bosonic layer. 

\end{itemize}

 \smallskip
 \noindent
$\boldsymbol{\mathcal H^g_\mathrm{R}}$. This corresponds to $[\text{NS},\text R]$ boundary conditions on the fermions. There are $\nu$ zero modes from the right-movers and none from the left-movers. The zero mode algebra is thus a Clifford algebra of rank $\nu$
\begin{equation}
 \{\psi_R^a,\psi_R^b\}=2\delta^{ab}\qquad a,b=1,2,\ldots,\nu\,.
\end{equation}
 
\begin{itemize}
\item $\nu$ odd. Verbatim our discussion above: this system has a $\mathbb Z_2$ anomaly associated to the Arf layer, diagnosed by the lack of a proper graded Hilbert space of states. This can also be seen from the lack of integral expansion of the torus partition function with $(\text R,\text{NS})$ boundary conditions along the two cycles of the torus 
\begin{equation}
 Z^{(g,1)}_{(\text R,\text{NS})}= (\chi_1+\chi_\epsilon)^\nu(\sqrt{2}\overline \chi_\sigma)^\nu
\label{eq:parttwistb}
\end{equation}
which again does not have a properly quantized expansion.
 
\item $\nu$ even. For $\nu$ even the theory has a well-defined Hilbert space $\mathcal H^g_\text{R}$ and well defined $(-1)^F$ and $(-1)^{F_L}$ operators obeying $[(-1)^{F_L},(-1)^F]=0$ in the Hilbert space. 

%The partition function $Z^{(g,1)}_{(\text R,\text{NS})}$ also exhibits a bosonic $\mathbb Z_2$ anomaly $\nu=4 \mod 8$ as $\mathcal H^g_\text{R}$ has states with spin $1/4+\mathbb Z/2$. 

\end{itemize}

Twisting by the symmetry eliminates the zero modes of one chirality. So effectively the twisting halves the number of zero-modes, and the system moves one step up the ladder of layers. In the untwisted Hilbert spaces $\mathcal H_X$, $\nu=1\mod2 $ means that $g$ anticommutes with $(-1)^F$. For twisted Hilbert spaces $\mathcal H^g_X$, $\nu=1\mod2$ means lack of graded Hilbert spaces.

\paragraph{Projective rotations.} In the discussion so far, we argued that the $\nu=1\mod 2$ layer of the anomaly can be detected by the symmetry algebra of $g$ and $(-1)^F$. It is noteworthy that, if we look at a more general group of symmetries, the \emph{full} anomaly $\nu\mod8$ can be detected.

The detection of the $\nu\in\mathbb Z_8$ anomaly is well-understood by now. This anomaly measures the anomalous spin of operators on the twisted Hilbert spaces $\mathcal H^g_X$. From our point of view, this statement is understood as follows. When we compactify the theory on a circle, the model acquires an extra flavor symmetry: translations along the compact direction become internal symmetries of the effective $0+1$ dimensional quantum mechanical model. Therefore, the Hilbert spaces on the circle realize a representation of a larger symmetry group: the initial $G_f=\mathbb Z_2\times\mathbb Z_2^F$ symmetry is enhanced to $G_f\times U(1)$, with $U(1)$ the group of rotations around the spatial circle. While the bottom layer $\nu\mod2$ only measures a projective action of $G_f$, the full anomaly $\nu\mod8$ measures a projective realization of the extended group $G_f\times U(1)$.

In a non-anomalous theory, the Hilbert space realizes a double-cover of the symmetry group $U(1)$, because of the presence of fermions --- the system is invariant under $4\pi$ rotations. In an anomalous theory, the Hilbert space realizes a higher cover of $U(1)$, because of the presence of operators with fractional spin. We claim that in a system with anomaly $\nu\in\mathbb Z_8$, the twisted Hilbert spaces $\mathcal H^g_X$ realize a $2\times (8/\gcd(\nu,8))$-cover of $U(1)$.

This claim can be established by looking again at a free fermion system. In this system, the anomalous spin comes from the ``zero-point momentum'' of the fermions, namely $1/16$ units for each Ramond zero-mode. In the untwisted Hilbert space there is no fractional momentum: if both chiralities are NS, there are no zero-modes, and if they are both R, the zero-point momenta cancel out. In the twisted Hilbert space, one of the two chiralities is NS and the other is R, so there is no cancellation, and there are $\nu/16$ units of fractional momentum. Therefore, the eigenvalues of the momentum operator are half-integers (from the oscillator modes), plus $\nu/16$ (from the zero-modes). In other words, $2P\pm\nu/8\in\mathbb Z$. The momentum operator $P$ is precisely the generator of $U(1)$ rotations; therefore, instead of two circles, we need to perform $2\times (8/\gcd(\nu,8))$ full turns instead in order to compensate for this fractional momentum. This proves the claim.

This also admits a path-integral interpretation. In a system with $\nu\in\mathbb Z_8$ units of anomaly, the partition function twisted by the $g$ symmetry picks up a phase $e^{\pm 2\pi i \nu/8}$ under a $4\pi$ rotation, which signals the presence of operators with fractional momentum. We show this by looking at the free fermion system, whose partition function over the twisted Hilbert space $\mathcal H_X^g$ is~\eqref{eq:parttwist},~\eqref{eq:parttwistb}
\begin{equation}
%Z^{(1,g)}_{(\text{NS},\text{NS})}= (\chi_1+\chi_\epsilon)^\nu (\sqrt2\overline\chi_\sigma)^\nu
\begin{alignedat}{3}
\mathcal H^g_\text{NS}&\colon\quad Z^{(g,1)}_{(\text{NS},\text{NS})}&&=(\sqrt{2}\chi_\sigma)^\nu (\overline\chi_1+\overline\chi_\epsilon)^\nu\\
\mathcal H^g_\text{R}&\colon\quad Z^{(g,1)}_{(\text R,\text{NS})}&&= (\chi_1+\chi_\epsilon)^\nu(\sqrt{2}\overline \chi_\sigma)^\nu
\end{alignedat}
\end{equation}
A $4\pi$ rotation corresponds to a $T^2$ modular transformation, which induces the phase $e^{\pm 4\pi i \nu h_\sigma}\equiv e^{\pm 2\pi i \nu/8}$, with $h_\sigma=1/16$ the conformal weight of $\chi_\sigma$.
%\begin{equation}
%Z^{(1,g)}_{(\text{NS},\text{NS})}\mapsto e^{-2\pi i \nu/8}(\chi_1+\chi_\epsilon)^\nu (\sqrt2\overline\chi_\sigma)^\nu
%\end{equation}
%as required.

To summarize, the effect of the anomaly $\nu\in\mathbb Z_8$ is that the twisted Hilbert spaces realize a projective representation of the rotation symmetry group $U(1)$, and from this projective action one can read off the anomaly. For a given value of $\nu$, the Hilbert space realizes a $2\times (8/\gcd(\nu,8))$ cover of $U(1)$, as opposed to a double-cover, and the generator of rotations has fractional momentum $\nu/16$. In a conformal theory, this can also be detected by performing a $T^2$ modular transformation on the twisted partition function. A very similar philosophy was used in~\cite{Tachikawa:2016cha} to derive the time-reversal anomaly in a $2+1d$ dimensional system.

\section{Anomalies in spin TQFT Hilbert space}\label{sec:t_reversal}

In this section we demonstrate the existence of anomalies in fermionic TQFTs by looking directly at their Hilbert space. We follow the construction of the Hilbert space of a fermionic TQFT in~\cite{Delmastro:2020dkz}; see~\cite{Gaiotto:2015zta,Lan_2017,Bhardwaj:2016clt,usher2016fermionic,Aasen:2017ubm,Lou:2020gfq} for related work. Here we summarize the main ingredients, leaving most details to appendix~\ref{sec:condensation}.

Given an arbitrary spin TQFT, one may sum over all spin structures in order to yield a bosonic TQFT. We refer to this theory as the \emph{bosonic parent/shadow} of the original spin TQFT. This process of summing over spin structures corresponds to gauging the zero-form symmetry generated by fermion parity $\mathbb Z_2^F=\langle(-1)^F\rangle$. Given such bosonic theory, one may undo the gauging, i.e., we can recover the original spin TQFT by gauging a dual $\mathbb Z_2$ symmetry, this time a one-form symmetry~\cite{Gaiotto:2014kfa}. This symmetry is generated by a certain fermionic line operator $\psi$, i.e., $\mathbb Z_2^{(1)}=\langle \psi\rangle$. Gauging this one-form symmetry is also known as \emph{condensing} the anyon $\psi$.

\begin{figure}[!h]
\centering
\begin{tikzpicture}%[baseline=0]

\draw[thick] (0,0) ellipse (2cm and 1cm);
\draw[thick,domain=180:360,smooth,variable=\x] plot ({.9*cos(\x)},{.2+.4*sin(\x)});
\draw[thick,domain=23:157,smooth,variable=\x] plot ({.9*cos(\x)},{-.1+.4*sin(\x)});
\draw[thick,blue] (0,.1) ellipse (1.4cm and .6cm);

\node at (-1.6,.1) {\footnotesize$\color{blue}{\beta}$};

\begin{scope}[rotate=-60,shift={(.75,.2)}]
\draw[thick,red,domain=10:150,smooth,variable=\x] plot ({.45*cos(\x)},{.3*sin(\x)});
\draw[thick,red,dashed,domain=180:320,smooth,variable=\x] plot ({.05+.45*cos(\x)},{.1+.3*sin(\x)});
\end{scope}

\node at (1.1,-.65) {\footnotesize$\color{red}{\alpha}$};

%\filldraw (0,.7) circle (1pt);
\filldraw[black!20!green] (0,.85) circle (1pt);
\node at (.2,.83) {\footnotesize$\color{black!20!green}{\mathsf g}$};

\node at (3,0) {$=$};

\begin{scope}[shift={(5,0)}]

\draw[thick,blue,domain=-16:333,smooth,variable=\x] plot ({.8*cos(\x)},{.8*sin(\x)});
\draw[thick,red,domain=108:87+360,smooth,variable=\x] plot ({.8+.5*cos(\x)},{.3*sin(\x)});

\draw[thick,black!20!green] (0,.8) -- (0,1.5);
%\draw[thick,blue] (0,0) circle (.8cm);
\filldraw[thick,black!20!green] (0,.8) circle (1pt);

\node at (0,0) {\footnotesize$\times$};

\node at (.2,1.4) {\footnotesize$\color{black!20!green}{\mathsf g}$};
\node at (1.5,0) {\footnotesize$\color{red}{\alpha}$};
\node at (0,-1.05) {\footnotesize$\color{blue}{\beta}$};

\end{scope}

\end{tikzpicture}
\caption{Schematic notation for an arbitrary configuration of anyons on the torus, in the presence of a puncture $\mathsf g$. The green line represents the vertical (time) direction, orthogonal to the torus. We insert an anyon $\mathsf g$ along this direction, which from the point of view of the torus becomes a puncture (a marked point). The red line represents the $\mathsf a$-cycle, and the blue one the $\mathsf b$-cycle. We insert Wilson lines with anyons $\alpha,\beta$ along these cycles, respectively. The cross $\times$ represents the hole. The states in the Hilbert space $\mathcal H$ are created by wrapping anyons around the $\mathsf b$-cycle. The states in the \emph{twisted} Hilbert space $\mathcal H^{\mathsf g}$ are created by wrapping anyons around the $\mathsf b$-cycle, in presence of a vertical anyon $\mathsf g$.}
\label{fig:torus_lines_schematic}
\end{figure}
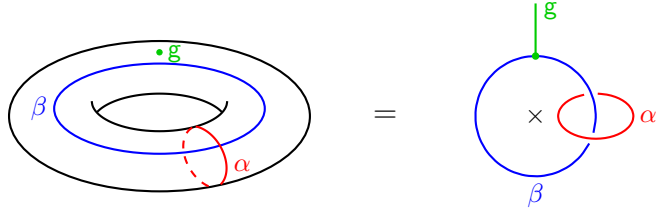

With this in mind, the Hilbert space of the spin TQFT is easily obtained in terms of the Hilbert space of the bosonic parent, by means of the standard procedure of gauging a symmetry. The Hilbert space of the bosonic parent, being a standard TQFT, is well-understood. Specifically, the torus Hilbert space has a basis of states labelled by the anyons~\cite{witten1989}:
\begin{equation}\label{eq:basis_boson_TQFT}
\mathcal H(T^2)=\underset{\beta\in\mathcal A}{\operatorname{Span}}\bigg[\,\begin{tikzpicture}[baseline=-3pt]
\draw[thick] (0,0) circle (.5cm);
\node[scale=.6] at (0,0) {$\times$};
\node[scale=.8] at (.7,0) {$\beta$};

\end{tikzpicture}
\bigg]
\end{equation}
Here $\mathcal A$ denotes the set of all anyons in the bosonic parent -- a finite set -- and the loop denotes a Wilson line labeled by $\beta$ wrapped along the $\mathsf b$-cycle of the torus (see figure~\ref{fig:torus_lines_schematic}).

The set $\mathcal A$ comes equipped with extra structure; for example, we have the modular matrix $S\colon\mathcal A\times\mathcal A\to \mathbb C$ that implements the large diffeomorphism $(\mathsf a,\mathsf b)\mapsto(\mathsf b,-\mathsf a)$. Similarly, we also have the modular matrix $T\colon\mathcal A\times\mathcal A\to \mathbb C$ that implements the large diffeomorphism $(\mathsf a,\mathsf b)\mapsto(\mathsf a,\mathsf a+\mathsf b)$; these two transformations $S,T$ generate the set of all large diffeomorphisms, the modular group $SL_2(\mathbb Z)$. By our choice of basis~\eqref{eq:basis_boson_TQFT} where the states are wrapped around the $\mathsf b$-cycle, the $T$-matrix is diagonal, with $T=\diag(e^{2\pi (h-c/24)})$, where $c$ is the central charge of the system and $h\colon\mathcal A\to\mathbb Q/\mathbb Z$ denotes the topological spin of the lines.

The Hilbert space of the fermionic theory, let us call it $\hat{\mathcal H}$, depends on the spin structure of the torus. We denote these tori as $T^2_{s^\mathsf a,s^\mathsf b}$, where $s^\mathsf a,s^\mathsf b=\pm1$ refers to the boundary condition around the cycle $\mathsf a$- and $\mathsf b$-cycles, respectively. We also denote the $s=-1$ boundary condition by NS and the $s=+1$ boundary condition by R. We claim that the corresponding Hilbert spaces are spanned by the following bases:
\begin{equation}\label{eq:hilbert_space_fermion_tqft}
\begin{aligned}
\hat{\mathcal H}(T^2_{\text{NS-NS}})&\colon
\hspace{12pt}\tikz[baseline=-3]{
\draw[thick] (0,0) circle (.5cm);
\node[scale=.6] at (0,0) {$\times$};
\node[scale=.8] at (0,-.7) {$a$};
\node[anchor=south] at (1,-.3) {$+$};
\begin{scope}[shift={(2,0)}]
\draw[thick] (0,0) circle (.5cm);
\node[scale=.6] at (0,0) {$\times$};
\node[scale=.8] at (0,-.7) {$a\times\psi$};
\end{scope}
}\\
\hat{\mathcal H}(T^2_{\text{NS-R}})&\colon
\hspace{12pt}\tikz[baseline=-3]{
\draw[thick] (0,0) circle (.5cm);
\node[scale=.6] at (0,0) {$\times$};
\node[scale=.8] at (0,-.7) {$a$};
\node[anchor=south] at (1,-.3) {$-$};
\begin{scope}[shift={(2,0)}]
\draw[thick] (0,0) circle (.5cm);
\node[scale=.6] at (0,0) {$\times$};
\node[scale=.8] at (0,-.7) {$a\times\psi$};
\end{scope}
}\\
\hat{\mathcal H}(T^2_{\text{R-NS}})&\colon
\begin{cases}
\tikz[baseline=-3]{
\draw[thick] (0,0) circle (.5cm);
\node[scale=.6] at (0,0) {$\times$};
\node[scale=.8] at (0,-.7) {$x$};
\node[anchor=south] at (1,-.3) {$+$};
\begin{scope}[shift={(2,0)}]
\draw[thick] (0,0) circle (.5cm);
\node[scale=.6] at (0,0) {$\times$};
\node[scale=.8] at (0,-.7) {$x\times\psi$};
\end{scope}
}\\
\tikz[baseline=-3]{
\draw[thick] (0,0) circle (.5cm);
\node[scale=.6] at (0,0) {$\times$};
\node[scale=.8] at (0,-.7) {$m$};
}
\end{cases}\\
\hat{\mathcal H}(T^2_{\text{R-R}})&\colon
\begin{cases}
\tikz[baseline=-3]{
\draw[thick] (0,0) circle (.5cm);
\node[scale=.6] at (0,0) {$\times$};
\node[scale=.8] at (0,-.7) {$x$};
\node[anchor=south] at (1,-.3) {$-$};
\begin{scope}[shift={(2,0)}]
\draw[thick] (0,0) circle (.5cm);
\node[scale=.6] at (0,0) {$\times$};
\node[scale=.8] at (0,-.7) {$x\times\psi$};
\end{scope}
}\\
\tikz[baseline=-3]{
\draw[thick] (0,0) circle (.5cm);
\draw[thick] (-1,0) -- (-.5,0);
\fill (-.5,0) circle (1.5pt);
\node[scale=.6] at (0,0) {$\times$};
\node[scale=.8] at (0,-.7) {$m$};
\node[scale=.8] at (-1.2,0) {$\psi$};
}
\end{cases}
\end{aligned}
\end{equation}
Here, $a$ takes values in the subset of lines of $\mathcal A$ with the property that $h_\alpha\equiv h_{\alpha\times\psi}\mod1$. On the other hand, both $x$ and $m$ denote the lines in $\mathcal A$ such that $h_\alpha\equiv h_{\alpha\times\psi}+1/2\mod1$; the difference between $x$ lines and $m$ lines is that the former satisfy $x\times\psi\neq x$ while the latter satisfy $m\times\psi=m$. Finally, the state with an open line denotes the anyon $\beta$ around the spatial torus and the anyon $\psi$ running along the time direction, cf.~figure~\ref{fig:torus_lines_schematic}. (From the point of view of the spatial torus, the line operator $\psi$ looks like a puncture, i.e., a local operator; it is essentially a constant spinor, a zero-mode, which explains why it only exists in the odd spin structure).

The rationale behind the structure above is the following. Given the bosonic theory, gauging the one-form symmetry generated by $\psi$ means inserting this operator in all possible ways; summing over insertions along the spatial cycles projects the spectrum into the invariant states, and summing over insertions along the time circle introduces twisted sectors. One can check that the states in~\eqref{eq:hilbert_space_fermion_tqft} are indeed invariant under insertion of $\psi$ along any of the spatial cycles. The twisted sectors are precisely the states with a puncture, which do not live in $\mathcal H$ but in the defect Hilbert space $\mathcal H^\psi$ instead. The details of this construction are elaborated upon in appendix~\ref{sec:condensation}.

The explicit geometric structure of states in~\eqref{eq:hilbert_space_fermion_tqft} also allows us find how the operators of the spin TQFT act on the different states. For example, the Wilson lines act by inserting an anyon around the cycle they are supported on, and therefore they act as
\begin{equation}
\begin{aligned}
W^{(\mathsf a)}(\alpha)\colon\,\tikz[baseline=-2pt]{
\draw[thick] (0,0) circle (.5cm);
\node[scale=.6] at (0,0) {$\times$};
\node[scale=.8] at (0,-.7) {$\beta$};}\ &\mapsto\ 
\tikz[baseline=-2pt]{
\draw[thick] (0,0) circle (.5cm);
\fill[white] (-.48,-.14) circle (1.2pt);
\draw[thick,rotate=-45] (-.1,-.5) ellipse (.15cm and .2cm);
\fill[white] (-.32,-.38) circle (1.2pt);
\node[scale=.6] at (0,0) {$\times$};
\node[scale=.8] at (-.85,-.5) {$\alpha$};
\begin{scope}
\clip (-.32,-.38) circle (2pt);
\draw[thick] (0,0) circle (.5cm);
\end{scope}

\node[scale=.8] at (0,-.7) {$\beta$};}\ \equiv\ \frac{S_{\alpha,\beta}}{S_{\boldsymbol1,\beta}}\ 
\tikz[baseline=-2pt]{
\draw[thick] (0,0) circle (.5cm);
\node[scale=.6] at (0,0) {$\times$};
\node[scale=.8] at (0,-.7) {$\beta$};}\\
W^{(\mathsf b)}(\alpha)\colon\,\tikz[baseline=-2pt]{
\draw[thick] (0,0) circle (.5cm);
\node[scale=.6] at (0,0) {$\times$};
\node[scale=.8] at (0,-.7) {$\beta$};}\ &\mapsto\hspace{20pt}
\tikz[baseline=-2pt]{

\draw[thick] (0,0) circle (.5cm);
\draw[thick] (0,0) circle (.4cm);

\node[scale=.6] at (0,0) {$\times$};
\node[scale=.8] at (0,-.23) {$\alpha$};

\node[scale=.8] at (0,-.7) {$\beta$};}\ \equiv\quad
\tikz[baseline=-2pt]{
\draw[thick] (0,0) circle (.5cm);
\node[scale=.6] at (0,0) {$\times$};
\node[scale=.8] at (0,-.7) {$\alpha\times\beta$};}
\end{aligned}
\end{equation}
on states without puncture, and as~\cite{Li:1989hs}
\begin{equation}
W^{(\mathsf b)}(\alpha)\colon\,\tikz[baseline=-2pt]{

\draw[thick] (-1,0) -- (-.5,0);
\draw[thick] (0,0) circle (.5cm);
\fill (-.5,0) circle (1.5pt);

\node[scale=.6] at (0,0) {$\times$};
\node[scale=.8] at (-1.2,0) {$\psi$};

\node[scale=.8] at (0,-.7) {$\beta$};}\ \mapsto\quad
\tikz[baseline=-2pt]{

\draw[thick] (-1,0) -- (-.5,0);
\draw[thick] (0,0) circle (.5cm);
\draw[thick] (0,0) circle (.4cm);
\fill (-.5,0) circle (1.5pt);

\node[scale=.6] at (0,0) {$\times$};
\node[scale=.8] at (0,-.23) {$\alpha$};
\node[scale=.8] at (-1.2,0) {$\psi$};

\node[scale=.8] at (0,-.7) {$\beta$};}\ \equiv\quad F_{\beta,\alpha\times\beta}\begin{bmatrix}\psi&\alpha\times\beta\\\beta&\alpha\end{bmatrix}\ 
\tikz[baseline=-2pt]{
\draw[thick] (-1,0) -- (-.5,0);
\fill (-.5,0) circle (1.5pt);
\draw[thick] (0,0) circle (.5cm);
\node[scale=.6] at (0,0) {$\times$};
\node[scale=.8] at (0,-.7) {$\alpha\times\beta$};
\node[scale=.8] at (-1.2,0) {$\psi$};}\ ,
\end{equation}
on states with a puncture. Here $S\colon\mathcal A\times\mathcal A$ denotes the modular matrix of the bosonic parent, and $F$ its $F$-symbols. From these expressions it is trivial to write down the Wilson operators as matrices acting on $\hat{\mathcal H}$ (with respect to the basis~\eqref{eq:hilbert_space_fermion_tqft}).

Similarly, one can also write down how modular transformations map the different Hilbert spaces $\hat{\mathcal H}(T^2_{XY})$ into each other. For example, an $S$-transformation acts as
\begin{equation}
S\colon\,\tikz[baseline=-2pt]{
\draw[thick] (0,0) circle (.5cm);
\node[scale=.6] at (0,0) {$\times$};
\node[scale=.8] at (0,-.7) {$\beta$};}\ \mapsto\ 
\sum_{\alpha\in\mathcal A} S_{\alpha,\beta}\ 
\tikz[baseline=-2pt]{
\draw[thick] (0,0) circle (.5cm);
\node[scale=.6] at (0,0) {$\times$};
\node[scale=.8] at (0,-.7) {$\alpha$};}
\end{equation}
on states without puncture, and as
\begin{equation}
S\colon\,\tikz[baseline=-2pt]{

\draw[thick] (-1,0) -- (-.5,0);
\draw[thick] (0,0) circle (.5cm);
\fill (-.5,0) circle (1.5pt);

\node[scale=.6] at (0,0) {$\times$};
\node[scale=.8] at (-1.2,0) {$\psi$};

\node[scale=.8] at (0,-.7) {$\beta$};}\ \mapsto\ \sum_{\alpha\in\mathcal A}S_{\alpha,\beta}(\psi)\ 
\tikz[baseline=-2pt]{
\draw[thick] (-1,0) -- (-.5,0);
\fill (-.5,0) circle (1.5pt);
\draw[thick] (0,0) circle (.5cm);
\node[scale=.6] at (0,0) {$\times$};
\node[scale=.8] at (0,-.7) {$\alpha\times\beta$};
\node[scale=.8] at (-1.2,0) {$\psi$};}\ ,
\end{equation}
on states with a puncture. Here $S(\psi)$ is the $S$-matrix of the bosonic parent in the once-punctured torus (cf.~\eqref{eq:puncture_S_matrix}). Given these two expressions one can easily check that $S$-transformations reshuffle the different spin structures as expected, namely $(X,Y)\mapsto(Y,X)$. Identical considerations hold for $T$-transformations (these are actually simpler because they do not see the puncture, so the formula $T=\diag(e^{2\pi (h-c/24)})$ is still valid for states in $\mathcal H^\psi$).

The final important remark concerning the fermionic Hilbert space $\hat{\mathcal H}$ is that it is a super-vector space, i.e., its states are either bosons or fermions. Given that $(-1)^F$ is, by definition, dual to the gauged one-form symmetry $\mathbb Z_2^{(1)}$, it is clear that the states charged under the former correspond to the states coming from the twisted sector, i.e., the fermions in $\hat{\mathcal H}$ are precisely those that include a $\psi$-puncture. In this sense, $(-1)^F$ is trivial in the even spin structure Hilbert spaces, and in the odd spin structure Hilbert space it equals $+1$ on $x$-lines and $-1$ on $m$-lines. In a fermionic theory the modular group is no longer $SL_2(\mathbb Z)$, but rather a $\mathbb Z^F_2$ extension, known as the metaplectic group $Mp_1(\mathbb Z)$, defined by the relations
\begin{equation}
S^2=(ST)^3,\qquad S^4=(-1)^F\,.
\end{equation}
The torus Hilbert spaces of spin TQFTs realize a unitary representation of this group.

We next illustrate all these considerations by explicitly working out several specific examples of spin TQFTs. We show that time-reversal invariant theories with $\mathsf T^2=(-1)^F$ with $\mathbb Z_{16}$
anomalies $\nu=2\mod 4$ have time-reversal in the Hilbert space realized as
\begin{equation}
\mathsf T^2=(-1)^F\times(-1)^{\Arf(T^2)}\,,
\end{equation}
thus exhibiting the anomalies associated with the Arf layer. We then show that spin TQFTs with $\nu$ odd have a time-reversal symmetry that anticommutes with $(-1)^F$ on $\mathcal H_\text{R-R}$
\begin{equation}
\{\mathsf T,(-1)^F\}=0\,,
\end{equation}
thus exhibiting the anomalies associated with the $\psi$ layer

\subsection{$\nu=2\mod4$: Arf layer}\label{sec:semion_fermion}

In this section we consider time-reversal invariant theories with $\nu=2\mod 4$ and show that they realize the expected behavior associated with the Arf layer. We work out in detail here the example of the semion-fermion theory which has $\nu=2$, although the same behaviour is observed in any theory with $\nu=2\mod4$. Other common examples of time-reversal invariant spin TQFTs are $Sp(n)_n$ and $SO(n)_n$, which have $\nu=2n$ and $\nu=n$, respectively. One can check that e.g.~$Sp(3)_3$ and $SO(6)_6$, which have $\nu=6$, exhibit the same behaviour. We do not reproduce the explicit computation here because the matrices are very large and the details do not contain any new ingredients.

The semion-fermion theory is a fermionic TQFT with four anyons: the vacuum $\boldsymbol1$, a semion $s$, a transparent fermion $\psi$, and the composite $s\times\psi$. A Chern-Simons realization of this theory is $U(1)_2\times \{\boldsymbol 1,\psi\}$, where $U(1)_2=\{\boldsymbol1,s\}$ is the TQFT of a single semion, and $\psi$ denotes a transparent fermion. The invertible factor can be written as $\{\boldsymbol 1,\psi\}=SO(n)_1$ for any $n\in\mathbb Z$; the most convenient choice is $n=-2$, so that the theory has vanishing central charge (as required by time-reversal). In other words, we shall consider $U(1)_2\times U(1)_{-1}$, whose Lagrangian reads
\begin{equation}
\mathcal L=\frac{1}{4\pi}(2 a\,\mathrm da-b\,\mathrm db)\,,
\end{equation}
where $a,b$ are $U(1)$ connections. The time-reversal symmetry of this Lagrangian acts as follows~\cite{Seiberg:2016rsg,Delmastro:2019vnj}:
\begin{equation}\label{eq:t_rev_SF}
\mathsf T(a)=a-b,\qquad \mathsf T(b)=2a-b\,.
\end{equation}

One way of constructing the Hilbert space of this theory is to take $U(1)_2\times U(1)_{-4}$, which is bosonic, and condense the line $\psi=(0,2)$; this is a fermionic quotient, and so the result is a spin TQFT, where $\psi$ becomes transparent.%because $h_\psi=1/2$. (More generally, $U(1)_{nk^2}/\mathbb Z_k=U(1)_n$, where the one-form symmetry is generated by the line with charge $nk$, and whose spin is $h_{nk}=n/2$).

We begin by constructing the Hilbert space of the bosonic parent, $U(1)_2\times U(1)_{-4}$. This is a $2\times4=8$-dimensional space, whose basis can be taken to be
\begin{equation}
|\alpha,\beta\rangle=W^{(\mathsf b)}(\alpha,\beta)|0\rangle,\qquad (\alpha,\beta)\in\mathbb Z_2\times\mathbb Z_4\,,
\end{equation}
where $|0\rangle$ denotes the vacuum state -- the empty torus -- and $W$ denotes a Wilson line:
\begin{equation}
W^{(\mathsf c)}(\alpha,\beta):=\exp i\oint_{\mathsf c} (\alpha a+\beta b)\,,
\end{equation}
for any $\mathsf c\in H_1(T^2,\mathbb Z)=\mathbb Z[\mathsf a]\oplus\mathbb Z[\mathsf b]$.

The Wilson lines along the $\mathsf b$-cycle act on a generic state as follows:
\begin{equation}\label{eq:SF_fusion}
W^{(\mathsf b)}(\alpha,\beta)|\alpha',\beta'\rangle=|\alpha+\alpha'\mod2,\ \beta+\beta'\mod4\rangle\,.
\end{equation}
The action of the Wilson lines associated to other cycles can be obtained using the modular operations. For example, on the $\mathsf a$-cycle, one has
\begin{equation}
W^{(\mathsf a)}(\alpha)|\alpha'\rangle=\frac{S_{\alpha\alpha'}}{S_{\boldsymbol1\alpha'}}|\alpha'\rangle\,,
\end{equation}
where $S$ denotes the $S$-matrix of the system. In the semion-fermion theory, this matrix reads $S_{(\alpha,\beta),(\alpha',\beta')}=e^{\pi i(\beta \beta'/2-\alpha\alpha')}/4$.

We now condense the fermion $\psi=(0,2)$. The braiding phase of a generic line $(\alpha,\beta)$ with respect to the fermion is $B((\alpha,\beta),\psi)=e^{-i\pi \beta}$, and so the NS- and R-lines are as follows:
\begin{equation}
\begin{aligned}
\text{NS}&\colon\quad\mathcal A_\text{NS}=\{(\alpha,\beta)\colon \beta=0,2\}\\
\text{R}&\colon\quad\mathcal A_{\hspace{2pt}\text{R}\hspace{3pt}}=\{(\alpha,\beta)\colon \beta=1,3\}
\end{aligned}
\end{equation}

Furthermore, there are no fixed-points under fusion with $\psi$. Indeed, the lines are all in two-dimensional orbits, paired as follows:
\begin{equation}
\begin{aligned}
\text{NS}&\colon\quad(\alpha,0)\overset{\times \psi}{\longleftrightarrow}(\alpha,2)\\
\text{R}&\colon\quad(\alpha,1)\overset{\times \psi}{\longleftrightarrow}(\alpha,3)\,.
\end{aligned}
\end{equation}

The lack of fixed-points indicates that there are no Majorana lines in the theory, i.e., all states are bosonic. In the terminology of section~\ref{sec:fermion_condensation}, all lines $(\alpha,\beta)$ are $a$-type or $x$-type, depending on whether $\beta$ is even or odd; and there are no $m$-lines.

\paragraph{Hilbert space.} With these preliminaries in mind, we now construct the torus Hilbert space(s) of the fermionic theory. As in section~\ref{sec:fermion_condensation}, the states of the condensed phase are expressed as linear combinations of those of the bosonic parent, and the specific combinations are determined by the spin structure (cf.~\eqref{eq:spin_torus_basis}):
\begin{itemize}
\item If we take \textbf{NS-NS} boundary conditions, the two states are
\begin{equation}
\begin{aligned}
|0;\text{NS-NS}\rangle&=\frac{1}{\sqrt2}(|0,0\rangle+|0,2\rangle)\\
|1;\text{NS-NS}\rangle&=\frac{1}{\sqrt2}(|1,0\rangle+|1,2\rangle)\,.
\end{aligned}
\end{equation}

\item If we take \textbf{NS-R} boundary conditions, the two states are
\begin{equation}
\begin{aligned}
|0;\text{NS-R}\rangle&=\frac{1}{\sqrt2}(|0,0\rangle-|0,2\rangle)\\
|1;\text{NS-R}\rangle&=\frac{1}{\sqrt2}(|1,0\rangle-|1,2\rangle)\,.
\end{aligned}
\end{equation}

\item If we take \textbf{R-NS} boundary conditions, the two states are
\begin{equation}
\begin{aligned}
|0;\text{R-NS}\rangle&=\frac{1}{\sqrt2}(|0,1\rangle+|0,3\rangle)\\
|1;\text{R-NS}\rangle&=\frac{1}{\sqrt2}(|1,1\rangle+|1,3\rangle)\,.
\end{aligned}
\end{equation}

\item If we take \textbf{R-R} boundary conditions, the two states are
\begin{equation}
\begin{aligned}
|0;\text{R-R}\rangle&=\frac{1}{\sqrt2}(|0,1\rangle-|0,3\rangle)\\
|1;\text{R-R}\rangle&=\frac{1}{\sqrt2}(|1,1\rangle-|1,3\rangle)\,.
\end{aligned}
\end{equation}

\end{itemize}

\paragraph{Modularity.} As a consistency check, we can study how modular transformations move us around these four Hilbert spaces. Take for example the $S$-transformation. In the bosonic parent, this operation acts as
\begin{equation}
S|\alpha,\beta\rangle=\sum_{\substack{\alpha'\in\mathbb Z_2\\ \beta'\in\mathbb Z_4}}S_{(\alpha,\beta),(\alpha',\beta')}|\alpha',\beta'\rangle
\end{equation}
with $S_{(\alpha,\beta),(\alpha',\beta')}=e^{i\pi(\beta \beta'/2-\alpha \alpha')}/4$. Using this we obtain the action of $S$ on the fermionic theory. For example, it acts on the NS-NS states as follows:
\begin{equation}
\begin{aligned}
S|0;\text{NS-NS}\rangle&=\frac{1}{\sqrt2}\bigg[S|0,0\rangle+S|0,2\rangle\bigg]\\
&=\frac{1}{\sqrt2}\frac14\bigg[|0, 0\rangle + |0, 1\rangle + |0, 2\rangle + |0, 3\rangle + |1, 0\rangle + 
 |1, 1\rangle + |1, 2\rangle + |1, 3\rangle+\\
&\hphantom{=\frac{1}{\sqrt2}\frac14\bigg[\ }|0, 0\rangle - |0, 1\rangle + |0, 2\rangle - |0, 3\rangle + |1, 0\rangle - 
 |1, 1\rangle + |1, 2\rangle - |1, 3\rangle\bigg]\\
 &=\frac{1}{\sqrt2}\frac12 (|0, 0\rangle + |0, 2\rangle + |1, 0\rangle + |1, 2\rangle)\,.
\end{aligned}
\end{equation}
We recognize this state as $\frac12(|0;\text{NS-NS}\rangle+|1;\text{NS-NS}\rangle)$. Through an identical computation one can show that $S$ maps $|1;\text{NS-NS}\rangle$ into $\frac12(|0;\text{NS-NS}\rangle-|1;\text{NS-NS}\rangle)$. In both cases we see that an $S$-transformation maps states in $\hat{\mathcal H}_\text{NS-NS}$ into $\hat{\mathcal H}_\text{NS-NS}$, precisely as expected (cf.~\eqref{eq:S_T_torus_spin}); and, moreover, the specific matrix that realizes this transformation is
\begin{equation}
\hat S_{\text{NS-NS}\to\text{NS-NS}}=\frac12\begin{pmatrix}1&1\\1&-1\end{pmatrix}\,.
\end{equation}
By performing $S$-transformations on the other three Hilbert spaces we see that they are permuted exactly as they should, namely $S\colon\hat{\mathcal H}_{s^\mathsf a,s^\mathsf b}\to \hat{\mathcal H}_{s^\mathsf b,s^\mathsf a}$; and that they act as the following matrices:
\begin{equation}
\hat S_{\text{NS-R}\to\text{R-NS}}=\hat S_{\text{R-NS}\to\text{NS-R}}=-i\hat S_{\text{R-R}\to\text{R-R}}=\frac12\begin{pmatrix}1&1\\1&-1\end{pmatrix}\,.
\end{equation}
A $T$-transformation, on the other hand, acts in the bosonic parent as
\begin{equation}
T|\alpha,\beta\rangle=e^{i \pi (\alpha^2/2-\beta^2/4)}|\alpha,\beta\rangle\,,
\end{equation}
which induces the following transformation in the fermionic quotient: $T\colon\hat{\mathcal H}_{s^\mathsf a,s^\mathsf b}\to \hat{\mathcal H}_{s^\mathsf a,s^\mathsf as^\mathsf b}$, with matrices
\begin{equation}
\hat T_{\text{NS-NS}\to\text{NS-R}}=\hat T_{\text{NS-R}\to\text{NS-NS}}=\hat T_{\text{R-NS}\to\text{R-R}}=\hat T_{\text{R-R}\to\text{NS-R}}=\begin{pmatrix}
1&0\\0&i
\end{pmatrix}\,.
\end{equation}
(The semion-fermion theory is rather degenerate, not least due to the fact that it factorizes into a bosonic TQFT and a trivial spin TQFT; in the general case, the matrices $\hat S_{s^\mathsf a,s^\mathsf b},\hat T_{s^\mathsf a,s^\mathsf b}$ are all typically distinct.)

A final ingredient as regards modularity is the charge-conjugation operation, which acts in homology as $\mathsf C\colon(\mathsf a,\mathsf b)\mapsto (-\mathsf a,-\mathsf b)$. Unlike the $S$- and $T$-operations, charge-conjugation fixes all spin structures: $\mathsf C\colon\hat{\mathcal H}_{s^\mathsf a,s^\mathsf b}\to \hat{\mathcal H}_{s^\mathsf a,s^\mathsf b}$. This operation acts on the $U(1)$ connection as $a\mapsto -a$ or, equivalently, on the anyons as $\alpha\mapsto \bar\alpha=-\alpha$. The semion and the fermion are both self-conjugate (cf.~$-1=1\mod2$), which means that $\mathsf C$ acts trivially on all the anyons of the theory. That being said, this operator need not act trivially on the Hilbert space. Its action is easily computed given the expression of the fermionic Hilbert space in terms of the bosonic parent, namely
\begin{equation}
\mathsf C|\alpha,\beta\rangle=|{}-\alpha\mod2,{}-\beta\mod 4\rangle\,.
\end{equation}
For example, this action induces the following action on the quotient theory:
\begin{equation}
\begin{aligned}
\mathsf C|0;\text{NS-NS}\rangle&=\frac{1}{\sqrt2}(\mathsf C|0,0\rangle+\mathsf C|0,2\rangle)\\
&=\frac{1}{\sqrt2}(|0,0\rangle+|0,2\rangle)\\
&=|0;\text{NS-NS}\rangle\,.
\end{aligned}
\end{equation}
Repeating this operation for the rest of basis vectors, we arrive at
\begin{equation}
\hat{\mathsf C}_{\text{NS-NS}\to\text{NS-NS}}=\hat{\mathsf C}_{\text{NS-R}\to\text{NS-R}}=\hat {\mathsf C}_{\text{R-NS}\to\text{R-NS}}=-\hat {\mathsf C}_{\text{R-R}\to\text{NS-R}}=\boldsymbol 1_2
\end{equation}
or, more succinctly, $\hat{\mathsf C}=(-1)^{\Arf(s)}$.

Given the explicit expressions for the $\hat S,\hat T,\hat C$ matrices, we can check that they realize a unitary representation of the modular group. In this case, the lack of Majorana lines means that $(-1)^F$ is trivial, which means that the modular algebra is just that of the regular torus with no punctures. In other words, the modular transformations satisfy $S^2=(ST)^3=\mathsf C$, with $\mathsf C^2=1$. The matrices $\hat S,\hat T,\hat C$ calculated above indeed satisfy this algebra, as expected. In checking this one must keep in mind that $\hat S,\hat T$ do not live in $\mathrm{End}(\hat{\mathcal H}_s)$ (unlike in the bosonic case) but rather in $\mathrm{Hom}(\hat{\mathcal H}_s,\hat{\mathcal H}_{s'})$ (cf.~\eqref{eq:MCG_algebra_spin}).

\paragraph{Wilson lines.} An arbitrary element $|v\rangle\in\hat{\mathcal H}_{s^{\mathsf a},s^{\mathsf b}}\cong \mathbb C^{2|0}$ can be written as $|v\rangle=c_0|0;s^{\mathsf a}s^{\mathsf b}\rangle+c_1|1;s^{\mathsf a}s^{\mathsf b}\rangle$ for some coefficients $c_0,c_1\in\mathbb C$. Furthermore, all operators $\mathcal O\in\operatorname{End}(\mathcal H_{s^{\mathsf a},s^{\mathsf b}})$ can be represented as $2\times2$ complex matrices. The Wilson lines themselves, in particular, are represented as $2\times2$ matrices. Their explicit form can be inferred from the expression for our basis vectors in terms of those of the bosonic parent, and the fact that the Wilson lines in the bosonic theory act as in~\eqref{eq:SF_fusion}. For example,
\begin{equation}
\begin{aligned}
W^{(\mathsf a)}(1,0)|1;\text{NS-NS}\rangle&=\frac{1}{\sqrt2}W^{(\mathsf a)}(1,0)(|1,0\rangle+|1,2\rangle)\\
&=\frac{1}{\sqrt2}e^{-i\pi}(|1,0\rangle+|1,2\rangle)\\
&=-|1;\text{NS-NS}\rangle\,,
\end{aligned}
\end{equation}
and
\begin{equation}
\begin{aligned}
W^{(\mathsf b)}(1,0)|1;\text{NS-NS}\rangle&=\frac{1}{\sqrt2}W^{(\mathsf b)}(1,0)(|1,0\rangle+|1,2\rangle)\\
&=\frac{1}{\sqrt2}(|2,0\rangle+|2,2\rangle)\\
&=+|0;\text{NS-NS}\rangle\,.
\end{aligned}
\end{equation}

The rest of matrix elements are computed using the same idea. Denoting the semion by $\varsigma=(1,0)$, and the fermion by $\psi=(0,2)$, the end result is the Pauli matrices
\begin{equation}
W^{(\mathsf a)}_{s^{\mathsf a},s^{\mathsf b}}(\varsigma)=\sigma^3,\qquad W^{(\mathsf b)}_{s^{\mathsf a},s^{\mathsf b}}(\varsigma)=\sigma^1,\qquad W^{(\mathsf c)}_{s^{\mathsf a},s^{\mathsf b}}(\psi)=-s^{\mathsf c}\boldsymbol 1_2\,.
\end{equation}
(As before, the fact that $W^{(\mathsf c)}(\varsigma)$ is independent of $s^\mathsf a,s^\mathsf b$ is rather particular to this simple system; in generic spin TQFTs these matrices depend non-trivially on the spin structure.)

One can easily check that the Wilson lines satisfy the fusion rules of the theory, namely $\varsigma^2=\psi^2=\boldsymbol1$.

\paragraph{Time-reversal.} Finally, we implement time-reversal as an explicit operator in $\hat{\mathcal H}_{s^{\mathsf a},s^{\mathsf b}}$. We write $\mathsf T=\tau K$, where $K$ denotes complex conjugation and $\tau\in\mathbb C^2\times\mathbb C^2$; this factorisation is not canonical, in the sense that $\tau$ and $K$ are separately convention-dependent -- only their product is meaningful. We shall fix them by declaring that our basis is real, $K|\alpha;s^{\mathsf a}s^{\mathsf b}\rangle=|\alpha;s^{\mathsf a}s^{\mathsf b}\rangle$, where $\alpha=0,1$. In other words, $K$ acts by complex-conjugating the coefficients:
\begin{equation}
K(c_0|0;s^{\mathsf a}s^{\mathsf b}\rangle+c_1|1;s^{\mathsf a}s^{\mathsf b}\rangle)\equiv c^*_0|0;s^{\mathsf a}s^{\mathsf b}\rangle+c^*_1|1;s^{\mathsf a}s^{\mathsf b}\rangle\,.
\end{equation}
Naturally, this definition is not basis-independent. But $\mathsf T$, which is the object we care about, is, so this is enough for our purposes.

We recall that $\mathsf T$ acts on the $U(1)$ fields as $\mathsf T(a)=a-b,\ \mathsf T(b)=2a-b$ (cf.~\eqref{eq:t_rev_SF}). This induces the following transformation on the Wilson lines:
\begin{equation}
\begin{aligned}
W^{(\mathsf c)}(\varsigma)&=\exp i\oint_{\mathsf c} a\mapsto \exp i\oint_{\mathsf c} (a-b)\equiv W^{(\mathsf c)}(\varsigma)W^{(\mathsf c)}(\psi)\\
W^{(\mathsf c)}(\psi)&=\exp i\oint_{\mathsf c} b\mapsto \exp i\oint_{\mathsf c} (2a-b)\equiv W^{(\mathsf c)}(\psi)
\end{aligned}
\end{equation}
where we have used $W(\varsigma)^2=W(\varsigma^2)=\boldsymbol1_2$. More to the point, time-reversal acts on the anyons by fixing the vacuum and the fermion, and by exchanging $\varsigma\leftrightarrow \varsigma\times \psi$.

With this, time-reversal acts on the Hilbert space as follows:
\begin{equation}
\mathsf T\, W^{(\mathsf c)}_{s^\mathsf a,s^\mathsf b}(\alpha)\mathsf T^{-1}=W^{(\mathsf c)}_{s^\mathsf a,s^\mathsf b}(\mathsf T(\alpha))\qquad \mathsf c\in\{\mathsf a,\mathsf b\},
\end{equation}
where $\alpha\in\{\boldsymbol1,\varsigma,\psi,\psi\times \varsigma\}$. As $W(\psi)\propto \boldsymbol1_2$, the only non-trivial equation corresponds to the semion, $W(\varsigma)$, which in our basis reads
\begin{equation}
\tau_{s^{\mathsf a},s^{\mathsf b}} (W^{(\mathsf c)}_{s^{\mathsf a},s^{\mathsf b}}(\varsigma))^*\tau_{s^{\mathsf a},s^{\mathsf b}}^{-1}=-s^{\mathsf c}W^{(\mathsf c)}_{s^{\mathsf a},s^{\mathsf b}}(\varsigma)\qquad \mathsf c\in\{\mathsf a,\mathsf b\}.
\end{equation}
This is nothing but a set of linear equations in the components of $\tau$, with solution
\begin{equation}
(\tau_{s^{\mathsf a},s^{\mathsf b}})_{\alpha,\beta}=(-s^{\mathsf b})^\alpha \delta _{\alpha+\beta+\frac{1}{2} (s^{\mathsf a}+1)}
\end{equation}
up to an inconsequential global phase, and where $\alpha,\beta=0,1$ and $\delta_x=1$ if $x$ is even, and $\delta_x=0$ if odd.

We next note that
\begin{equation}
\begin{aligned}
(\tau\tau^*)_{\alpha,\beta}&=\sum_\gamma \tau_{\alpha,\gamma}\tau^*_{\gamma,\beta}\\
&=\sum_\gamma (-s^{\mathsf b})^{\alpha+\gamma} \delta _{\alpha+\gamma+\frac{1}{2} (s^{\mathsf a}+1)}\delta _{\beta+\gamma+\frac{1}{2} (s^{\mathsf a}+1)}\\
&=(-s^{\mathsf b})^{\frac{1}{2} (s^{\mathsf a}+1)} \delta_{\alpha+\beta}\,.
\end{aligned}
\end{equation}

Finally, observe that the expression above can also be written as $\tau\tau^*=(-1)^{\Arf(s)}$, which means that time-reversal satisfies
\begin{equation}
\mathsf T^2=\tau\tau^*\equiv (-1)^{\Arf(s)}\,.
\end{equation}

In this theory, fermion parity is trivial, $(-1)^F\equiv 1$. Therefore, the equation above means that the time-reversal algebra $\mathsf T^2=(-1)^F$ is deformed when acting on the Hilbert space, in the form
\begin{equation}
\mathsf T^2=(-1)^F\times(-1)^{\Arf(s)}
\end{equation} 
which is precisely what we expected, given that the theory has $\nu=2\mod4$.

%{\color{red}
%here we should explain why arf has nothing to do with fermion parity. Indeed, there are no fermions in the torus hilbert space, and so $(-1)^F$ is invisible. We would need to quantise the theory on some other 3fold to see it. On the other hand, arf is singalling an anomaly. indeed, $\mathsf T^2=(-1)^F$ is realised projectively. This is interpreted as $\nu=2\mod4$.
%
%although strictly speaking this does suggest that $(-1)^F$ is in the algebra. the reason is that there are no anomalies associated to $\mathsf T^2=1$, and so the projective phase implies that $\mathsf T^2$ is non-trivial. The only reasonable candidate is $(-1)^F$, because this is the only operator in the theory that acts trivially on the lines, so if something is on the r.h.s.~of $\mathsf T^2$ it must be this operator.
%}

\subsection{$\nu$ odd: fermion layer}\label{sec:other}

In this section we consider time-reversal invariant theories with $\nu$ odd and show that they realize the expected behavior associated with the fermion layer. We work out in detail here the example of $SO(3)_3$ Chern-Simons theory that has $\nu=3$. One can repeat the exercise for other theories with $\nu$ odd, such as $SO(5)_5$, which has $\nu=5$; the main conclusions are identical.

With this in mind, we next study the spin TQFT $SO(3)_3$. This is the smallest intrinsically fermionic topological theory, in the sense that it supports both bosonic and fermionic states, unlike the previous section, where all states were bosonic. The presence of fermionic states is directly related to the presence of Majorana lines, i.e., of fixed-points of the condensing fermion. These will introduce new ingredients into the picture.

The bosonic parent of the theory is $Spin(3)_3=SU(2)_6$, which becomes $SO(3)_3$ upon gauging the $\mathbb Z_2$ center. So we first construct the bosonic theory. This theory has seven states, labelled by their isospin: $|j\rangle$, where $j=0,\frac12,1,\dots,3$. Under modular transformations, these states transform as
\begin{equation}
\begin{aligned}
S|j\rangle&=\sum_{j'}S_{j,j'}|j'\rangle,\qquad S_{j,j'}=\frac12\sin\frac\pi8(2j+1)(2j'+1)\\
T|j\rangle&=e^{\pi i(2j (2 j+2)-3)/16}|j\rangle\,.
\end{aligned}
\end{equation}

The condensing fermion $\psi$ corresponds to the line $j=3$. The braiding phase of a generic line $j$ with respect to $\psi$ is $B(j,\psi)=(-1)^{2j}$, which means that the NS-lines are those with integral isospin, and the R-lines are those with half-integral isospin:
\begin{equation}
\begin{aligned}
\text{NS}&\colon\quad\mathcal A_\text{NS}=\{j=0,1,2,3\}\\
\text{R}&\colon\quad\mathcal A_{\hspace{2pt}\text{R}\hspace{3pt}}=\{j=\tfrac12,\tfrac32,\tfrac52\}\,.
\end{aligned}
\end{equation}

The NS-lines are all in two-dimensional orbits, paired up as follows:
\begin{equation}
0\overset{\times\psi}\longleftrightarrow3,\qquad1\overset{\times\psi}\longleftrightarrow2\,.
\end{equation}
On the other hand, in the R sector there is one two-dimensional orbit, and one fixed-point:
\begin{equation}
\frac12\overset{\times\psi}\longleftrightarrow\frac52,\qquad \frac32\tikz[baseline=-4pt]{\node[rotate=-90,scale=1.1] at (0,0) {$\curvearrowleft$};}\!\!\tikz[baseline=-2pt]{\node at (0,0) {${}^{\times\psi}$};}\,.
\end{equation}
In other words, $0,1,2,3$ are all $a$-type; $\frac12,\frac52$ are both $x$-type; and $\frac32$ is $m$-type.

\paragraph{Hilbert space.} With the information above we have all we need in order to construct the Hilbert space of the fermionic theory. As usual, the states of the condensed phase $SO(3)_3$ are expressed in terms of those of its parent, the specific expression being determined by the choice of spin structure:
\begin{itemize}
\item If we take \textbf{NS-NS} boundary conditions, the two states are
\begin{equation}
\begin{aligned}
|0;\text{NS-NS}\rangle&=\frac{1}{\sqrt2}(|0\rangle+|3\rangle)\\
|1;\text{NS-NS}\rangle&=\frac{1}{\sqrt2}(|1\rangle+|2\rangle)\,.
\end{aligned}
\end{equation}

\item If we take \textbf{NS-R} boundary conditions, the two states are
\begin{equation}
\begin{aligned}
|0;\text{NS-R}\rangle&=\frac{1}{\sqrt2}(|0\rangle-|3\rangle)\\
|1;\text{NS-R}\rangle&=\frac{1}{\sqrt2}(|1\rangle-|2\rangle)\,.
\end{aligned}
\end{equation}

\item If we take \textbf{R-NS} boundary conditions, the two states are
\begin{equation}
\begin{aligned}
|0;\text{R-NS}\rangle&=\frac{1}{\sqrt2}(|1/2\rangle+|5/2\rangle)\\
|1;\text{R-NS}\rangle&=|3/2\rangle\,.
\end{aligned}
\end{equation}

\item If we take \textbf{R-R} boundary conditions, the two states are
\begin{equation}
\begin{aligned}
|0;\text{R-R}\rangle&=\frac{1}{\sqrt2}(|1/2\rangle-|5/2\rangle)\\
|1;\text{R-R}\rangle&=|3/2;\psi\rangle\,,
\end{aligned}
\end{equation}
where, we remind the reader, $|\alpha;\beta\rangle$ denotes the anyon $\alpha$ in presence of a $\beta$ puncture (cf.~\eqref{eq:torus_states_puncture}).

\end{itemize}

\paragraph{Modularity.} As a check of the formalism so far, let us construct the modular data associated to these states, and check that it behaves as expected from general considerations. The even-spin-structure Hilbert spaces do not contain punctures, which means their modular data is computed in exactly the same way as in the previous section. For example, performing an $S$-transformation on an NS-R state we expect to obtain an R-NS state, which is easily confirmed:
\begin{equation}
\begin{aligned}
S&|0;\text{NS-R}\rangle\\
&=\frac{1}{\sqrt2}(S|0\rangle-S|3\rangle)\\
&=\frac{1}{4}\bigg[\\
&+\sqrt{1-\xi }|0\rangle+|1/2\rangle+\sqrt{1+\xi}|1\rangle+\sqrt{2} |3/2\rangle+\sqrt{1+\xi}|2\rangle+|5/2\rangle+\sqrt{1-\xi }|3\rangle\\
&-\sqrt{1-\xi }|0\rangle+|1/2\rangle-\sqrt{1+\xi}|1\rangle+\sqrt{2} |3/2\rangle-\sqrt{1+\xi}|2\rangle+|5/2\rangle-\sqrt{1-\xi }|3\rangle\\
&\qquad\bigg]\\
&=\frac{1}{2}(|1/2\rangle+\sqrt2|3/2\rangle+|5/2\rangle)\\
&=\frac{1}{\sqrt2}(|0;\text{R-NS}\rangle+|1;\text{R-NS}\rangle)\,,
\end{aligned}
\end{equation}
where in the second line we have denoted $\xi=\sin\pi/4=1/\sqrt2$.

Acting with $S$ on the rest of even-spin-structure Hilbert spaces we see that they are indeed permuted as $S\colon\hat{\mathcal H}_{s^\mathsf a,s^\mathsf b}\to \hat{\mathcal H}_{s^\mathsf b,s^\mathsf a}$; and, moreover, this action is effected by the following matrices:
\begin{equation}
\begin{aligned}
\hat S_{\text{NS-NS}\to\text{NS-NS}}&=\frac12\begin{pmatrix}
+\sqrt{2-\sqrt2}&+\sqrt{2+\sqrt2}\\
+\sqrt{2+\sqrt2}&-\sqrt{2-\sqrt2}
\end{pmatrix}\\
\hat S_{\text{NS-R}\to\text{R-NS}}&=\hat S_{\text{R-NS}\to\text{NS-R}}=\frac{1}{\sqrt2}\begin{pmatrix}
+1&+1\\
+1&-1
\end{pmatrix}\,.
\end{aligned}
\end{equation}

$T$-transformations work similarly. For example, they should map states in the NS-NS sector into the NS-R sector, which is indeed what happens:
\begin{equation}
\begin{aligned}
T|0;\text{NS-NS}\rangle&=\frac{1}{\sqrt2}(T|0\rangle+T|3\rangle)\\
&=e^{-3\pi i/16}\frac{1}{\sqrt2}(|0\rangle-|3\rangle)\\
&=e^{-3\pi i/16}|0;\text{NS-R}\rangle\,.
\end{aligned}
\end{equation}
Acting with $T$ on the rest of basis vectors, one confirms that $T$-transformations map $T\colon\hat{\mathcal H}_{s^\mathsf a,s^\mathsf b}\to \hat{\mathcal H}_{s^\mathsf a,s^\mathsf as^\mathsf b}$, through the following matrices:
\begin{equation}
\begin{aligned}
\hat T_{\text{NS-NS}\to\text{NS-R}}&=\hat T_{\text{NS-R}\to\text{NS-NS}}=e^{-3\pi i/16}\begin{pmatrix}
1&0\\
0&i
\end{pmatrix}\\
\hat T_{\text{R-NS}\to\text{R-NS}}&=\begin{pmatrix}
1&0\\
0&e^{3\pi i/4}
\end{pmatrix}\,.
\end{aligned}
\end{equation}

One can easily check that all the expected properties of the modular group (i.e.,~\eqref{eq:MCG_algebra_spin}) are satisfied, where $(-1)^F=+\boldsymbol1_2$, as all states are bosonic.

The odd-spin-structure Hilbert space $\hat{\mathcal H}_\text{R-R}$ is much more interesting. The state $|0;\text{R-R}\rangle\sim |1/2\rangle-|5/2\rangle$ contains no punctures, so it is a boson, whereas the state $|1;\text{R-R}\rangle=|3/2;\psi\rangle$ has a $\psi$-puncture, so it is a fermion. In other words, in the R-R sector the fermion parity operator is non-trivial, $(-1)^F=\sigma_z$. This makes the analysis of modular transformations more involved. In particular, these transformations should not mix these two states, and they should not take us outside $\hat{\mathcal H}_\text{R-R}$ (as the $s=(+1,+1)$ spin structure is fixed by all of the mapping class group). These expectations are confirmed by direct computation. For example, acting on $|0;\text{R-R}\rangle$ with an $S$-transformation we get
\begin{equation}
\begin{aligned}
S|0;\text{R-R}\rangle&=\frac{1}{\sqrt2}(S|1/2\rangle-S|5/2\rangle)\\
&=\frac{1}{\sqrt2}(|1/2\rangle-|5/2\rangle)\\
&=|0;\text{R-R}\rangle
\end{aligned}
\end{equation}
which is indeed in $\hat{\mathcal H}_\text{R-R}$ (and has not mixed with the fermion, as it never could: modular transformations do not mix configurations with different punctures).

The action of $S$ on $|1;\text{R-R}\rangle=|3/2;\psi\rangle$ is more subtle, because the state contains a puncture, so we need the $S$-matrix in the once-punctured torus. This matrix is given by~\eqref{eq:puncture_S_matrix}
\begin{equation}
\begin{aligned}
S_{3/2,3/2}(\psi)&=\sum_{j=0}^3\frac{\theta_j}{\theta_{3/2}^2}S_{0,j}F_{3/2,3/2}\begin{bmatrix}\psi&3/2\\3/2&j\end{bmatrix}\\
&=\sum_{j=0}^3e^{\pi i(2j (j+1)-15)/8}\times\frac{1}{2} \sin\frac{\pi}{8} (2 j+1)\times(-1)^j\\
&=(-1)^{7/4}\,,
\end{aligned}
\end{equation}
which means that, altogether,
\begin{equation}
\hat S_{\text{R-R}\to\text{R-R}}=\begin{pmatrix}
1&0\\0&(-1)^{7/4}
\end{pmatrix}\,.
\end{equation}

$T$-transformations, on the other hand, do not care about the puncture, so they are just given by the spin of the states:
\begin{equation}
\hat T_{\text{R-R}\to\text{R-R}}=\begin{pmatrix}
1&0\\
0&(-1)^{3/4}
\end{pmatrix}\,.
\end{equation}

These two matrices are also easily seen to satisfy the expected modular properties, namely they are unitary and obey the algebra of $Mp_1(\mathbb Z)$, to wit, $\hat S^2=(\hat S\hat T)^3$, $\hat S^4=(-1)^F$.

\paragraph{Wilson lines.} Given the choice of basis for the different Hilbert spaces as above, one can express the operators of the theory -- the Wilson lines -- as $2\times2$ complex matrices. The even-spin-structure Hilbert spaces contain no punctures, so the basic idea is identical to the previous sections, namely we induce the action on the quotient theory from that of its bosonic parent. In the bosonic parent the Wilson lines act as
\begin{equation}
W^{(\mathsf b)}(j)|j'\rangle=|j\times j'\rangle\equiv \sum_{j''=|j-j'|}^{\min(j+j',6-j-j')}|j''\rangle\,,
\end{equation}
which means that, for example,
\begin{equation}
\begin{aligned}
W^{(\mathsf b)}(1)|0;\text{NS-NS}\rangle&=\frac{1}{\sqrt2}(W^{(\mathsf b)}(1)|0\rangle+W^{(\mathsf b)}(1)|3\rangle)\\
&=\frac{1}{\sqrt2}(|1\rangle+|2\rangle)\,,
\end{aligned}
\end{equation}
which we identify as $|1;\text{NS-NS}\rangle$. Similarly, $W^{(\mathsf b)}(1)|1;\text{NS-NS}\rangle=|0;\text{NS-NS}\rangle+2|1;\text{NS-NS}\rangle$. The $\mathsf a$-cycle computation is analogous: in the bosonic parent Wilson lines act as
\begin{equation}
W^{(\mathsf a)}(j)|j'\rangle=\frac{S_{j,j'}}{S_{0,j'}}|j'\rangle\,,
\end{equation}
which implies that
\begin{equation}
\begin{aligned}
W^{(\mathsf a)}(1)|0;\text{NS-NS}\rangle&=\frac{1}{\sqrt2}(W^{(\mathsf a)}(1)|0\rangle+W^{(\mathsf a)}(1)|3\rangle)\\
&=\frac{1}{\sqrt2}(1 +\sqrt2)(|0\rangle+|3\rangle)\,,
\end{aligned}
\end{equation}
which equals $(1+\sqrt2)|0;\text{NS-NS}\rangle$. Repeating this calculation on all even-spin-structure Hilbert spaces, we obtain the following collection of matrices:
%\begin{equation}
%\begin{alignat}{3}
%W^{(\mathsf a)}_{\text{NS-NS}}(1)&=\begin{pmatrix}1+\sqrt2&0\\0&1-\sqrt2\end{pmatrix},\quad &&
%W^{(\mathsf a)}_{\text{NS-R}}(1)=\begin{pmatrix}1+\sqrt2&0\\0&1-\sqrt2\end{pmatrix},\quad &&
%W^{(\mathsf a)}_{\text{R-NS}}(1)=\begin{pmatrix}1&0\\0&-1\end{pmatrix}\\
%W^{(\mathsf b)}_{\text{NS-NS}}(1)&=\begin{pmatrix}0&1\\1&2\end{pmatrix},\quad &&
%W^{(\mathsf b)}_{\text{NS-R}}(1)=\begin{pmatrix}0&1\\1&0\end{pmatrix},\quad &&
%W^{(\mathsf b)}_{\text{R-NS}}(1)=\begin{pmatrix}1&1\\2&1\end{pmatrix}\,,
%\end{alignat}
%\end{equation}
\begin{equation}
\begin{split}
W^{(\mathsf a)}_{\text{NS-NS}}(1)&=\begin{pmatrix}1+\sqrt2&0\\0&1-\sqrt2\end{pmatrix}\\
W^{(\mathsf a)}_{\text{NS-R}}(1)&=\begin{pmatrix}1+\sqrt2&0\\0&1-\sqrt2\end{pmatrix}\\
W^{(\mathsf a)}_{\text{R-NS}}(1)&=\begin{pmatrix}1&0\\0&-1\end{pmatrix}
\end{split}\hspace{40pt}
\begin{split}
W^{(\mathsf b)}_{\text{NS-NS}}(1)&=\begin{pmatrix}0&1\\1&2\end{pmatrix}\\
W^{(\mathsf b)}_{\text{NS-R}}(1)&=\begin{pmatrix}0&1\\1&0\end{pmatrix}\\
W^{(\mathsf b)}_{\text{R-NS}}(1)&=\begin{pmatrix}1&\sqrt2\\\sqrt2&1\end{pmatrix}\,,
\end{split}
\end{equation}
The same computation for the rest of NS-lines yields $W^{(\mathsf c)}_{s^\mathsf a,s^\mathsf b}(3)=-s^\mathsf c\boldsymbol1_2$ for the transparent fermion, and $W^{(\mathsf c)}_{s^\mathsf a,s^\mathsf b}(2)=W^{(\mathsf c)}_{s^\mathsf a,s^\mathsf b}(1)W^{(\mathsf c)}_{s^\mathsf a,s^\mathsf b}(3)$ (which is the expected relation given the fusion rule $2=1\times3$, i.e., that the lines $j=1,2$ are paired up through fusion with $j=3$). One can also check that these matrices satisfy the algebra required by the fusion rule $1\times1=0+1+2$, namely $W^{(\mathsf c)}_{s^\mathsf a,s^\mathsf b}(1)^2=\boldsymbol1_2+(1-s^\mathsf c)W^{(\mathsf c)}_{s^\mathsf a,s^\mathsf b}(1)$. Finally, it is also checked that, under modular transformations, these matrices are permuted as they should, e.g. $S_{s^\mathsf a,s^\mathsf b}W_{s^\mathsf a,s^\mathsf b}^{(\mathsf a)}(\alpha)(S_{s^\mathsf a,s^\mathsf b})^\dagger=W^{(\mathsf b)}_{s^\mathsf b,s^\mathsf a}(\bar\alpha)$.

We now move on to the odd-spin-structure sector, the R-R Hilbert space. The bosonic state $|0;\text{R-R}\rangle\sim |1/2\rangle-|5/2\rangle$ contains no punctures, so it behaves in the same manner as the states in the even-spin-structure sector, for example
\begin{equation}
\begin{aligned}
W^{(\mathsf b)}(1)|0;\text{R-R}\rangle&=\frac{1}{\sqrt2}(W^{(\mathsf b)}(1)|1/2\rangle-W^{(\mathsf b)}(1)|5/2\rangle)\\
&=\frac{1}{\sqrt2}(|1/2\rangle+|3/2\rangle-|3/2\rangle-|5/2\rangle)\\
&=|0;\text{R-R}\rangle\,.
\end{aligned}
\end{equation}
The fermionic state $|1;\text{R-R}\rangle=|3/2;\psi\rangle$, on the other hand, requires using the data of the once-punctured torus, cf.~\eqref{eq:Wilson_puncture}:
\begin{equation}
\begin{aligned}
W^{(\mathsf b)}(1)|1;\text{R-R}\rangle&=W^{(\mathsf b)}(1)|3/2;\psi\rangle\\
%&=\sum_{j=1/2,3/2,5/2}F_{3/2,j}\begin{bmatrix}3&j\\3/2&1\end{bmatrix}|j;\psi\rangle\\
&=F_{3/2,3/2}\begin{bmatrix}3&3/2\\3/2&1\end{bmatrix}|3/2;\psi\rangle
\end{aligned}
\end{equation}
which evaluates to $-|3/2;\psi\rangle$.

The $\mathsf a$-cycle does not see the puncture (cf.~\eqref{eq:braid_general_state}), and so its evaluation is straightforward. All in all, the Wilson lines in the R-R sector read
\begin{equation}
W^{(\mathsf a)}_{\text{R-R}}(1)=W^{(\mathsf b)}_{\text{R-R}}(1)=\begin{pmatrix}1&0\\0&-1\end{pmatrix}\,,
\end{equation}
together with $W^{(\mathsf c)}_{\text{R-R}}(3)=-\boldsymbol1_2$ for the transparent fermion, and $W^{(\mathsf c)}_{\text{R-R}}(2)=-W^{(\mathsf c)}_{\text{R-R}}(1)$ (as expected from the fusion rule $2=1\times3$). It is easily checked that these matrices behave properly under modular transformations, e.g.~$S_{\text{R-R}}W_{\text{R-R}}^{(\mathsf a)}(\alpha)(S_{\text{R-R}})^\dagger=W^{(\mathsf b)}_{\text{R-R}}(\bar\alpha)$.

\paragraph{Time-reversal.} Finally, we discuss the behaviour of the theory under time-reversal. Recall that $SO(3)_3$ is time-reversal symmetric thanks to the level-rank duality~\cite{Aharony:2016jvv}
\begin{equation}
SO(3)_3\ \longleftrightarrow\ SO(3)_{-3}\times SO(9)_1
\end{equation}

A key aspect of this duality is that time-reversal is not really a symmetry of $SO(3)_3$, but rather a map $\mathsf T\colon SO(3)_3\mapsto SO(3)_{3}\times SO(9)_{-1}$. The factor $SO(9)_1$ is invertible, so we should think of $SO(3)_3$ being time-reversal invariant only if we mod out by SPTs. In the strict sense, it is not.

In the $U(1)_k$ case this obstruction was easily circumvented: the duality $U(1)_k\leftrightarrow U(1)_{-k}\times SO(4)_1$ could be rewritten as $U(1)_k\times U(1)_{-1}\leftrightarrow U(1)_{-k}\times U(1)_{+1}$, i.e., we could break up the invertible factor into two, and split them symmetrically into the two theories. In this situation, we would say time-reversal is not a symmetry of $U(1)_k$, but rather of $U(1)_k\times U(1)_{-1}$: this theory is identical to its conjugate, even taking into account SPTs.

In the $SO(3)_3$ case, no such solution is possible: the SPT $SO(9)_1$ cannot be split into two equal factors; such splitting would require fractional levels $SO(9/2)_1^2$, which is not well-defined as a $3d$ theory.

An equivalent way to phrase this discussion is by thinking of the central charge -- indeed, this number is what classifies $3d$-SPTs with no symmetry. Time-reversal always maps $c$ into $-c$, which means a theory can only be time-reversal invariant, in the strict sense, if $c=0$. If $c\neq0$, we may be able to correct this by multiplying by a suitable SPT, but this is not always possible. Indeed, the SPT $SO(n)_1$ has $c=n/2$, which means we can only correct the central charge in multiples of $1/2$. In other words, the minimal SPT has $c=1/2$, corresponding to a single edge Majorana fermion. Any other SPT will consist of an integral number of copies of this system.

In the $U(1)_k$ case, the central charge takes value $c=1$, so this obstruction is avoidable: we just have to tensor the theory with two copies of the Majorana fermion, i.e., $SO(1)_1^2=U(1)_1$. This makes the central charge of the product theory, $U(1)_k\times U(1)_{-1}$, vanish, making it a valid candidate for a time-reversal invariant theory. In the $SO(3)_3$ case, $c=9/4$, which is not a multiple of $1/2$, which means no redefinition can correct the central charge. The theory is not, and cannot be made, time-reversal invariant in the strict sense. Only in the relative sense when we think of QFTs as absolute theories, modulo invertible ones.

More generally, if a given theory $A$ is known to be time-reversal invariant in the relative sense, then necessarily $c\propto 1/4$. Indeed, time-reversal maps $A$ into $\bar A$, modulo some SPT, and so $A\leftrightarrow \bar A\times SO(n)_1$ for some $n$. The central charge of $A$ therefore satisfies $c(A)=-c(A)+n/2$, which means that $c(A)=n/4$, as claimed. If $c=0\mod 1/4$, i.e., if $c\propto 1/2$, then the theory can be made time-reversal invariant in the strict sense, by considering $A\times SO(2c(A))_{-1}$, whose central charge vanishes. If $c\neq0\mod1/4$, this is not possible. In other words, $c\mod1/4$ measures the most primitive obstruction to being time-reversal invariant as a pure $3d$ theory. This is the $\nu_1$ layer.\footnote{See~\cite{Kobayashi:2021hqp} for a very similar recent discussion.}%Needless to say, this is not the end of the story: if $c=0\mod1/4$, there can be further, more subtle anomalies.

The discussion above is set up in the framework of spin TQFTs, but an analogous situation happens in other families of theories. For example, the minimal \emph{bosonic} SPT is $(E_8)_1$, which has $c=8$, which means that bosonic time-reversal invariant theories always have $c\propto 4$, and that $c\mod4$ is the first layer in the anomaly of time-reversal invariance.

%Going back to our example of $SO(3)_3$, we see that time-reversal cannot be implemented as an operator in $\mathrm{End}(\hat{\mathcal H}_s)$, but rather is to be thought of as an operator in
%\begin{equation}
%\mathrm{Hom}(\hat{\mathcal H}_s(SO(3)_3)\to \hat{\mathcal H}_s(SO(3)_3\times SO(9)_{-1}))
%\end{equation}

Going back to our example of $SO(3)_3$, let us see what we can say about time-reversal, neglecting the fact that it is not an operator acting on $SO(3)_3$, but rather a map from this theory into $SO(3)_{3}\times SO(9)_{-1}$. Time-reversal acts on the lines of $SO(3)_3$ as $1\leftrightarrow 2\equiv 1\times \psi$. If we ignore the SPT, this descends to the Hilbert space action
\begin{equation}
\begin{aligned}
\tau\, (W_{s^\mathsf a,s^\mathsf b}^{(\mathsf c)}(1))^*\,\tau^{-1}&=W_{s^\mathsf a,s^\mathsf b}^{(\mathsf c)}(2)\\
&=-s^\mathsf cW_{s^\mathsf a,s^\mathsf b}^{(\mathsf c)}(1)
\end{aligned}
\end{equation}
where $\mathsf T=\tau K$. The solution to this matrix equation is
\begin{equation}
\begin{aligned}
\tau_{\text{NS-NS}}&\propto\boldsymbol1_2\\
\tau_{\text{NS-R}}&\propto\sigma^z\\
\tau_{\text{R-NS}}&\propto\sigma^x
\end{aligned}
\end{equation}
for the even spin structures, and
\begin{equation}
\tau_\text{R-R}=\begin{pmatrix}
0&z_1\\z_2&0
\end{pmatrix}
\end{equation}
for the odd spin structure, where $z_{1,2}\in \mathbb C$ are some arbitrary coefficients.

Finally, recall that $(-1)^F=\boldsymbol1_2$ for even spin structure, and $(-1)^F=\sigma^z$ for the odd spin structure. It is clear from these expressions that $\tau$ commutes with $(-1)^F$ for even spin structures, and anti-commutes for the odd spin structure 
\begin{equation}
\{\mathsf T,(-1)^F\}=0\,,
\end{equation}
precisely as expected from a system with $\nu$ odd and associated with the $\psi$ layer. We have also established that this behavior is present in other time-reversal invariant theories with $\nu$ odd.

\section*{Acknowledgments}

We would like to thank Theo Johnson-Freyd, Nathan Seiberg, Ryan Thorngren, and Alex Turzillo for useful discussions. Research at Perimeter Institute is supported in part by the Government of Canada through the Department of Innovation, Science and Economic Development Canada and by the Province of Ontario through the Ministry of Colleges and Universities. Any opinions, findings, and conclusions or recommendations expressed in this material are those of the authors and do not necessarily reflect the views of the funding agencies.

\vfill\eject

\appendix

\section{Spin TQFTs and anyon condensation}\label{sec:condensation}

In this section we outline the construction of TQFTs that depend on the spin structure of the underlying manifold. The strategy we will pursue is the following. Given one such theory, one may sum over all spin structures to yield a bosonic TQFT. This corresponds to gauging the zero-form symmetry generated by fermion parity $\mathbb Z_2=\langle(-1)^F\rangle$. This gauging generates a dual $\mathbb Z_2$ $(d-2)$-form symmetry, whose gauging takes us back to the original spin TQFT. Therefore, any spin TQFT can be constructed by gauging a certain $(d-2)$-form symmetry in a bosonic TQFT,\footnote{For example, in $d=2$ this corresponds to a zero-form symmetry. This has been studied recently, see e.g.~\cite{Karch:2019lnn,Hsieh:2020uwb,Kulp:2020iet}. In $d=4$ one gauges a two-form symmetry, cf.~e.g.~\cite{Chen:2018nog}. One should keep in mind that, potentially, an anomaly could make these gaugings ill-defined, e.g.~if summing over spin structures leads to an identically vanishing partition function. This subtlety shall play no role in this work.} and we reduce the problem of constructing spin TQFTs to the more familiar problem of gauging a higher-form symmetry in regular (bosonic) TQFTs. We shall follow this strategy in $d=3$ spacetime dimensions, where one can be quite explicit, thanks to the powerful formalism of modular tensor categories and two-dimensional chiral algebras.

With this in mind, we begin by reviewing known facts about $3d$ TQFTs, and the gauging of one-form symmetries. From the $2d$ point of view this corresponds to extending the chiral algebra by a simple current, and in the condensed-matter language to (abelian) anyon condensation. %In order for the gauging to be consistent, the one-form symmetry must be anomaly-free, but it may have a mixed anomaly with other symmetries. Of particular interest is the case where there is in fact such a mixed anomaly with the Lorentz symmetry, in which case the theory acquires a dependence on the spin structure. This allows us to construct spin TQFTs by condensing a fermion in a bosonic TQFT.

Consider a $3d$ bosonic TQFT. The most basic observable of the theory is the partition function $Z(M)$, where $M$ is a compact 3-manifold. For example, if the manifold takes the form $M=\mathbb S^1\times\Sigma$, with $\mathbb S^1$ a circle representing the time direction, and $\Sigma$ a compact surface, then the partition function computes the dimension of the Hilbert space assigned, by canonical quantization, to the spatial slice:
\begin{equation}\label{eq:Z_dim}
Z(\mathbb S^1\times\Sigma)=\dim(\mathcal H(\Sigma))\,.
\end{equation}

%In writing this we are neglecting the fact that TQFTs typically require a choice of 2-framing (i.e., there is a gravitational anomaly related to the central charge). The formula above is correct only in the so-called canonical framing (cf.~Atiyah); if we choose a different framing, the partition function picks up a certain phase. In what follows we shall ignore this subtlety altogether.

The observables of the TQFT depend only on the topology of $M$, and therefore diffeomorphisms of $\Sigma$ must act unitarily in $\mathcal H(\Sigma)$. Transformations that are continuously connected to the identity act trivially, so effectively we get a unitary representation of the mapping class group, the group of (equivalence classes of) large diffeomorphisms. If one understands the Hilbert space $\mathcal H(\Sigma)$, and the action of the MCG on it, one can compute -- via surgery -- the partition function on an arbitrary 3-manifold $M$.

With this in mind, our main task is to understand the Hilbert space assigned by a TQFT to a compact Riemann surface $\Sigma$, and how Dehn twists act on it. The basic data of the TQFT that determines this information is the following:
\begin{itemize}
\item The set of anyons $\mathcal A$, a finite set. This set contains a distinguished anyon, the vacuum $\boldsymbol1$.

\item The modular matrix $S\colon\mathcal A\times\mathcal A\to\mathbb C$.

\item The topological spin $\theta=e^{2\pi i h}\colon\mathcal A\to U(1)$.

\end{itemize}

The full data of the TQFT involves other, more subtle objects, the so-called $F$- and $R$-symbols. These will play a role later on; for now, the $S$-matrix is enough. By a key result of Verlinde, the dimension of $\mathcal H(\Sigma)$ is determined by $S$ as follows~\cite{VERLINDE1988360,moore1989}:
\begin{equation}
\dim(\mathcal H(\Sigma))=\sum_{\alpha\in\mathcal A}S_{\boldsymbol1,\alpha}^{\chi(\Sigma)}\,,
\end{equation}
where $\chi$ denotes the Euler characteristic ($\chi(\Sigma_g)=2-2g$ for a genus $g$ surface $\Sigma_g$). In particular, the torus has $\chi(\Sigma_1)=0$, which means that $\mathcal H(\Sigma_1)\cong \mathbb C[\mathcal A]$, i.e., a basis of states of the torus Hilbert space is labelled by the anyons of the TQFT. The MCG of the torus, $SL_2(\mathbb Z)=\langle S,T\rangle$, is generated by $S$ and $T:=e^{-2\pi i c/24}\diag(\theta)$.

The theory also admits line defect operators, also labelled by $\mathcal A$. Namely, we can wrap an anyon $\alpha\in\mathcal A$ around the time circle $\mathbb S^1$, which produces a defect Hilbert space $\mathcal H(\Sigma^\alpha)$. From the point of view of the spatial surface, the anyon $\alpha$ looks like a marked point. Given a family of such punctures $\alpha_1,\dots,\alpha_n$, the generalization of the Verlinde formula is~\cite{moore1989}
\begin{equation}\label{eq:general_verlinde_boson}
\dim(\mathcal H(\Sigma^{\alpha_1\cdots\alpha_n}))=\sum_{\alpha\in\mathcal A}S_{\boldsymbol1,\alpha}^{\chi(\Sigma)}\prod_{i=1}^nS_{\alpha_i,\alpha}\,,
\end{equation}
where $\chi(\Sigma_g^{\alpha_1\cdots\alpha_n})=2-2g-n$ for a surface with $g$ handles and $n$ boundary components. The most fundamental surface is the so-called \emph{trinion}, i.e., a sphere with three punctures. This surface defines the \emph{fusion coefficients}:
\begin{equation}\label{eq:trinion_verlinde_boson}
\mathcal N_{\alpha_1,\alpha_2,\alpha_3}:=\dim(\mathcal H(\Sigma_0^{\alpha_1\alpha_2\alpha_3}))\equiv \sum_{\beta\in\mathcal A}\frac{S_{\alpha_1,\beta}S_{\alpha_2,\beta}S_{\alpha_3,\beta}}{S_{\boldsymbol1,\beta}}\,,
\end{equation}
which endows $\mathcal A$ with a product structure, leading to the \emph{fusion ring} of the TQFT. The partition function on an arbitrary surface can be computed by gluing trinions (cf.~the ``pants decomposition''). Using unitarity of $S$ one can recover the general case~\eqref{eq:general_verlinde_boson} from the trinion~\eqref{eq:trinion_verlinde_boson}.

An explicit basis of states on $\mathcal H(\Sigma_g^{\alpha_1\cdots\alpha_n})$ can be written as follows:
\begin{equation}
\begin{tikzpicture}[baseline=0pt]

\draw[thick] (-.34,.34) -- (-.7,.7);
\draw[thick] (-.34,-.34) -- (-.7,-.7);
\fill (-.36,.36) circle (1.5pt);
\fill (-.36,-.36) circle (1.5pt);

\fill (-.65,0) circle (.6pt);
\fill (-.62,.15) circle (.6pt);
\fill (-.62,-.15) circle (.6pt);

\node[scale=.8] at (-.95,.7) {$\alpha_1$};
\node[scale=.8] at (-.95,-.7) {$\alpha_n$};

\draw[thick] (0,0) circle (.5cm);
\node[scale=.6] at (0,0) {$\times$};
%\node[scale=.8] at (0,-.7) {$\beta_1$};
\draw[thick] (.5,0) -- (1.5,0);
\fill (.5,0) circle (1.5pt);
%\node[scale=.8] at (1,.4) {$\beta'_1$};

\fill (1.5,0) circle (1.5pt);
\draw[thick] (2,0) circle (.5cm);
\node[scale=.6] at (2,0) {$\times$};
%\node[scale=.8] at (2,-.7) {$\beta_2$};
\draw[thick] (2.5,0) -- (3.5,0);
\fill (2.5,0) circle (1.5pt);
%\node[scale=.8] at (3,.4) {$\beta'_2$};

\fill (3.5,0) circle (1.5pt);
\draw[thick] (4,0) circle (.5cm);
\node[scale=.6] at (4,0) {$\times$};
%\node[scale=.8] at (4,-.7) {$\beta_3$};
\draw[thick] (4.5,0) -- (5,0);
\fill (4.5,0) circle (1.5pt);

\node[scale=.8] at (5.5,0) {$\cdots$};

\fill (6.5,0) circle (1.5pt);
\draw[thick] (6,0) -- (6.5,0);
\draw[thick] (7,0) circle (.5cm);
\node[scale=.6] at (7,0) {$\times$};
%\node[scale=.8] at (7,-.7) {$\beta_g$};

\end{tikzpicture}\ ,
\end{equation}
where each cross $\times$ represents a handle of $\Sigma_g$. Each segment carries an orientation and an anyon label (which we omit to simplify the notation). Each trivalent vertex with incoming anyons $\alpha,\beta,\gamma$ carries an internal vector index, taking values in $1,2,\dots,\mathcal N_{\alpha,\beta,\gamma}$, which we always leave implicit.

In short, the states of $\mathcal H(\Sigma_g^{\alpha_1\cdots\alpha_n})$ can be represented as labelled oriented graphs with $g$ cycles; leaves labelled by the punctures $\alpha_1,\dots,\alpha_n$; edges labelled by anyons $\alpha\in\mathcal A$; and trivalent vertices labelled by internal vector indices taking $\mathcal N_{\alpha,\beta,\gamma}$ values, as determined by the incident edges $\alpha,\beta,\gamma$.

Different bases of the Hilbert space are related to the one above by the $F$- and $R$-moves, effected by the aforementioned $F$- and $R$-symbols. These are in correspondence with the different pants decompositions of the surface.

%Question: How does Torelli act? Does the basis of the Hilbert space only depend on the homology basis of $\Sigma$?

Of particular relevance is the case of the torus with a single puncture, whose states we label as $|\beta;\alpha\rangle\in\mathcal H(\Sigma_1^\alpha)$, corresponding to the configuration
\begin{equation}\label{eq:torus_states_puncture}
\begin{tikzpicture}[baseline=0pt]

\node at (-2.4,0) {$|\beta;\alpha\rangle\ =$};

\draw[thick] (-.5,0) -- (-1,0);
\draw[thick] (0,0) circle (.5cm);
\node[scale=.6] at (0,0) {$\times$};
\node[scale=.8] at (0,-.7) {$\beta$};
\fill (-.5,0) circle (1.5pt);
\node[scale=.8] at (-1.2,0) {$\alpha$};

\end{tikzpicture}\ .
\end{equation}
There is one such state for each possible vertex, i.e., the degeneracy of $|\beta;\alpha\rangle$ is given by the fusion coefficient $\mathcal N_{\alpha,\beta,\bar\beta}$. In particular, the diagram is allowed only if $\alpha\times \beta\propto\beta+\cdots$, i.e., if $\beta$ may ``absorb'' the puncture $\alpha$. The case of no punctures corresponds to the vacuum anyon $\alpha=\boldsymbol1$, so that all $\beta\in\mathcal A$ are allowed, and they all carry degeneracy $\mathcal N_{\boldsymbol1,\beta,\bar\beta}=1$. For non-trivial $\alpha$, some $\beta\in\mathcal A$ may not contribute to $\mathcal H(\Sigma_1^\alpha)$, and some other $\beta\in\mathcal A$ may contribute more than one state.

Given a basis of states for $\mathcal H(\Sigma_1)$ one can write down the operators acting on this space as matrices. In particular, the Wilson loop operators admit such a representation. Let $W^{(\mathsf c)}(\alpha)$ denote the Wilson loop labelled by the anyon $\alpha\in\mathcal A$ running through the cycle $\mathsf c\in H_1(\Sigma_1,\mathbb Z)=\mathbb Z[\mathsf a]\oplus\mathbb Z[\mathsf b]$, where $\mathsf a,\mathsf b$ are the standard homology cycles. These operators act by inserting $\alpha$ along the given cycle, e.g.
\begin{equation}
\begin{aligned}
W^{(\mathsf a)}(\alpha)|\beta\rangle\ &=\ 
\tikz[baseline=-2pt]{
\draw[thick] (0,0) circle (.5cm);
\fill[white] (-.48,-.14) circle (1.2pt);
\draw[thick,rotate=-45] (-.1,-.5) ellipse (.15cm and .2cm);
\fill[white] (-.32,-.38) circle (1.2pt);
\node[scale=.6] at (0,0) {$\times$};
\node[scale=.8] at (-.85,-.5) {$\alpha$};
\begin{scope}
\clip (-.32,-.38) circle (2pt);
\draw[thick] (0,0) circle (.5cm);
\end{scope}

\node[scale=.8] at (0,-.7) {$\beta$};}\ =\ \frac{S_{\alpha,\beta}}{S_{\boldsymbol1,\beta}}\ 
\tikz[baseline=-2pt]{
\draw[thick] (0,0) circle (.5cm);
\node[scale=.6] at (0,0) {$\times$};
\node[scale=.8] at (0,-.7) {$\beta$};}\\
W^{(\mathsf b)}(\alpha)|\beta\rangle\ &=\quad
\tikz[baseline=-2pt]{

\draw[thick] (0,0) circle (.5cm);
\draw[thick] (0,0) circle (.4cm);

\node[scale=.6] at (0,0) {$\times$};
\node[scale=.8] at (0,-.23) {$\alpha$};

\node[scale=.8] at (0,-.7) {$\beta$};}\ =\quad
\tikz[baseline=-2pt]{
\draw[thick] (0,0) circle (.5cm);
\node[scale=.6] at (0,0) {$\times$};
\node[scale=.8] at (0,-.7) {$\alpha\times\beta$};}
\end{aligned}
\end{equation}
whence
\begin{equation}
\begin{aligned}
\langle \beta'|W^{(\mathsf a)}(\alpha)|\beta\rangle&=\delta_{\beta \beta'}\frac{S_{\alpha \beta}}{S_{\boldsymbol1,\beta}}\\
\langle \beta'|W^{(\mathsf b)}(\alpha)|\beta\rangle&=\mathcal N^{\beta'}{}_{\alpha\beta}
\end{aligned}
\end{equation}
Naturally, given that $S$ interchanges the cycles $\mathsf a$ and $\mathsf b$ (up to a sign), one has
\begin{equation}\label{eq:modular_torus_wilson}
W^{(\mathsf a)}(\alpha)=S W^{(\mathsf b)}(\alpha)S^\dagger,\qquad W^{(\mathsf b)}(\alpha)=S W^{(\mathsf a)}(\bar \alpha)S^\dagger
\end{equation}
which is just the statement that the characters $S_{\alpha \beta}/S_{\boldsymbol 1,\beta}$ diagonalize the fusion rules.

The higher-genus case is analogous. Given a basis of $\mathcal H(\Sigma_g^{\alpha_1\dots\alpha_n})$ one can express the Wilson lines as matrices. As above, a Wilson loop $W^{(\mathsf c)}(\alpha)$ inserts the anyon $\alpha$ along the cycle $\mathsf c\in H_1(\Sigma^{\alpha_1\dots\alpha_n}_g,\mathbb Z)$. For example, the $\mathsf a$-cycles are identical to the torus, inasmuch as wrapping an anyon in the orthogonal cycle is a local operation: one can shrink it to a point. The value of $W^{(\mathsf a^i)}$ acting on a given state only depends on the line running through the segment orthogonal to $\mathsf a^i$, irrespective of what the rest of the state is doing:
%\begin{equation}\label{eq:braid_general_state}
%\tikz[baseline=-2pt]{
%\draw[thick] (0,0) circle (.5cm);
%\fill[white] (-.48,-.14) circle (1.2pt);
%\draw[thick,rotate=-45] (-.1,-.5) ellipse (.15cm and .2cm);
%\fill[white] (-.32,-.38) circle (1.2pt);
%\node[scale=.6] at (0,0) {$\times$};
%\node[scale=.8] at (-.85,-.5) {$\alpha$};
%\begin{scope}
%\clip (-.32,-.38) circle (2pt);
%\draw[thick] (0,0) circle (.5cm);
%\end{scope}
%\draw[thick] (0,.5) -- (0,1);
%\draw[thick] ({.5*cos(70)},{.5*sin(70)}) -- ({cos(70)},{sin(70)});
%\draw[thick] ({.5*cos(110)},{.5*sin(110)}) -- ({cos(110)},{sin(110)});
%
%\fill (0,.5) circle (1.5pt);
%\fill ({.5*cos(70)},{.5*sin(70)}) circle (1.5pt);
%\fill ({.5*cos(110)},{.5*sin(110)}) circle (1.5pt);
%
%\node[scale=.8] at (0,-.7) {$\beta$};}\ =\ \frac{S_{\alpha,\beta}}{S_{\boldsymbol1,\beta}}\ 
%\tikz[baseline=-2pt]{
%\draw[thick] (0,0) circle (.5cm);
%\node[scale=.6] at (0,0) {$\times$};
%\node[scale=.8] at (0,-.7) {$\beta$};
%\draw[thick] (0,.5) -- (0,1);
%\draw[thick] ({.5*cos(70)},{.5*sin(70)}) -- ({cos(70)},{sin(70)});
%\draw[thick] ({.5*cos(110)},{.5*sin(110)}) -- ({cos(110)},{sin(110)});
%
%\fill (0,.5) circle (1.5pt);
%\fill ({.5*cos(70)},{.5*sin(70)}) circle (1.5pt);
%\fill ({.5*cos(110)},{.5*sin(110)}) circle (1.5pt);}\ .
%\end{equation}
\begin{equation}\label{eq:braid_general_state}
\tikz[baseline=-2pt]{
\begin{scope}
\clip (-1,-1) rectangle (0,1);
\draw[thick] (0,0) circle (.5cm);
\fill[white] (-.49,.15-.075) circle (1.2pt);
\draw[thick] (-.5,-.075) ellipse (.2cm and .15cm);
\fill[white] (-.45,-.15-.075) circle (1.2pt);
\node[scale=.8] at (-.85,-.05) {$\alpha$};
\begin{scope}
\clip (-.45,-.15-.075) circle (2pt);
\draw[thick] (0,0) circle (.5cm);
\end{scope}
\end{scope}

%\node[scale=.8] at (.5,-.51) {$\cdots$};
\draw[thick,dashed] (0,-.5) -- (.8,-.5);
\draw[thick,dashed] (0,.5) -- (.8,.5);

\node[scale=.8] at (-.35,-.65) {$\beta$};}\ \ =\ \ \frac{S_{\alpha,\beta}}{S_{\boldsymbol1,\beta}}\ 
\tikz[baseline=-2pt]{
\begin{scope}
\clip (-1,-1) rectangle (0,1);
\draw[thick] (0,0) circle (.5cm);
\end{scope}

%\node[scale=.8] at (.5,-.51) {$\cdots$};
\draw[thick,dashed] (0,-.5) -- (.8,-.5);
\draw[thick,dashed] (0,.5) -- (.8,.5);

\node[scale=.8] at (-.35,-.65) {$\beta$};}
\end{equation}

The $\mathsf b$-cycles, on the other hand, cannot be shrunk, and so depend on the entire configuration around such cycle: the lines running therein, and the punctures going in and out. For example, the once-punctured torus has~\cite{Li:1989hs}
\begin{equation}\label{eq:Wilson_puncture}
W^{(\mathsf b)}(\alpha)|\beta;\gamma\rangle\ =\quad
\tikz[baseline=-2pt]{

\draw[thick] (-1,0) -- (-.5,0);
\draw[thick] (0,0) circle (.5cm);
\draw[thick] (0,0) circle (.4cm);
\fill (-.5,0) circle (1.5pt);

\node[scale=.6] at (0,0) {$\times$};
\node[scale=.8] at (0,-.23) {$\alpha$};
\node[scale=.8] at (-1.2,0) {$\gamma$};

\node[scale=.8] at (0,-.7) {$\beta$};}\ =\quad F_{\beta,\alpha\times\beta}\begin{bmatrix}\gamma&\alpha\times\beta\\\beta&\alpha\end{bmatrix}\ 
\tikz[baseline=-2pt]{
\draw[thick] (-1,0) -- (-.5,0);
\fill (-.5,0) circle (1.5pt);
\draw[thick] (0,0) circle (.5cm);
\node[scale=.6] at (0,0) {$\times$};
\node[scale=.8] at (0,-.7) {$\alpha\times\beta$};
\node[scale=.8] at (-1.2,0) {$\gamma$};}\ .
\end{equation}
Configurations with more punctures carry more factors of $F$. The matrix elements of arbitrary configurations of Wilson lines, on surfaces with arbitrary genus and arbitrary punctures, is entirely determined in terms of the TQFT data of the theory. Generalizing~\eqref{eq:modular_torus_wilson}, the lines around the different cycles are unitarily related through the MCG of the surface.

The invertible defects -- the abelian punctures -- correspond to group symmetries of the theory. These are line operators, so the symmetry is a higher-form symmetry, in this case a one-form symmetry. Gauging the symmetry corresponds to summing over all possible insertions of the defect. This produces a new TQFT, whose set of anyons $\hat{\mathcal A}$ and modular data $\hat S$ are fixed in terms of the data of the ungauged theory. Making this procedure precise is the goal of the rest of this appendix.

The one-form symmetry group is always a finite abelian group, i.e., a product of cyclic groups. Each abelian anyon in $\mathcal A$ generates a cyclic subgroup; condensing this anyon means gauging this subgroup. If we are interested in gauging a product of cyclic groups, we can always condense a single generator at a time, iteratively. We can therefore assume without loss of generality that the group to be gauged is cyclic, say, $\mathbb Z^{(1)}_n=\langle \mathsf g\rangle$, with $\mathsf g\in\mathcal A$ a certain abelian anyon. The $\mathbb Z^{(1)}_n$ symmetry partitions the spectrum $\mathcal A$ into $n$ equivalence classes, according to their braiding with respect to $\mathsf g$:
\begin{equation}\label{eq:braid_q}
\mathcal A=\bigsqcup_{q=0}^{n-1}\mathcal A_q,\qquad \mathcal A_q:=\{\alpha\in\mathcal A\,\mid\, B(\mathsf g,\alpha)=e^{2\pi i q/n}\}\,,
\end{equation}
where $B(\mathsf g,\alpha):=S_{\mathsf g,\alpha}/S_{\boldsymbol1,\alpha}\equiv\theta(\mathsf g\times \alpha)/\theta(\mathsf g)\theta(\alpha)$ is the braiding phase with respect to $\mathsf g$. The modular data of the theory behaves properly with respect to this grading, e.g.~\cite{Schellekens:1990xy}
\begin{equation}\label{eq:grading_Zn_S}
S_{\mathsf g^i\times \alpha,\mathsf g^j\times\beta}=B(\mathsf g,\mathsf g)^{ij}B(\mathsf g,\alpha)^jB(\mathsf g,\beta)^iS_{\alpha,\beta}
\end{equation}

The 't Hooft anomaly of $\mathbb Z^{(1)}_n$ is given by $B(\mathsf g,\mathsf g)$, which must equal $1$ if the symmetry is to be gauged. In this situation, one can prove that $\theta(\mathsf g)=\pm1$, i.e., the generator is either a boson or a fermion. In the former case, the gauging yields another bosonic TQFT, while in the latter case the theory acquires a dependence on the spin structure, i.e., it becomes a spin TQFT. For now, we assume that $\mathsf g$ is a boson, and return to the more interesting case of fermionic quotients later on.

\subsection{Boson anyon condensation}

We begin with some bosonic TQFT with anyons $\mathcal A$ and modular matrix $S$, and wish to condense some boson $\mathsf g\in\mathcal A$, to produce some other bosonic TQFT, with anyons $\hat{\mathcal A}$ and new modular matrix $\hat S$. The standard lore of this procedure is as follows. First, in order to construct $\hat{\mathcal A}$, one performs the following three steps~\cite{MOORE1989422}:
\begin{enumerate}
\item Select the set of neutral lines, $\mathcal A_0$ (cf.~\eqref{eq:braid_q}), i.e., those with trivial braiding with respect to $\mathsf g$.

\item Identify any two lines that are in the same $\mathbb Z^{(1)}_n$-orbit, i.e., if they differ by the action of $\mathsf g^j$ for some $j\in\mathbb Z_n$.

\item If a given $\mathbb Z^{(1)}_n$-orbit has less than $n$ elements, it splits into several different anyons in $\hat{\mathcal A}$. Specifically, if the length is $\ell$, then the orbit descends to $n/\ell$ copies in the condensed theory.

\end{enumerate}

%Given the set $\hat{\mathcal A}$, there are several competing proposals for how the matrix $\hat S$ is to be computed. To the best of our knowledge, all of them proceed by postulating a reasonable ansatz, and fixing the arbitrary coefficients through consistency conditions, such as modularity. In what follows we shall describe a different way to understand the condensation, which gives a satisfying picture that further justifies the three-step procedure above, and allows us to compute the modular data of the condensed theory from first principles, with no need to introduce any ans\"atze. It also admits a natural extension to spin TQFTs which shines a new light -- and goes beyond -- what is currently understood about such theories.

In what follows we shall describe the geometric interpretation of these rules, which will allow us to compute the modular data of the condensed theory from first principles, with no need to introduce any ans\"atze. It also admits a natural extension to spin TQFTs which shines a new light -- and goes beyond -- what is currently understood about such theories.

The main idea to obtain $\hat{\mathcal A}$ is to find the torus Hilbert space of the condensed theory, from which one can read off the set of anyons by writing down a basis of vectors (recall that $\mathcal H(\Sigma_1)\cong \mathbb C[\mathcal A]$). The condensed theory is obtained by gauging the $\mathbb Z_n$ one-form symmetry, which means we are to insert the generator $\mathsf g$ in all possible ways. Summing over all insertions $\mathsf g^j$, $j=0,1,\dots,n-1$, along the spatial cycles project the Hilbert space into the invariant states. Insertions along the temporal cycles introduce twisted sectors. We will next see that insertions along the three cycles in $M=\mathbb S^1\times\Sigma_1$ indeed reproduce the three steps above.

Let us begin with the time circle. Inserting $\mathsf g^j$ along the time direction means taking the torus with a puncture labelled by $\mathsf g^j$. Therefore, the states of the condensed theory are generically of the form $|\alpha;\mathsf g^j\rangle$ for some $\alpha\in\mathcal A$. In other words, the Hilbert space of the condensed theory must be a subspace of the Hilbert space of the original theory, in the presence of an arbitrary puncture:
\begin{equation}
\hat{\mathcal H}(\Sigma_1)\subseteq \bigoplus_{j=0}^{n-1}\mathcal H(\Sigma_1^{\mathsf g^j})\,.
\end{equation}

The states of $\mathcal H(\Sigma_1^{\mathsf g^j})$ are labelled by anyons $\alpha\in\mathcal A$ with the property $\mathsf g^j\times \alpha=\alpha$. In particular, $j$ must be proportional to the length of the $\mathbb Z^{(1)}_n$-orbit of $\alpha$. We shall denote this orbit by $[\alpha]$, and its length $\ell_\alpha:=|[\alpha]|$ equals the minimal integer such that $\mathsf g^{\ell_\alpha}\times\alpha=\alpha$. This integer divides $n$, and any other integer $j$ with $\mathsf g^j\times \alpha=\alpha$ is of the form $j=\ell_\alpha k$, for some integer $k=0,1,\dots,n/\ell_\alpha-1$. This reproduces the third condition above, namely if a given orbit is shorter than $\ell_\alpha=n$, it descends to $n/\ell_\alpha$ copies in the condensed theory. The copies just label the number of insertions of the symmetry defect we use to create the state.

Let us now move on to the spatial circles; insertions of the symmetry elements along these circles shall project into the invariant subspace. In this case, the meaning of \emph{invariant} depends on which cycle we insert the symmetry element on; a symmetry along the $\mathsf a$-cycle acts via braiding, and along the $\mathsf b$-cycle via fusion. In the end, we must have states that are invariant under $\mathsf g$, both with respect to braiding and to fusion. Let us discuss these two cases in turn.
\begin{itemize}
\item Take first the $\mathsf a$-cycle, which is the circle that is orthogonal to the one we use to create states. Given a state created by a line $\alpha\in\mathcal A$ running along the $\mathsf b$-cycle, the configuration we obtain by inserting $\mathsf g$ is $B(\alpha,\mathsf g)|\alpha;\mathsf g^{\ell_\alpha k}\rangle$ (cf.~\eqref{eq:braid_general_state}). The phase $B(\alpha,\mathsf g)$ equals $e^{2\pi i q/n}$ for $\alpha\in\mathcal A_q$ (cf.~\eqref{eq:braid_q}). Summing over all insertions $\mathsf g^j$ produces the phase
\begin{equation}
\begin{aligned}
\sum_{j=0}^{n-1}B(\alpha,\mathsf g^j)&=\sum_{j=1}^{n-1}e^{2\pi i qj/n}\\
&=n\delta_{q,0}
\end{aligned}
\end{equation}
which indeed projects to the states with $q=0$, i.e., to $\alpha\in\mathcal A_0$. We thus reproduce the first condition, namely the states in the condensed theory must be neutral under $\mathbb Z_n^{(1)}$, i.e., taken from $\mathcal A_0$.

\item Finally, if we now consider the second spatial circle, the $\mathsf b$-cycle, and insert $\mathsf g^j$, we obtain $|\mathsf g^j\times \alpha;\mathsf g^{\ell_\alpha k}\rangle$. Summing over all $j$ (and normalizing to have unit norm) leads to
\begin{equation}\label{eq:diag_basis}
|[\alpha],k\rangle:=\frac{1}{\sqrt{\ell_\alpha}}\sum_{j=0}^{\ell_\alpha-1}|g^j\times\alpha;g^{k\ell_\alpha}\rangle,\qquad [\alpha]\in \mathcal A_0/\sim,\ k\in\mathbb Z_{n/\ell_\alpha}\,.
\end{equation}
which is indeed invariant under $\mathsf g$, where now this symmetry acts via fusion (i.e., $\alpha\mapsto \mathsf g\times\alpha$). We thus reproduce the second condition, namely the fact that the anyons of the condensed theory $\hat{\mathcal A}$ are labelled by $\mathbb Z^{(1)}_n$ orbits.

\end{itemize}

We thus see that, as expected, the insertions along the three circles indeed reproduce the three conditions we are used to. All in all, a basis of states is labelled by a pair of indices: $[\alpha]$, denoting a $\mathbb Z_n^{(1)}$ orbit of neutral lines $\alpha\in\mathcal A_0$, plus a degeneracy label taking values in $k=0,1,\dots,n/\ell_\alpha$. This degeneracy label, arguably the most subtle ingredient so far, is in fact quite natural from the point of view of gauging $\mathbb Z_n^{(1)}$: $k\ell_\alpha$ just denotes how many copies of the $\mathsf g$-puncture we insert in order to create the state, i.e., from which twisted Hilbert space the state comes from.

The presentation of $\hat{\mathcal H}(\Sigma_1)$ above also gives us a natural way to compute the modular data of the condensed theory, in particular, the modular matrix $\hat S$. Specifically, a modular transformation acting on a state $|[\alpha];k\rangle$ is nothing but the $S$-matrix of the uncondensed theory, in the presence of a puncture $g^{k\ell_\alpha}$:
\begin{equation}
\langle [\alpha];k|\hat S|[\alpha'];k'\rangle=\delta_{k\ell_\alpha,k'\ell_{\alpha'}}\sqrt{\ell_\alpha\ell_{\alpha'}} S_{\alpha,\alpha'}(\mathsf g^{k\ell_{\alpha}})\,.
\end{equation}

The modular matrix in the once-punctured torus can be expressed in terms of the $F$-symbols of the parent theory, namely~\cite{Li:1989hs}
\begin{equation}\label{eq:puncture_S_matrix}
S_{\alpha,\alpha'}(\mathsf g^j)=\sum_{\beta\in\alpha\times\alpha'}\frac{\theta(\beta)}{\theta(\alpha)\theta(\alpha')}S_{\boldsymbol1,\beta}F_{\alpha,\alpha'}\begin{bmatrix}
\mathsf g^j&\alpha'\\\alpha&\beta
\end{bmatrix}\,.
\end{equation}

The basis~\eqref{eq:diag_basis} of $\hat{\mathcal H}(\Sigma_1)$ is natural because it makes $\hat S$ block-diagonal, but it does not correspond to the anyon basis. The most obvious way to see this is that the would-be quantum dimension $d_{[\alpha];k}=\hat S_{[\alpha];k, [\boldsymbol1];0}/\hat S_{[\boldsymbol1];0, [\boldsymbol1];0}$ vanishes for $k\neq0$.

In order to identify the anyon basis we can look at the dual $\mathbb Z_n^{(0)}$ symmetry. The charged states are those with the puncture. More precisely, in the diagonal basis the states transform as $\mathbb Z_n^{(0)}\colon|[\alpha];k\rangle\mapsto e^{2\pi i k/\ell_\alpha}|[\alpha];k\rangle$. In the anyon basis, this symmetry should act as a cyclic permutation of the anyons, that is, as $\mathbb Z_n^{(0)}\colon|[\alpha];\hat k\rangle\mapsto |[\alpha];\hat k+1\rangle$ for some label $\hat k$. We conclude that the anyon basis is in fact nothing but the Fourier transform (Pontryagin dual) of the diagonal basis~\eqref{eq:diag_basis}:
\begin{equation}
\begin{aligned}
|[\alpha],\hat k\rangle:&=\frac{1}{\sqrt{n/\ell_\alpha}}\sum_{k=0}^{n/\ell_\alpha-1}e^{2\pi i \hat kk \ell_\alpha/n}|[\alpha];k\rangle \\
&=\frac{1}{\sqrt{n}}\sum_{k=0}^{n/\ell_\alpha-1}\sum_{j=0}^{\ell_\alpha-1}e^{2\pi i \hat kk \ell_\alpha/n}%|g^j\times\alpha;g^{k\ell_\alpha}\rangle
\quad\begin{tikzpicture}[baseline=0pt]

\draw[thick] (0,.5) -- (0,1);
\draw[thick] (0,0) circle (.5cm);
\node[scale=.6] at (0,0) {$\times$};
\node[scale=.8] at (0,-.7) {$\mathsf g^j\times\alpha$};
\fill (0,.5) circle (1.5pt);
\node[scale=.8] at (0,1.2) {$\mathsf g^{k\ell_\alpha}$};

\end{tikzpicture}\,,
\end{aligned}
\end{equation}
where $\hat k\in\mathbb Z^*_{n/\ell_\alpha}$. In this basis, the quantum dimension takes the expected value $d_{[\alpha];\hat k}=(\ell_\alpha/n)S_{\alpha,\boldsymbol1}/S_{\boldsymbol1,\boldsymbol1}=(\ell_\alpha/n)d_\alpha$. %This follows from $\hat S_{[\alpha];\hat k,[\boldsymbol1];0}=\ell_\alpha S_{\alpha,\boldsymbol1}$.
 The anyons of the quotient $\hat{\mathcal A}$ create the states $|[\alpha],\hat k\rangle$ by acting on the vacuum.

Given the matrix $\hat S$ in the fusion basis, one can use the Verlinde formula to compute the dimension of the Hilbert space of the condensed theory, for an arbitrary Riemann surface, with an arbitrary number of punctures. In the particular case of no external punctures, the formula only involves matrix elements with the vacuum, in which case the matrix with punctures $S(\mathsf g^j)$ does not contribute except for the vacuum insertion, that is, the regular $S$ matrix of the uncondensed theory. In other words, the dimension of the Hilbert space of the condensed theory, in the case of no punctures, can conveniently be computed using only the $S$ matrix of the parent theory, without the need to know the $F$-symbols:
\begin{equation}
\begin{aligned}
\dim(\hat{\mathcal H}(\Sigma_g))&=\sum_{\alpha\in\hat{\mathcal A}}\hat S_{\boldsymbol1,\alpha}^{\chi(\Sigma)}\\
&\equiv \sum_{\alpha\in\mathcal A_0}\frac{n}{\ell_\alpha^{2g}}S_{\boldsymbol1,\alpha}^{\chi(\Sigma)}\,.
\end{aligned}
\end{equation}

It is possible to generalize the expressions above to the case where there is a non-trivial background flux for the $\mathbb Z_n^{(0)}$ magnetic symmetry dual to the gauged $\mathbb Z_n^{(1)}$ symmetry. For example, if the flux of such background is $q\in\mathbb Z_n$, then the states are created from the subset $\mathcal A_q$ instead of $\mathcal A_0$. Summing over all such backgrounds, i.e., gauging the $\mathbb Z_n^{(0)}$ symmetry, takes us back to the original ungauged theory $\mathcal A$. We shall not need this generalization in this paper.

\subsection{Fermion anyon condensation}\label{sec:fermion_condensation}

We now move on to the more interesting case of fermion condensation: we have some bosonic TQFT, and we wish to condense a certain abelian fermion, which we denote by $\psi\in\mathcal A$. We can assume without loss of generality that this line generates a $\mathbb Z_2^{(1)}$ symmetry, i.e.~$\psi^2=\boldsymbol 1$, for otherwise we can first condense the boson $\mathsf g=\psi^2$ (as in the previous section), and then condense the resulting fermion, which will satisfy $\psi^2=\boldsymbol 1$.

The philosophy underlying fermion condensation is essentially the same as in boson condensation: we construct the Hilbert space of the condensed theory from the states in the parent theory, perhaps in presence of $\psi$-punctures. Roughly speaking, the configurations with non-trivial background flux can be thought of as the different spin structures on $\Sigma$.

Before actually constructing the spin TQFT by condensing a fermion in a bosonic TQFT, let us discuss what we are to expect from this condensation in the first place. A spin TQFT should assign to manifolds of the form $\mathbb S^1\times \Sigma$ a super-vector space $\hat{\mathcal H}(\Sigma)$, which depends only on the topology of $\Sigma$, together with its spin structure $s$. (We use a hat to denote the Hilbert space of the condensed theory, and reserve the notation $\mathcal H(\Sigma)$ for that of the bosonic parent). Depending on the spin structure on the time circle $\mathbb S^1$, the partition function computes either the regular trace over $\hat{\mathcal H}(\Sigma)$, or the super-trace (i.e., the trace weighted by fermion parity). Specifically, the spin generalization of~\eqref{eq:Z_dim} is
\begin{equation}
\begin{aligned}
Z(\mathbb S^1_\text{NS}\times\Sigma)&=\tr_{\hat{\mathcal H}(\Sigma)}(\mathrm{id})\\
Z(\mathbb S^1_\text{R}\times\Sigma)&=\tr_{\hat{\mathcal H}(\Sigma)}(-1)^F
\end{aligned}
\end{equation}
where $\mathbb S^1_\text{NS}$ denotes the circle with anti-periodic boundary conditions, and $\mathbb S^1_\text{R}$ the circle with periodic boundary conditions. Therefore, if the super-vector space $\hat{\mathcal H}(\Sigma)$ is $\mathbb C^{b|f}$, then Neveu-Schwarz boundary conditions compute $b+f$, and Ramond boundary conditions compute $b-f$.

We shall denote a compact surface with genus $g$ and spin structure $s$ by $\Sigma_{g;s}$. As in the bosonic case, large diffeomorphisms act unitarily in $\hat{\mathcal H}(\Sigma_{g;s})$. The MCG as a spin surface is a subgroup of the MCG as a surface, $\text{MCG}(\Sigma_{g;s})\subseteq \text{MCG}(\Sigma_g)$. The reason for this is that some diffeomorphisms that leave $\Sigma$ invariant as a topological space, actually change the spin structure $s\mapsto s'$, and so do not constitute elements of the MCG as a spin surface. The canonical example is the $T$-transformation on the torus, which performs a Dehn twist around the $\mathsf a$-cycle. As such, it maps $( s^\mathsf a,s^\mathsf b)\mapsto (s^\mathsf a,s^\mathsf as^\mathsf b)$. This is an element of the spin MCG if $s^\mathsf a=+1$, but it is not if $s^\mathsf a=-1$. On the other hand, $T^2$ is in the spin MCG for any spin structure.

Elements of the spin MCG act unitarily in the Hilbert space, namely,
\begin{equation}
\text{MCG}(\Sigma_{g;s})\colon\hat{\mathcal H}(\Sigma_{g;s})\to \hat{\mathcal H}(\Sigma_{g;s})\,.
\end{equation}
On the other hand, elements of the regular MCG induce isomorphisms of (generically distinct) super-vector spaces,
\begin{equation}
\text{MCG}(\Sigma_{g})\colon\hat{\mathcal H}(\Sigma_{g;s})\to \hat{\mathcal H}(\Sigma_{g;s'})\,.
\end{equation}
This means, for example, that the partition function $Z(\mathbb S^2\times\Sigma_{g;s})$ is invariant under $\text{MCG}(\Sigma_{g})$; and, more generally, observables only depend on the equivalence class of $s$ under the regular MCG. It is known that there are only two equivalence classes of spin structures modulo MCG, the so-called \emph{even} and \emph{odd} spin structures. These are distinguished by the \emph{Arf invariant}~\cite{Arf1941UntersuchungenQ}. If two spin structures have the same Arf parity, then there exists some MCG element that maps one into the other. If they have different Arf parity, no such MCG element exists. In conclusion, observables of spin TQFTs depend on $s$ only through $\Arf(s)$.

For fixed spin structure, $\text{MCG}(\Sigma_{g;s})$ is represented by a unitary operator in $\hat{\mathcal H}(\Sigma_{g;s})$. That being said, due to the $\mathbb Z_2$ grading of this vector space, this action typically gets extended. Namely, the Hilbert space of spin TQFTs realize a unitary representation of a certain non-trivial $\mathbb Z_2$ extension of the spin MCG. To be explicit, (modding out by Torelli, i.e., working in homology) the MCG of $\Sigma_g$ is the integral symplectic group $Sp_g(\mathbb Z)$, and the spin MCG is some subgroup thereof. The Hilbert space of the theory realizes a unitary representation of the so-called \emph{metaplectic group} $Mp_g(\mathbb Z)$, which is defined as the (essentially unique) $\mathbb Z_2$ extension of the symplectic group
\begin{equation}
\mathbb Z_2\hookrightarrow Mp_g(\mathbb Z)\twoheadrightarrow Sp_g(\mathbb Z)\,.
\end{equation}
This extension corresponds to the fact that a $2\pi$ rotation is represented by the trivial element in $\text{MCG}(\Sigma_{g;s})$, while it lifts to $(-1)^F$ in $\hat{\mathcal H}(\Sigma_{g;s})$.

In order to illustrate these ideas it proves useful to focus on the torus, $\Sigma_1$. There are four spin tori, depending on the boundary conditions on the two spatial circles $\Sigma_{1;s^\mathsf a, s^\mathsf b}=\mathbb S^1_{s^\mathsf a}\times\mathbb S^1_{ s^\mathsf b}$:
\begin{equation}
\begin{aligned}
\Arf(s^\mathsf a,s^\mathsf b)=+1&\colon\quad\begin{cases}
\mathbb S^1_\text{NS}\times\mathbb S^1_\text{NS}\\
\mathbb S^1_\text{NS}\times\mathbb S^1_\text{R}\\
\mathbb S^1_\text{R}\times\mathbb S^1_\text{NS}
\end{cases}\\
\Arf(s^\mathsf a,s^\mathsf b)=-1&\colon\quad \quad \mathbb S^1_\text{R}\times\mathbb S^1_\text{R}
\end{aligned}
\end{equation}

The MCG of the torus is the modular group $SL_2(\mathbb Z)=\langle \hat S,\hat T\rangle$, acting as
\begin{equation}\label{eq:S_T_torus_spin}
\begin{aligned}
\hat S\colon \Sigma_{1;s^\mathsf a,s^\mathsf b}&\to \Sigma_{1;s^\mathsf b,s^\mathsf a}\\
\hat T\colon \Sigma_{1;s^\mathsf a,s^\mathsf b}&\to \Sigma_{1;s^\mathsf a, s^\mathsf as^\mathsf b}
\end{aligned}
\end{equation}
Therefore, the subgroup that fixes each spin structure is
\begin{equation}
\begin{aligned}
\text{MCG}(\Sigma_{1;--})&=\langle \hat S,\hat T^2\rangle\\
\text{MCG}(\Sigma_{1;-+})&=\langle \hat S\hat T\hat S,\hat T^2\rangle\\
\text{MCG}(\Sigma_{1;+-})&=\langle \hat S\hat T^2\hat S,\hat T\rangle\\
\text{MCG}(\Sigma_{1;++})&=\langle \hat S,\hat T\rangle
\end{aligned}
\end{equation}
Needless to say, the first three groups are all isomorphic, as they are related through $SL_2(\mathbb Z)$ conjugation:
\begin{equation}
\langle \hat S\hat T\hat S,\hat T^2\rangle=\hat T\langle \hat S,\hat T^2\rangle \hat T^{-1},\qquad \langle \hat S\hat T^2\hat S,\hat T\rangle=(\hat S\hat T)\langle \hat S,\hat T^2\rangle(\hat S\hat T)^{-1}\,.
\end{equation}
This group is a congruence subgroup of $SL_2(\mathbb Z)$ of index $3$, usually denoted by $\Gamma_0(2)$. The fourth group, on the other hand, is $SL_2(\mathbb Z)$ itself.

The diffeomorphism $\hat S^4$ corresponds to a $2\pi$ rotation and, as such, acts trivially in a bosonic theory and so is represented by the identity element in $SL_2(\mathbb Z)$; conversely, in a fermionic theory it is represented by $(-1)^F$. Thus, modular transformations in spin theories satisfy
\begin{equation}
\hat S^2=(\hat S\hat T)^3,\qquad \hat S^4=(-1)^F\,,
\end{equation}
with $(-1)^F$ an order $2$ central element. These relations define the group $Mp_1(\mathbb Z)$.\footnote{One should keep in mind that $Mp_1(\mathbb Z)$ does not act faithfully in $\hat{\mathcal H}(\Sigma_{1;s})$ (in fact, the metaplectic group is not a matrix group; it does not admit faithful finite-dimensional representations). This fact is most drastic when the theory, for whatever reason, has no fermionic states at all: in such cases, the actions of $Sp_1(\mathbb Z)$ and $Mp_1(\mathbb Z)$ are indistinguishable, inasmuch as $(-1)^F$ is trivial. For example, the theory lacks fermionic states if $s$ is an even spin structure, or if the theory is secretly bosonic (through some non-trivial duality). In such cases, one can think of the modular group as being $SL_2(\mathbb Z)$ instead of $Mp_1(\mathbb Z)$: their difference is invisible in the Hilbert space anyway. Similar considerations hold in higher genus.}

The Hilbert spaces $\hat{\mathcal H}(\Sigma_{g;s})$ are best understood by giving an explicit basis for them. As in the previous section, the Hilbert space of the fermionic theory can be constructed by condensing a fermion in the bosonic parent. Indeed, the Hilbert space of the spin theory is a subspace of the Hilbert space of the bosonic parent, together with the space with a $\psi$-puncture,
\begin{equation}
\hat{\mathcal H}(\Sigma_{g;s})\subseteq \mathcal H(\Sigma_g)\oplus\mathcal H(\Sigma_g^\psi)\,.
\end{equation}

Let us write down a basis for $\hat{\mathcal H}(\Sigma_{1;s})$ in terms of the states of the bosonic parent. Recall that the states on the torus in the bosonic theory are labelled by the anyons $\mathcal A$. The $\mathbb Z_2^{(1)}$ symmetry generated by $\psi$ partitions the spectrum $\mathcal A$ into two equivalence classes, distinguished by the braiding phase $B(\psi,\,\cdot\,)=\pm1$. In the case of boson condensation we denoted these two equivalence classes by $\mathcal A_0$ and $\mathcal A_1$; in the present context it is more natural to denote them by $\mathcal A_\text{NS}$ and $\mathcal A_\text{R}$:
\begin{equation}
\mathcal A=\mathcal A_\text{NS}\sqcup\mathcal A_\text{R},\qquad\begin{cases}
\mathcal A_\text{NS}:=\{\alpha\in\mathcal A\,\mid\, B(\alpha,\psi)=+1\}\\
\mathcal A_\text{R}\hspace{5pt}:=\{\alpha\in\mathcal A\,\mid\, B(\alpha,\psi)=-1\}
\end{cases}
\end{equation}
The two equivalence classes are further partitioned according to the length of the orbits. For a generic $\mathbb Z_n^{(1)}$ symmetry, the orbits come in lengths that divide $n$; for $n=2$, we have two-dimensional orbits and one-dimensional ones. We refer to the latter as \emph{Majorana lines}. It is easy to convince oneself that these can only appear in $\mathcal A_\text{R}$. We shall use the label $\alpha$ to denote generic lines of $\mathcal A$; on the other hand, lines of $\mathcal A_\text{NS}$ will be denoted by the more specific label $a$, while two-dimensional orbits of $\mathcal A_\text{R}$ by $x$ and one-dimensional ones by $m$. We will say that $\alpha$ is $a$-type, $x$-type, or $m$-type, according to this classification:
\begin{equation}
\begin{aligned}
a&\in\mathcal A_\text{NS}\\
x&\in\mathcal A_\text{R}\quad \&\quad |\,x\,|=2\\
m&\in\mathcal A_\text{R}\quad \&\quad |m|=1\,.
\end{aligned}
\end{equation}
In other words, $x$-lines satisfy $x\times \psi\neq x$, while $m$-lines satisfy $m\times\psi=m$.

As in the previous section, we take the bosonic theory, and condense a fermion $\psi$. In the condensed theory, $\psi$ becomes almost trivial: it should be represented by the identity operator, up to a sign, depending on the spin structure around the cycle it is supported on. This determines how the states in the condensed phase are obtained in terms of those of the uncondensed one. %More precisely, if $\psi$ runs around a cycle $\mathsf c\in H_1(\Sigma,\mathbb Z)$, then it should be represented by the operator $-s_\mathsf c\ \mathrm{id}$, where $s$ denotes the spin structure around $\mathsf c$. The negative sign is due to the fact that $\psi$ should be trivial on contractible cycles, and those have $s_\mathsf c=-1$.
We claim that the basis of $\hat{\mathcal H}(\Sigma_{1;s})$ can be taken as
\begin{equation}\label{eq:spin_torus_basis}
\begin{aligned}
\hat{\mathcal H}(\Sigma_{1;--})&\colon%\operatorname{Span}_{\mathbb C}\bigg[\bigg\{\frac{1}{\sqrt2}\bigg(|a\rangle+|\psi\times a\rangle\bigg)\bigg\}\bigg]\\
\tikz[baseline=-3]{
\node[anchor=south] at (-1.2,-.65) {$\frac{1}{\sqrt2}\ \bigg($};
\draw[thick] (-.5,-.4) -- (-.5,.4);
\draw[thick] (.5,.4) -- (.65,0) -- (.5,-.4);
\draw[thick] (0,0) circle (.25 cm);
\node[anchor=south] at (0,-.8) {\footnotesize$a$};
\node[anchor=south] at (1.3,-.3) {$+$};
\node[anchor=south] at (3.5,-.65) {$\vphantom{\frac{1}{\sqrt2}}\bigg)$};
\begin{scope}[shift={(2.5,0)}]
\draw[thick] (-.5,-.4) -- (-.5,.4);
\draw[thick] (.5,.4) -- (.65,0) -- (.5,-.4);
\draw[thick] (0,0) circle (.25 cm);
\node[anchor=south] at (0,-.8) {\footnotesize$a\times\psi$};
\end{scope}
}\\
\hat{\mathcal H}(\Sigma_{1;-+})&\colon%\operatorname{Span}_{\mathbb C}\bigg[\bigg\{\frac{1}{\sqrt2}\bigg(|a\rangle-|\psi\times a\rangle\bigg)\bigg\}\bigg]\\
\tikz[baseline=-3]{
\node[anchor=south] at (-1.2,-.65) {$\frac{1}{\sqrt2}\ \bigg($};
\draw[thick] (-.5,-.4) -- (-.5,.4);
\draw[thick] (.5,.4) -- (.65,0) -- (.5,-.4);
\draw[thick] (0,0) circle (.25 cm);
\node[anchor=south] at (0,-.8) {\footnotesize$a$};
\node[anchor=south] at (1.3,-.3) {$-$};
\node[anchor=south] at (3.5,-.65) {$\vphantom{\frac{1}{\sqrt2}}\bigg)$};
\begin{scope}[shift={(2.5,0)}]
\draw[thick] (-.5,-.4) -- (-.5,.4);
\draw[thick] (.5,.4) -- (.65,0) -- (.5,-.4);
\draw[thick] (0,0) circle (.25 cm);
\node[anchor=south] at (0,-.8) {\footnotesize$a\times\psi$};
\end{scope}
}\\
\hat{\mathcal H}(\Sigma_{1;+-})&\colon%\operatorname{Span}_{\mathbb C}\bigg[\bigg\{\frac{1}{\sqrt2}\bigg(|x\rangle+|\psi\times x\rangle\bigg)\bigg\}\cup\big\{|m\rangle\big\}\bigg]\\
\begin{cases}
\tikz[baseline=-3]{
\node[anchor=south] at (-1.2,-.65) {$\frac{1}{\sqrt2}\ \bigg($};
\draw[thick] (-.5,-.4) -- (-.5,.4);
\draw[thick] (.5,.4) -- (.65,0) -- (.5,-.4);
\draw[thick] (0,0) circle (.25 cm);
\node[anchor=south] at (0,-.8) {\footnotesize$x$};
\node[anchor=south] at (1.3,-.3) {$+$};
\node[anchor=south] at (3.5,-.65) {$\vphantom{\frac{1}{\sqrt2}}\bigg)$};
\begin{scope}[shift={(2.5,0)}]
\draw[thick] (-.5,-.4) -- (-.5,.4);
\draw[thick] (.5,.4) -- (.65,0) -- (.5,-.4);
\draw[thick] (0,0) circle (.25 cm);
\node[anchor=south] at (0,-.8) {\footnotesize$x\times\psi$};
\end{scope}
}\\
\tikz[baseline=-3]{
\node[anchor=south] at (-1.2,-.65) {$\hphantom{\frac{1}{\sqrt2}\ \bigg(}$};
\draw[thick] (-.5,-.4) -- (-.5,.4);
\draw[thick] (.5,.4) -- (.65,0) -- (.5,-.4);
\draw[thick] (0,0) circle (.25 cm);
\node[anchor=south] at (0,-.8) {\footnotesize$m$};
}
\end{cases}\\
\hat{\mathcal H}(\Sigma_{1;++})&\colon%\operatorname{Span}_{\mathbb C}\bigg[\bigg\{\frac{1}{\sqrt2}\bigg(|x\rangle-|\psi\times x\rangle\bigg)\bigg\}\cup\big\{|m;\psi\rangle\big\}\bigg]
\begin{cases}
\tikz[baseline=-3]{
\node[anchor=south] at (-1.2,-.65) {$\frac{1}{\sqrt2}\ \bigg($};
\draw[thick] (-.5,-.4) -- (-.5,.4);
\draw[thick] (.5,.4) -- (.65,0) -- (.5,-.4);
\draw[thick] (0,0) circle (.25 cm);
\node[anchor=south] at (0,-.8) {\footnotesize$x$};
\node[anchor=south] at (1.3,-.3) {$-$};
\node[anchor=south] at (3.5,-.65) {$\vphantom{\frac{1}{\sqrt2}}\bigg)$};
\begin{scope}[shift={(2.5,0)}]
\draw[thick] (-.5,-.4) -- (-.5,.4);
\draw[thick] (.5,.4) -- (.65,0) -- (.5,-.4);
\draw[thick] (0,0) circle (.25 cm);
\node[anchor=south] at (0,-.8) {\footnotesize$x\times\psi$};
\end{scope}
}\\
\tikz[baseline=-3]{
\node[anchor=south] at (-1.2,-.65) {$\hphantom{\frac{1}{\sqrt2}\ \bigg(}$};
\draw[thick] (-.5,-.4) -- (-.5,.4);
\draw[thick] (.5,.4) -- (.65,0) -- (.5,-.4);
\draw[thick] (.2,0) circle (.25 cm);
\draw[thick] (-.2,0) -- (-.05,0);
\node[anchor=south] at (.1,-.8) {\footnotesize$m$};
\node[anchor=south] at (-.3,-.2) {\footnotesize$\psi$};
\fill (-.05,0) circle (1.5pt);
}
\end{cases}
\end{aligned}
\end{equation}

The reasoning behind the construction of this basis is the same as in the case of boson condensation. Namely, the gauged theory is obtained by inserting $\psi^j$ in all possible ways. Here $\psi$ is of order two, so there are only two possible blocks: $j=0$ or $j=1$, i.e., no insertion, or a single $\psi$-insertion. Furthermore, the specific linear combination of states is decided by the spin structure. For example, inserting $\psi$ along the $\mathsf a$-cycle inserts the phase $|\alpha;\psi^j\rangle\mapsto B(\alpha,\psi)|\alpha;\psi^j\rangle$. This should reproduce the sign $s^\mathsf a$, which means that $\mathcal A_\text{NS}$ lines create states in $s^\mathsf a=-1$ boundary conditions, and $\mathcal A_\text{R}$ lines create states in $s^\mathsf a=+1$ boundary conditions. This explains why the basis is constructed using $a$-type lines in $\hat{\mathcal H}(\Sigma_{1;-,\bullet})$, and $x$- and $m$-type lines in $\hat{\mathcal H}(\Sigma_{1;+,\bullet})$.

Similarly, inserting $\psi$ along the $\mathsf b$-cycle fuses the state into $|\alpha;\psi^j\rangle\mapsto (-1)^j|\psi\times\alpha;\psi^j\rangle$. If this is to reproduce the boundary condition $s^\mathsf b$, we are required to consider the linear combination $|\alpha\rangle-s^\mathsf b|\psi\times\alpha\rangle$ for two-dimensional orbits; and, for Majorana lines, the puncture should be present if and only if $s^\mathsf b=+1$.

Note that, unlike in the case of bosonic condensation, here the multiple copies associated to the short orbits live in different spaces. Moreover, there are no short orbits in the NS sector, so the observables of the theories (the Wilson lines associated to the NS anyons) do not require fixed-point resolution. In this sense, the fusion rules of the condensed theory are inherited from those of the parent in a straightforward manner, without the need of knowing the once-punctured $S$-matrix. In the bosonic case, the fusion rules of the short orbits do require this extra structure.

As a consistency check for the basis above, we can easily show that modular transformations map the different Hilbert spaces as expected. For example, take $\hat{\mathcal H}(\Sigma_{1;--})$, and apply an $S$-transformation:
\begin{equation}
\frac{1}{\sqrt{2}}(|a\rangle+|\psi\times a\rangle)\overset S\mapsto \sum_{\alpha'\in\mathcal A} \frac{1}{\sqrt2}(S_{a,\alpha'}+S_{\psi\times a,\alpha'})|\alpha'\rangle
\end{equation}
As in equation~\eqref{eq:grading_Zn_S}, we have $S_{\psi\times a,\alpha'}=B(\psi,\alpha')S_{a,\alpha'}$, which means that we may restrict the sum over $\alpha'$ to $a$-type lines, for the Ramond ones do not contribute -- they cancel out pairwise. With this,
\begin{equation}
\frac{1}{\sqrt{2}}(|a\rangle+|\psi\times a\rangle)\overset S\mapsto \sum_{a'\in\mathcal A_\text{NS}/\sim} 2S_{a,a'}\frac{1}{\sqrt2}(|a'\rangle+|\psi\times a'\rangle)
\end{equation}
which shows that $S$ maps $\hat{\mathcal H}(\Sigma_{1;--})$ to itself, as expected (cf.~\eqref{eq:S_T_torus_spin}). Similarly, $T$-transformations map
\begin{equation}
\frac{1}{\sqrt{2}}(|a\rangle+|\psi\times a\rangle)\overset T\mapsto e^{-2\pi i c/24}\frac{1}{\sqrt{2}}(\theta(a)|a\rangle+\theta(\psi\times a)|\psi\times a\rangle)
\end{equation}
Noting that $\theta(\psi\times a)=-\theta(a)$, this becomes
\begin{equation}
\frac{1}{\sqrt{2}}(|a\rangle+|\psi\times a\rangle)\overset T\mapsto e^{-2\pi i c/24}\theta(a)\frac{1}{\sqrt{2}}(|a\rangle-|\psi\times a\rangle)
\end{equation}
which shows that $T$ maps $\hat{\mathcal H}(\Sigma_{1;--})$ into $\hat{\mathcal H}(\Sigma_{1;-+})$, again as expected (cf.~\eqref{eq:S_T_torus_spin}).

The other three Hilbert spaces can also be seen to transform into each other in the expected manner. Not only that, but the exercise gives us the explicit expression for the $\hat S$ and $\hat T$ matrices of the condensed theory:
\begin{equation}
\begin{aligned}
\hat S\colon\hat{\mathcal H}(\Sigma_{1;--})\to\hat{\mathcal H}(\Sigma_{1;--})\ &\Longrightarrow\ \begin{cases}
\hat S_{a,a'}=2S_{a,a'}\end{cases}\\
\hat S\colon\hat{\mathcal H}(\Sigma_{1;-+})\to\hat{\mathcal H}(\Sigma_{1;+-})\ &\Longrightarrow\ \begin{cases}
\hat S_{a,x}=2S_{a,x}\\
\hat S_{a,m}=\sqrt2S_{a,m}\end{cases}\\
\hat S\colon\hat{\mathcal H}(\Sigma_{1;+-})\to\hat{\mathcal H}(\Sigma_{1;-+})\ &\Longrightarrow\ \begin{cases}
\hat S_{x,a}=2S_{x,a}\\
\hat S_{m,a}=\sqrt2S_{m,a}\end{cases}\\
\hat S\colon\hat{\mathcal H}(\Sigma_{1;++})\to\hat{\mathcal H}(\Sigma_{1;++})\ &\Longrightarrow\ \begin{cases}
\hat S_{x,x'}=2S_{x,x'}\\
\hat S_{x,m}=0\\
\hat S_{m,m'}=S_{m,m'}(\psi)
\end{cases}
\end{aligned}
\end{equation}
where $S_{m,m'}(\psi)$ denotes the $S$-matrix of the bosonic parent, in the once-punctured torus (cf.~\eqref{eq:puncture_S_matrix}). 
%One can summarise the expressions above as $\hat S_{\alpha,\alpha'}=\sqrt{\ell_\alpha\ell_{\alpha'}}S_{\alpha,\alpha'}$, with $S$ the modular matrix of the bosonic parent, and $\ell_\alpha$ the length of the orbit. In writing this it is understood that the possible values taken by $\alpha,\alpha'$ depend on the spin structure, to wit, $(\alpha,\alpha')\in\mathcal A_{s_\mathsf a}\times\mathcal A_{s_\mathsf b}$. Similarly, the 
The $\hat T$-matrix is given by a similar expression:
\begin{equation}
\begin{aligned}
\hat T\colon\hat{\mathcal H}(\Sigma_{1;--})\to\hat{\mathcal H}(\Sigma_{1;-+})\ &\Longrightarrow\ \begin{cases}
\hat T_{a,a'}=e^{-2\pi i c/24}\theta(a)(\delta_{a,a'}-\delta_{a,\psi\times a'})\end{cases}\\
\hat T\colon\hat{\mathcal H}(\Sigma_{1;-+})\to\hat{\mathcal H}(\Sigma_{1;--})\ &\Longrightarrow\ \begin{cases}
\hat T_{a,a'}=e^{-2\pi i c/24}\theta(a)(\delta_{a,a'}+\delta_{a,\psi\times a'})\end{cases}\\
\hat T\colon\hat{\mathcal H}(\Sigma_{1;+-})\to\hat{\mathcal H}(\Sigma_{1;+-})\ &\Longrightarrow\ \begin{cases}
\hat T_{x,x'}=e^{-2\pi i c/24}\theta(a)(\delta_{x,x'}+\delta_{x,\psi\times x'})\\
\hat T_{x,m}=0\\
\hat T_{m,m'}=e^{-2\pi i c/24}\theta(m)\delta_{m,m'}\end{cases}\\
\hat T\colon\hat{\mathcal H}(\Sigma_{1;++})\to\hat{\mathcal H}(\Sigma_{1;++})\ &\Longrightarrow\ \begin{cases}
\hat T_{x,x'}=e^{-2\pi i c/24}\theta(a)(\delta_{x,x'}-\delta_{x,\psi\times x'})\\
\hat T_{x,m}=0\\
\hat T_{m,m'}=e^{-2\pi i c/24}\theta(m)\delta_{m,m'}
\end{cases}
\end{aligned}
\end{equation}
%In a more succinct notation, $\hat T_{\alpha,\alpha'}=e^{-2\pi i c/24}\theta(\alpha)(\delta_{\alpha,\alpha'}-s_\mathsf as_\mathsf b\delta_{\alpha,\psi\times \alpha'})$, with $(\alpha,\alpha')\in\mathcal A_{s_\mathsf a}\times\mathcal A_{s_\mathsf a}$.

Finally, we discuss the third generator of the spin modular group, fermion parity. This zero-form symmetry is the dual symmetry to the gauged $\mathbb Z_2^{(1)}$, which means that the states with odd fermion parity are those that carry the puncture. This means that $(-1)^F=1$ in all even spin structures, while in the R-R sector one has
\begin{equation}
\begin{aligned}
((-1)^F)_{x,x'}&=+\delta_{x,x'}\\
((-1)^F)_{m,m'}&=-\delta_{m,m'}
\end{aligned}
\end{equation}

These matrices are unitary, symmetric, and satisfy $\hat S^4=(-1)^F$ and $\hat S^2=(\hat S\hat T)^3$. It is important to remark that these properties are understood in the $\mathbb Z_2$-graded sense, i.e., taking into account~\eqref{eq:S_T_torus_spin}. In other words, the precise relations are
\begin{equation}\label{eq:MCG_algebra_spin}
\begin{aligned}
\hat S_{s^{\mathsf a},s^{\mathsf b}}^\dagger&=\hat S^{-1}_{s^{\mathsf a},s^{\mathsf b}}\\
\hat T_{s^{\mathsf a},s^{\mathsf b}}^\dagger&=\hat T^{-1}_{s^{\mathsf a},s^{\mathsf b}}\\
\hat S_{s^{\mathsf b},s^{\mathsf a}}^t&=\hat S_{s^{\mathsf a},s^{\mathsf b}}\\
(\hat S_{s^{\mathsf b},s^{\mathsf a}}\hat S_{s^{\mathsf a},s^{\mathsf b}})^2&=(-1)^F_{s^{\mathsf a},s^{\mathsf b}}\\
\hat S_{s^{\mathsf b},s^{\mathsf a}}\hat S_{s^{\mathsf a},s^{\mathsf b}}&=\hat S_{s^{\mathsf b},s^{\mathsf a}}\hat T_{s^{\mathsf b},s^{\mathsf a}s^{\mathsf b}}\hat S_{s^{\mathsf a}s^{\mathsf b},s^{\mathsf b}}\hat T_{s^{\mathsf a}s^{\mathsf b},s^{\mathsf a}}\hat S_{s^{\mathsf a},s^{\mathsf a}s^{\mathsf b}}\hat T_{s^{\mathsf a},s^{\mathsf b}} \,,
\end{aligned}
\end{equation}
where $\mathcal O_{s^\mathsf a,s^\mathsf b}$ denotes the operator $\mathcal O\in\{\hat S,\hat T,(-1)^F\}$ when acting on $\hat{\mathcal H}(\Sigma_{1;s^\mathsf a,s^\mathsf b})$.

In the appendix~\ref{sec:examples_spin_TQFT} we construct several examples of quotient TQFTs, namely $SO(N)_k$ with $k=1,2,3$. Some of these illustrate bosonic anyon condensation, and some others fermionic anyon condensation.

\section{Examples of anyon condensation}\label{sec:examples_spin_TQFT}

Here we collect some extra examples of boson and fermion condensation, using theories of the form $SO(n)_k$ for small values of $k$. In particular, $k=1$ and $k=3$ exemplify fermion condensation, and $k=2$ boson condensation. The case $k=1$, i.e., $SO(n)_1$, is the generator of fermionic SPTs with no symmetry, and so is a key theory in the study of fermionic TQFTs. The case $k=2$ will be related to $U(1)$ theories, through the level-rank duality $SO(n)_2\leftrightarrow SO(2)_{-n}\equiv U(1)_{-n}$. Finally, the case $k=3$ will be constructed through the $SU(2)$ theory, thanks to the level-rank duality $SO(n)_3\leftrightarrow SO(3)_{-n}\equiv SU(2)_{-2n}/\mathbb Z_2$. We also include the case of $U(1)_k$ separately, this time focusing on its time-reversal invariance.

\subsection{$SO(n)_1$}

This is the minimal spin TQFT, and it has central charge $n/2$, so corresponds to $n$ boundary Majorana fermions. A single fermion, $SO(1)_1$, is the generator of the group of fermionic SPTs with no extra symmetries, $\Omega^4_\mathrm{spin}=\mathbb Z$. In other words, any invertible fermionic phase is equivalent to $SO(n)_1$ for some $n$. The theory can also be written as $n$ copies of (the inverse of) the gravitational Chern-Simons theory.

The bosonic parent of this theory is $Spin(n)_1$. The details of this theory depend on the parity of $n$.

\paragraph{$\boldsymbol{n=2m+1}$.} One can construct $SO(n)_1$ by condensing the fermion in the Ising category. The modular data for the parent theory is that of $\mathrm{Ising}_m=Spin(2m+1)_1$, which has three anyons:
\begin{equation}
\begin{aligned}
\boldsymbol1&=[1,0,0,\dots,0,0]\\
\sigma&=[0,0,0,\dots,0,1]\\
\psi&=[0,1,0,\dots,0,0]
\end{aligned}
\end{equation}
where $[\lambda_0,\lambda_1,\dots,\lambda_m]$ denote the extended Dynkin labels of the representation. These lines fuse according to $\psi\times\sigma=\sigma$ and $\sigma^2=\boldsymbol1+\psi$, and transform under modular transformations as follows:
\begin{equation}
\begin{split}
S|\boldsymbol1\rangle&=\frac12|\boldsymbol 1\rangle+\frac{1}{\sqrt2}|\sigma\rangle+\frac12|\psi\rangle\\
S|\sigma\rangle&=\frac{1}{\sqrt2}|\boldsymbol 1\rangle-\frac{1}{\sqrt2}|\psi\rangle\\
S|\psi\rangle&=\frac12|\boldsymbol 1\rangle-\frac{1}{\sqrt2}|\sigma\rangle+\frac12|\psi\rangle
\end{split}\hspace{50pt}\begin{split}
T|\boldsymbol1\rangle&=\mathrm e^{2\pi i\bigl(0-\frac{n}{48}\bigr)}|\boldsymbol 1\rangle\vphantom{\frac12}\\
T|\sigma\rangle&=\mathrm e^{2\pi i\bigl(\frac{n}{16}-\frac{n}{48}\bigr)}|\sigma\rangle\vphantom{\frac12}\\
T|\psi\rangle&=\mathrm e^{2\pi i\bigl(\frac12-\frac{n}{48}\bigr)}|\psi\rangle\vphantom{\frac12}
\end{split}
\end{equation}

The lines are partitioned according to their braiding with respect to $\psi$ as
\begin{equation}
\begin{aligned}
\text{NS}&\colon\quad\mathcal A_\text{NS}=\{\boldsymbol1,\ \psi\}\\
\text{R}&\colon\quad\mathcal A_{\hspace{2pt}\text{R}\hspace{3pt}}=\{\sigma\}
\end{aligned}
\end{equation}
and they are paired-up under fusion as
\begin{equation}
\boldsymbol1\overset{\times\psi}\longleftrightarrow\psi,\qquad \sigma\tikz[baseline=-3pt]{\node[rotate=-90,scale=1.1] at (0,0) {$\curvearrowleft$};}\!\!\tikz[baseline=-2pt]{\node at (0,0) {${}^{\times\psi}$};}\,.
\end{equation}

Therefore, the four Hilbert spaces of the theory are
\begin{itemize}
\item If we take \textbf{NS-NS} boundary conditions, the state is
\begin{equation}
\begin{aligned}
|0;\text{NS-NS}\rangle&=\frac{1}{\sqrt2}(|\boldsymbol1\rangle+|\psi\rangle)\,.
\end{aligned}
\end{equation}

\item If we take \textbf{NS-R} boundary conditions, the state is
\begin{equation}
\begin{aligned}
|0;\text{NS-R}\rangle&=\frac{1}{\sqrt2}(|\boldsymbol1\rangle-|\psi\rangle)\,.
\end{aligned}
\end{equation}

\item If we take \textbf{R-NS} boundary conditions, the state is
\begin{equation}
\begin{aligned}
|0;\text{R-NS}\rangle&=|\sigma\rangle\,.
\end{aligned}
\end{equation}

\item If we take \textbf{R-R} boundary conditions, the state is
\begin{equation}
\begin{aligned}
|0;\text{R-R}\rangle&=|\sigma;\psi\rangle\,,
\end{aligned}
\end{equation}
where, we remind the reader, $|\alpha;\beta\rangle$ denotes the anyon $\alpha$ in presence of a $\beta$ puncture (cf.~\eqref{eq:torus_states_puncture}).

\end{itemize}

We see that these spaces are all one-dimensional, as expected from an invertible theory. Furthermore, all states are bosonic, except for the one with a puncture, $|\sigma;\psi\rangle$, which means that $(-1)^F=(-1)^{\Arf(s)}$.

The modular data of the quotient can be computed in a straightforward manner. The only non-trivial case is the $S$ matrix in the R-R sector, which has a puncture. We can compute this matrix element using the general formula~\eqref{eq:puncture_S_matrix}, namely
\begin{equation}
S_{\sigma,\sigma}(\psi)=\sum_{\alpha=\boldsymbol1,\psi}\frac{\theta(\alpha)}{\theta(\sigma)^2}S_{\boldsymbol1,\alpha}F_{\sigma,\sigma}\begin{bmatrix}
\psi&\sigma\\\sigma&\alpha
\end{bmatrix}
\end{equation}
The $F$-symbols of the Ising category are well-known, cf.~$F(\alpha=\boldsymbol1)\equiv+1$ and $F(\alpha=\psi)\equiv -1$. With this,
\begin{equation}\label{eq:S_punct_free_fermion}
\hat S_{\text{R-R}}=\frac12(-1)^{3/4}(-i)^m(F(\psi)-F(\boldsymbol1))\equiv e^{i \pi (6 m+7)/4}
\end{equation}

This result, together with $\hat T_{\text{R-R}}=\mathrm e^{\pi i(2m+1)/12}$, confirms that the theory satisfies the expected modularity relations, $S^2=(ST)^3$ and $S^4=(-1)^F$.

A different perspective yields the same answer. The CFT $SO(n)_1$ is identical to $n$ free Majorana fermions, and the once-puctured conformal block $|\sigma;\psi\rangle$ is nothing but the torus one-point function of $\psi$. The insertion of $\psi$ removes the Ramond zero-mode, and hence this one-point function is $\langle\psi\rangle=q^{1/24}\prod_{r=1}^\infty(1-q^r)\equiv \eta(\tau)$, the Dedekind eta function. The punctured $S$-matrix is nothing but the phase acquired by $\langle\psi\rangle$ under an $S$-transformation, namely $\eta(-1/\tau)=\sqrt{-i\tau}\eta(\tau)$. The factor of $\tau^{1/2}$ is the weight associated to a primary of spin $h=1/2$, while the factor of $\sqrt{-i}$ is the sought-after $S$-matrix. For a system of $n=2m+1$ fermions, the $S$-matrix is $(\sqrt{-i})^{2m+1}\equiv e^{i \pi (6 m+7)/4}$, in agreement with~\eqref{eq:S_punct_free_fermion}.

%{\color{red} There might be a consistency check. The once-punctured block is nothing but the one-point function of a free fermion with Ramond boundary conditions. This one-point function can be computed as $\langle\psi\rangle=q^{1/24}\prod_{r\ge1}(1-q^r)$, because the insertion removes the zero-mode. This expression is equal to $\eta(q)$, the Dedekind eta function. Under a modular transformation we have $\eta(q)\mapsto \sqrt{-i\tau}\eta(q)$. Note that $(\sqrt{-i})^{2m+1}\equiv e^{i \pi (6 m+7)/4}$, in agreement with $\hat S_\text{R-R}$. $\eta(q)^{2m+1}$ also transforms correctly under $T$ transformations.}

\paragraph{$\boldsymbol{n=2m}$.} Here the bosonic parent $Spin(2m)_1$ has four lines, the trivial representation, the vector representation, and the two spinor representations. The quotient is obtained by condensing the vector. The Lie algebra is simply-laced, which automatically implies that the fusion rules are abelian, and so there are no fixed-points under fusion. Therefore, all states are bosonic. The four lines are split into two NS-lines (the trivial and the vector) and two R-lines (the two spinors), and each pair belongs to a two-dimensional orbit. This means that each Hilbert space is one-dimensional, as expected from an invertible theory, and moreover all states have $(-1)^F=+1$. The modular data is trivially computed, given that there are no short orbits.

\subsection{$SO(n)_2$}
\label{app:sos}
%The co-marks are .

Here we illustrate the construction of the bosonic theory $SO(n)_2$, by condensing an abelian boson in $Spin(n)_2$. We focus in particular on the odd-$n$ case, where all the modular data -- especially the $F$-symbols -- is fully known~\cite{ardonne2016classification}. We follow the notation therein.

Consider the algebra $\mathfrak{so}_{2n+1}=B_n$. Its comarks are $a^\vee=1,2,2,\dots,2,1$, which means that the theory $Spin(2n+1)_2$ has $n+4$ lines. We denote them as $\boldsymbol 1,\epsilon,\phi_i,\psi_\pm$, with $i=1,2,\dots,n$. The corresponding affine Dynkin labels are as follows:
\begin{equation}
\begin{aligned}
\boldsymbol 1&=[2,0,0,\dots,0]\\
\epsilon&=[0,2,0,\dots,0]\\
\phi_1&=[1,1,0,\dots,0]\\
\phi_i&=[0,\dots,0,1,0,\dots,0]\qquad \text{at position $i+1$}\\
\phi_n&=[0,0,\dots,0,2]\\
\psi_+&=[1,0,0,\dots,0,1]\\
\psi_-&=[0,1,0,\dots,0,1]
\end{aligned}
\end{equation}

The $S$-matrix reads
\begin{equation}
\begin{aligned}
S_{\boldsymbol 1,\boldsymbol 1}&=S_{\boldsymbol 1,\epsilon}=S_{\epsilon,\epsilon}=\frac{1}{2\sqrt{2n+1}}\\
S_{\boldsymbol 1,\psi_\pm}&=+\frac12\\
S_{\epsilon,\psi_\pm}&=-\frac12\\
S_{\psi_s,\psi_{s'}}&=\frac12ss'\\
S_{\boldsymbol 1,\phi_i}&=S_{\epsilon,\phi_i}=\frac{1}{\sqrt{2n+1}}\\
S_{\psi_\pm,\phi_i}&=0\\
S_{\phi_i,\phi_j}&=\frac{2}{\sqrt{2 n+1}}\cos \frac{2 \pi i j}{2 n+1}
\end{aligned}
\end{equation}
and the spins are
\begin{equation}
\begin{aligned}
h_{\boldsymbol 1}&=0\\
h_\epsilon&=1\\
h_{\phi_i}&=\frac12\frac{i (2 n+1-i)}{2 n+1}\\
h_{\psi_+}&=\frac18n\\
h_{\psi_-}&=\frac18n+\frac12
\end{aligned}
\end{equation}

From this one derives the quantum dimensions
\begin{equation}
\begin{aligned}
d_{\boldsymbol 1}&=1\\
d_\epsilon&=1\\
d_{\phi_i}&=2\\
d_{\psi_\pm}&=\sqrt{2n+1}
\end{aligned}
\end{equation}
and fusion rules,
\begin{equation}
\begin{aligned}
\epsilon\times\epsilon=\boldsymbol 1,\quad\psi_\pm\times \psi_\pm=\boldsymbol 1+\sum_{j=1}^n\phi_j,\quad\psi_\pm\times \psi_\mp=\epsilon+\sum_{j=1}^n\phi_j\\
\epsilon\times\phi_i=\phi_i,\quad\epsilon\times\psi_\pm=\psi_\mp,\quad\phi_i\times\psi_\pm=\psi_\pm+\psi_\mp\\
\phi_i\times\phi_i=\boldsymbol 1+\epsilon+\phi_{g(2i)},\quad \phi_i\times\phi_j=\phi_{g(i-j)}\times\phi_{g(i+j)},\quad i>j
\end{aligned}
\end{equation}
where $g(i)=i$ if $1\le i\le n$ and $g(i)=2n+1-i$ otherwise.

We see that there are no multiplicities, and all anyons are self-conjugate. We also note that $\epsilon$ is condensable, which leads to the bosonic theory $SO(2n+1)_2$. Let us analyse the quotient explicitly.

By looking at the braiding phase $B(\alpha,\epsilon)$ one learns that the unscreened anyons are $\boldsymbol 1,\epsilon,\phi_i$, while the screened anyons are $\psi_\pm$. Moreover, $\boldsymbol 1$ and $\epsilon$ are in the same orbit, while all the $\phi_i$ are fixed points. Thus, a basis for the condensed Hilbert space is as follows:
\begin{equation}
\begin{aligned}
|0\rangle&=\frac{1}{\sqrt2}(|\boldsymbol 1\rangle+|\epsilon\rangle)\\
|i\rangle&=|\phi_i\rangle\\
|n+i\rangle&=|\phi_i;\epsilon\rangle
\end{aligned}
\end{equation}
for $i=1,2,\dots,n$, and where $|\,\cdot\,;\epsilon\rangle$ denotes a state in the once-punctured torus. The condensed $S$-matrix is
\begin{equation}
\begin{aligned}
\hat S_{0,0}&=2S_{\boldsymbol 1,\boldsymbol 1}=\frac{1}{\sqrt{2n+1}}\\
\hat S_{0,i}&=\sqrt2 S_{\boldsymbol 1,\phi_i}=\sqrt{\frac{2}{2n+1}}\\
\hat S_{0,n+i}&=0\\
\hat S_{i,j}&=S_{\phi_i,\phi_j}=\frac{2}{\sqrt{2 n+1}}\cos \frac{2 \pi i j}{2 n+1}\\
\hat S_{i,j+n}&=0\\
\hat S_{i+n,j+n}&=S_{\phi_i,\phi_j}(\epsilon)
\end{aligned}
\end{equation}
where $S_{\phi_i,\phi_j}(\epsilon)$ is the $S$-matrix of $Spin(2n+1)_2$ in the presence of a puncture. This matrix element can be obtain as in~\eqref{eq:puncture_S_matrix}. For example, we compute
\begin{equation}
\begin{aligned}
%=S_{\alpha,\alpha'}(\mathsf g^j)
S_{\phi_i,\phi_i}(\epsilon)&=\sum_{\beta\in\phi_i\times\phi_i}\frac{\theta(\beta)}{\theta(\phi_i)^2}S_{\boldsymbol1,\beta}F_{\phi_i,\phi_i}\begin{bmatrix}
\epsilon&\phi_i\\\phi_i&\beta\end{bmatrix}\\
&=e^{-2\pi i\frac{i (2 n+1-i)}{2 n+1}}\frac{1}{\sqrt{2n+1}}\times\\
&\bigg(\frac12\underbrace{F_{\phi_i,\phi_i}\begin{bmatrix}
\epsilon&\phi_i\\\phi_i&\boldsymbol1\end{bmatrix}}_{+1}+\frac12\underbrace{F_{\phi_i,\phi_i}\begin{bmatrix}
\epsilon&\phi_i\\\phi_i&\epsilon\end{bmatrix}}_{+1}+e^{2\pi i (\frac12\frac{2i (2 n+1-2i)}{2 n+1})}\underbrace{F_{\phi_i,\phi_i}\begin{bmatrix}
\epsilon&\phi_i\\\phi_i&\phi_{g(2i)}\end{bmatrix}}_{-1}\bigg)\\
&=\frac{2i}{\sqrt{2n+1}}\sin \frac{2\pi i^2}{2n+1}
\end{aligned}
\end{equation}
while for $i\neq j$,
\begin{equation}
\begin{aligned}
S_{\phi_i,\phi_j}(\epsilon)&=\sum_{\beta\in\phi_i\times\phi_j}\frac{\theta(\beta)}{\theta(\phi_i)\theta(\phi_j)}S_{\boldsymbol1,\beta}F_{\phi_i,\phi_j}\begin{bmatrix}
\epsilon&\phi_j\\\phi_i&\beta
\end{bmatrix}\\
&=\frac{1}{\sqrt{2n+1}}\bigg(
e^{\frac{2 i \pi i j}{2 n+1}} \underbrace{F_{\phi_i,\phi_j}\begin{bmatrix}
\epsilon&\phi_j\\\phi_i&\phi_{g(i-j)}
\end{bmatrix}}_{+1}+
e^{-\frac{2 i \pi i j}{2 n+1}}\underbrace{F_{\phi_i,\phi_j}\begin{bmatrix}
\epsilon&\phi_j\\\phi_i&\phi_{g(i+j)}
\end{bmatrix}}_{-1}\bigg)\\
&=\frac{2 i}{\sqrt{2 n+1}}\sin \frac{2 \pi i j}{2 n+1}
\end{aligned}
\end{equation}
%(Here the $F$-symbols are actually something like $(-1)^j$ and $(-1)^{j+1}$. The important fact is that the two symbols have opposite sign, and so the matrix element above is correct only up to a global sign. The global phase of $\hat S_{ij}$ is irrelevant for $i\neq j$, because we can always redefine $|i\rangle\to e^{i\mu_i}|i\rangle$ such that $\hat S_{ij}\mapsto e^{i(\mu_i-\mu_j)}\hat S_{ij}$.)

All in all, the $S$-matrix of the quotient takes the form
\begin{equation}\arraycolsep=7pt\def\arraystretch{1.5}
\hat S=\frac{1}{\sqrt{2n+1}}\left(\begin{array}{c|c|c}
1 & \sqrt2 & 0\\\hline
\sqrt2 & 2\cos\frac{2\pi ij}{2n+1} & 0\\\hline
0 & 0 & 2i\sin\frac{2\pi ij}{2n+1}
\end{array}\right)\begin{array}{l}
\leftarrow|0\rangle\\
\leftarrow|i\rangle\\
\leftarrow|i;\epsilon\rangle
\end{array}
\end{equation}

One can easily check that this matrix is unitary, and satisfies the algebra of the (bosonic) modular group, $S^2=(ST)^3$, $S^4=1$.

In order to write down the fusion rules of the quotient we have to switch into the fusion basis, namely
\begin{equation}
|\phi_i,\pm\rangle=\frac{1}{\sqrt2}(|\phi_i\rangle\pm|\phi_i;\epsilon\rangle)\,.
\end{equation}
In this basis, the $S$-matrix becomes $\hat S_{ij}\sim\frac{1}{\sqrt{2n+1}} e^{2\pi i \frac{ij}{2n+1}}$. This is in agreement with the level-rank duality $SO(n)_2\sim SO(2)_{-n}=U(1)_{-n}$, where the $S$-matrix of $U(1)_k$ is $e^{-2\pi i \alpha \beta/k}/\sqrt k$. (Here $\sim$ denotes duality modulo $\{\boldsymbol1,\psi\}$, since $U(1)_k$ is spin for odd $k$.)

\subsection{$SO(n)_3$}

We construct the theory using level-rank duality $SO(n)_3=SO(3)_{-n}=SU(2)_{-2n}/\mathbb Z_2$. So we consider $SU(2)_k$ first.

There are $k+1$ lines, which we label as $j=0,\frac12,1,\dots,\frac12k$. The $S$-matrix reads
\begin{equation}
S_{ij}=\sqrt{\frac{2}{k+2}}\sin\frac{\pi(2i+1)(2j+1)}{k+2}
\end{equation}
and the spins are $h_j=\frac{j(j+1)}{k+2}$. The fusion rules read
\begin{equation}
j_1\times j_2=\sum_{j=|j_1-j_2|}^{\min(J,k-J)}j,\qquad J=j_1+j_2
\end{equation}
The quantum dimensions are $d_j=\frac{\sin \frac{\pi (2 j+1)}{k+2}}{\sin \frac{\pi }{k+2}}$, and so $j=k/2$ is abelian. The corresponding $\mathbb Z_2$ symmetry acts as $j\mapsto \frac12k-j$. The spin of this line is $k/4$, and so the symmetry is condensable if and only if $k$ is even, $k=2n$. The quotient theory is $PSU(2)_{2n}=SO(3)_n$; it is spin if $n$ is odd.

The only fixed-point is $j=n/2$, whose $S$-matrix element is given by~\eqref{eq:puncture_S_matrix}
\begin{equation}\label{eq:s_punct_su_2}
\begin{aligned}
S_{n/2,n/2}(n)&=\sum_{\beta\in n/2\times n/2}\frac{\theta(\beta)}{\theta(n/2)^2}S_{\boldsymbol1,\beta}F_{n/2,n/2}\begin{bmatrix}
n&n/2\\n/2&\beta
\end{bmatrix}\\
&=\frac{\theta(n/2)^{-2}}{\sqrt{n+1}}\sum_{j=0}^n \theta(j)\sin\frac{\pi}{2}\frac{2j+1}{n+1}
F_{n/2,n/2}\begin{bmatrix}
n&n/2\\n/2&j
\end{bmatrix}
\end{aligned}
\end{equation}
which, using $F=(-1)^j$, becomes $S_{n/2,n/2}(n)=e^{-3i \pi n/4}$. One may check that modularity is satisfied, $S^2=(ST)^3$ and $S^4=(-1)^F\equiv (-1)^n$. This is indeed consistent with the quotient being bosonic if $n$ is even, and fermionic if odd.

Let us consider the case of even $n$, where the quotient is bosonic. The unscreened lines are those with integer $j$, and the only fixed point is $j=n/2$. Thus, a basis for the Hilbert space is
\begin{equation}
\begin{aligned}
|j\rangle&=\frac{1}{\sqrt2}(|j\rangle+|n-j\rangle),\qquad j=0,1,\dots,n/2-1\\
|a_1\rangle&=|n/2\rangle\\
|a_2\rangle&=|n/2;n\rangle
\end{aligned}
\end{equation}
where $|\,\cdot\,;\alpha\rangle$ denotes the corresponding state with an $\alpha$-puncture.

The $S$-matrix of the quotient is
\begin{equation}
\begin{aligned}
\hat S_{i,j}&=2S_{i,j}=2\sqrt{\frac{1}{n+1}}\sin\frac{\pi(2i+1)(2j+1)}{2n+2}\\
\hat S_{i,a_1}&=\sqrt2S_{i,n/2}=(-1)^i\sqrt{\frac{2}{n+1}}\\
\hat S_{a_1,a_1}&=S_{n/2,n/2}=(-1)^{n/2}\sqrt{\frac{1}{n+1}}\\
\hat S_{a_1,a_2}&=0\\
\hat S_{a_2,a_2}&=S_{n/2,n/2}(n)=i^{n/2}\,.
\end{aligned}
\end{equation}

The fusion basis is defined by
\begin{equation}
|a_\pm\rangle=\frac{1}{\sqrt2}(|a_1\rangle\pm|a_2\rangle)\,.
\end{equation}
One can easily compute the $S$-matrix in this basis, from where one can compute, for example, the fusion rules of the theory.

Consider now the case of odd $n$, where the quotient is spin. The NS lines are those with integral isospin, and the R lines with half-integral isospin. A basis of the quotient Hilbert space is\footnote{We would like to thank N.~Seiberg for spotting a few typos in the formulae below.}
\begin{itemize}
\item If we take \textbf{NS-NS} boundary conditions, the states are
\begin{equation}
|j;\text{NS-NS}\rangle=\frac{1}{\sqrt2}(|j\rangle+|n-j\rangle),\quad j=0,1,\dots,\tfrac12(n-1)\,.
\end{equation}

\item If we take \textbf{NS-R} boundary conditions, the states are
\begin{equation}
|j;\text{NS-R}\rangle=\frac{1}{\sqrt2}(|j\rangle-|n-j\rangle),\quad j=0,1,\dots,\tfrac12(n-1)\,.
\end{equation}

\item If we take \textbf{R-NS} boundary conditions, the states are
\begin{equation}
\begin{aligned}
|j;\text{R-NS}\rangle&=\frac{1}{\sqrt2}(|j\rangle+|n-j\rangle),\quad j=\tfrac12,\tfrac32,\dots,\tfrac12(n-2)\\
|n/2;\text{R-NS}\rangle&=|n/2\rangle\,.
\end{aligned}
\end{equation}

\item If we take \textbf{R-R} boundary conditions, the states are
\begin{equation}
\begin{aligned}
|j;\text{R-R}\rangle&=\frac{1}{\sqrt2}(|j\rangle-|n-j\rangle),\quad j=\tfrac12,\tfrac32,\dots,\tfrac12(n-2)\\
|n/2;\text{R-R}\rangle&=|n/2;n\rangle\,.
\end{aligned}
\end{equation}

\end{itemize}

The modular data easily follows from this decomposition, and the once-punctured torus matrix element~\eqref{eq:s_punct_su_2}.

\subsection{$U(1)_k$}

In this section we construct the Hilbert space of the spin TQFT $U(1)_k$. This generalizes the construction of the semion-fermion theory of section~\ref{sec:semion_fermion}. For some special values of $k$ this theory is time-reversal invariant. The semion-fermion theory is recovered by taking $k=2$. For $k>2$, the time-reversal symmetry (when present) satisfies a more exotic algebra~\cite{Delmastro:2019vnj}, namely $\mathsf T^2=\mathsf C$, where $\mathsf C$ denotes an order-$2$ unitary symmetry (charge conjugation).

The construction of $U(1)_k$ is slightly different depending on whether $k$ is even or odd. Indeed, for $k$ odd the theory is naturally spin; but, for $k$ even, it is bosonic, and so it has to be multiplied by the trivial factor $\{\boldsymbol1,\psi\}$ if we are interested in its spin version. The latter case is rather similar to the semion-fermion theory, so here we will focus on the $k$ odd case here, and sketch the main differences for $k$ even at the end.

Consider the theory $U(1)_k$ with $k$ odd. Its bosonic parent is $U(1)_{4k}$, whose anyons are labelled as $\alpha\in\mathbb Z_{4k}$. The spin theory is obtained by condensing the fermion $\psi=2k$. The braiding phase of an arbitrary line $\alpha$ with respect to the fermion is $B(\alpha,\psi)=e^{\pi i \alpha}$, which means that the anyons are split as
\begin{equation}
\begin{aligned}
\text{NS}&\colon \alpha=2\beta,\hspace{43.5pt} \beta=0,1,\dots,2k-1\\
\text{R}\,&\colon \alpha=2\beta+1,\qquad \beta=0,1,\dots,2k-1\,.
\end{aligned}
\end{equation}

These are all in two-dimensional orbits, paired up as
\begin{equation}
\alpha\overset{\times\psi}\longleftrightarrow \alpha+2k\,.
\end{equation}
As there are no fixed-points, all states are bosonic.

\paragraph{Hilbert space and modularity.} Given the knowledge of the Hilbert space of the bosonic parent, and the action of the modular group on it, we easily construct the same objects in the quotient theory. In particular, the Hilbert space is $\mathcal H\cong \mathbb C^{4k}$, with states $|\alpha\rangle$, and modular transformations act as
\begin{equation}
\begin{aligned}
S|\alpha\rangle&=\sum_{\alpha'\in\mathbb Z_{4k}}S_{\alpha,\alpha'}|\alpha'\rangle\\
T|\alpha\rangle&=e^{2\pi i(\alpha^2/4k-1/24)}|\alpha\rangle\\
\mathsf C|\alpha\rangle&=|{}-\alpha\mod 4k\rangle\,,
\end{aligned}
\end{equation}
where $S_{\alpha,\alpha'}=e^{-2\pi i \alpha\alpha'/4k}/2\sqrt k$, and the term $-1/24$ in the $T$-transformation refers to the central charge of the theory.

The quotient space is as follows:
\begin{itemize}

\item If we take \textbf{NS-NS} boundary conditions, the states are
\begin{equation}
|\alpha;\text{NS-NS}\rangle=\frac{1}{\sqrt2}\left[|2\alpha\rangle+|2\alpha+2k\rangle\right],\quad \alpha=0,\dots,k-1
\end{equation}
and one has
\begin{equation}
\begin{aligned}
\hat S|\alpha;\text{NS-NS}\rangle&=\frac{1}{\sqrt k}\sum_{\alpha'=0}^{k-1}\mathrm e^{2\pi i \alpha \alpha'/k}|{\alpha'};\text{NS-NS}\rangle\\
\hat T|\alpha;\text{NS-NS}\rangle&=\mathrm e^{2\pi i \big(\frac{\alpha^2}{2k}-\frac{1}{24}\big)}|\alpha;\text{NS-R} \rangle\\
\hat{\mathsf C}|\alpha;\text{NS-NS}\rangle&=\sum_{\alpha'=0}^{k-1}(\delta_{\alpha+\alpha'}+\delta_{\alpha+\alpha'-k})|{\alpha'};\text{NS-NS}\rangle
\end{aligned}
\end{equation}
where $\delta_x=1$ if $x\equiv 0\mod 2k$, and $\delta_x=0$ otherwise.

\item If we take \textbf{NS-R} boundary conditions, the states are
\begin{equation}
|\alpha;\text{NS-R}\rangle=\frac{1}{\sqrt2}\left[|2\alpha\rangle-|2\alpha+2k\rangle\right],\quad \alpha=0,\dots,k-1
\end{equation}
and one has
\begin{equation}
\begin{aligned}
\hat S|\alpha;\text{NS-R}\rangle&=\frac{1}{\sqrt k}\sum_{\alpha'=0}^{k-1}\mathrm e^{2\pi i (2\alpha+1) \alpha'/2k}|{\alpha'};\text{R-NS}\rangle\\
\hat T|\alpha;\text{NS-R}\rangle&=\mathrm e^{2\pi i \big(\frac{\alpha^2}{2k}-\frac{1}{24}\big)}|\alpha;\text{NS-NS}\rangle\\
\hat{\mathsf C}|\alpha;\text{NS-R}\rangle&=\sum_{\alpha'=0}^{k-1}(\delta_{\alpha+\alpha'}-\delta_{\alpha+\alpha'-k})|{\alpha'};\text{NS-R}\rangle
\end{aligned}
\end{equation}

\item If we take \textbf{R-NS} boundary conditions, the states are
\begin{equation}
|\alpha;\text{R-NS}\rangle=\frac{1}{\sqrt2}\left[|2\alpha+1\rangle+|2\alpha+1+2k\rangle\right],\quad \alpha=0,\dots,k-1
\end{equation}
and one has
\begin{equation}
\begin{aligned}
\hat S|\alpha;\text{R-NS}\rangle&=\frac{1}{\sqrt k}\sum_{\alpha'=0}^{k-1}\mathrm e^{2\pi i \alpha (2\alpha'+1)/2k}|{\alpha'};\text{NS-R}\rangle\\
\hat T|\alpha;\text{R-NS}\rangle&=\mathrm e^{2\pi i \big(\frac{(2\alpha+1)^2}{8k}-\frac{1}{24}\big)}|\alpha;\text{R-NS} \rangle\\
\hat{\mathsf C}|\alpha;\text{R-NS}\rangle&=\sum_{\alpha'=0}^{k-1}(\delta_{\alpha+\alpha'+1}+\delta_{\alpha+\alpha'+1-k})|{\alpha'};\text{R-NS}\rangle
\end{aligned}
\end{equation}

\item If we take \textbf{R-R} boundary conditions, the states are
\begin{equation}
|\alpha;\text{R-R}\rangle=\frac{1}{\sqrt2}\left[|2\alpha+1\rangle-|2\alpha+1+2k\rangle\right],\quad \alpha=0,\dots,k-1
\end{equation}
and one has
\begin{equation}
\begin{aligned}
\hat S|\alpha;\text{R-R}\rangle&=\frac{1}{\sqrt k}\sum_{\alpha'=0}^{k-1}\mathrm e^{2 \pi i (2 \alpha+1) (2 \alpha'+1)/4 k}|{\alpha'};\text{R-R}\rangle\\
\hat T|\alpha;\text{R-R}\rangle&=\mathrm e^{2\pi i \big(\frac{(2\alpha+1)^2}{8k}-\frac{1}{24}\big)}|\alpha;\text{R-R}\rangle\\
\hat{\mathsf C}|\alpha;\text{R-R}\rangle&=\sum_{\alpha'=0}^{k-1}(\delta_{\alpha+\alpha'+1}-\delta_{\alpha+\alpha'+1-k})|{\alpha'};\text{R-R}\rangle
\end{aligned}
\end{equation}

\end{itemize}

It is reassuring to see that these modular transformations map the different Hilbert spaces precisely as they should (cf.~\eqref{eq:S_T_torus_spin}). Moreover, these matrices are unitary, $\hat S$ is symmetric ($\hat S_{s^\mathsf a,s^\mathsf b}^t=\hat S_{s^\mathsf b,s^\mathsf a}$), and they satisfy the modular algebra $(\hat S\hat T)^3=\hat S^2=\hat{\mathsf C}$ with $\hat{\mathsf C}^2=1$.

\paragraph{Wilson lines.} The Wilson lines are given by
\begin{equation}
\begin{aligned}
W^{(\mathsf a)}(\alpha)|\gamma;s^\mathsf as^\mathsf b\rangle &=\mathrm e^{-2\pi i \alpha(\gamma+(1+s^\mathsf a)/4)/k}|\gamma;s^\mathsf as^\mathsf b\rangle\\
W^{(\mathsf b)}(\alpha)|\gamma;s^\mathsf as^\mathsf b\rangle &=|{\alpha+\gamma};s^\mathsf as^\mathsf b\rangle-s^\mathsf b |{\alpha+\gamma+k};s^\mathsf as^\mathsf b\rangle
\end{aligned}
\end{equation}
where $\alpha\in\mathbb Z_{2k}$. They satisfy the expected properties, e.g.,
\begin{equation}
\begin{aligned}
W_{s^\mathsf a,s^\mathsf b}^{(\mathsf c)}(\psi)&=W_{s^\mathsf a,s^\mathsf b}^{(\mathsf c)}(k)=-s^{\mathsf c}\boldsymbol1_k\\
W_{s^\mathsf a,s^\mathsf b}^{(\mathsf c)}(\alpha\times\alpha')&=W_{s^\mathsf a,s^\mathsf b}^{(\mathsf c)}(\alpha)W_{s^\mathsf a,s^\mathsf b}^{(\mathsf c)}(\alpha')\\
%[W^{(\mathsf c)}(\alpha),W^{(\mathsf c)}(\alpha')]&=0\\
S_{s^\mathsf a,s^\mathsf b}W_{s^\mathsf a,s^\mathsf b}^{(\mathsf a)}(\alpha)(S_{s^\mathsf a,s^\mathsf b})^\dagger&=W^{(\mathsf b)}_{s^\mathsf b,s^\mathsf a}(\bar\alpha)\\
S_{s^\mathsf a,s^\mathsf b}W_{s^\mathsf a,s^\mathsf b}^{(\mathsf b)}(\alpha)(S_{s^\mathsf a,s^\mathsf b})^\dagger&=W_{s^\mathsf b,s^\mathsf a}^{(\mathsf a)}(\alpha)
\end{aligned}
\end{equation}

\paragraph{Time-reversal.} We now implement time-reversal invariance. Recall that $U(1)_k$ is time-reversal invariant if and only if $q^2=-1\mod k$ is solvable for some $q\in\mathbb Z$, in which case time-reversal acts as $\alpha\mapsto q\alpha$~\cite{Delmastro:2019vnj}. This means that, given $\mathsf T=\tau K$, we require
\begin{equation}
\begin{aligned}
\tau (W_2^{(\mathsf c)})^*\tau^{-1}&=W_{2q}^{(\mathsf c)}\\
&=(W_2^{(\mathsf c)})^q
\end{aligned}
\end{equation}
with solution
\begin{equation}
\tau_{\alpha,\beta}=(-s^\mathsf b)^{\alpha +\beta} \delta_{2 \alpha q+2 \beta+\frac{1}{2}(s^\mathsf a+1)(q+1) }
\end{equation}
up to a global phase. One can check that
\begin{equation}
\tau\tau^*=(-1)^{\Arf(s)}\big(\delta_{\frac{1}{2} (s^\mathsf a+1)+\alpha +\beta}-s^\mathsf b\delta_{\frac{1}{2} (s^\mathsf a+1)+\alpha +\beta -k}\big)
\end{equation}
and so $\mathsf T^2=(-1)^{\Arf(s)}\hat{\mathsf C}$.

We see that the time-reversal algebra is deformed by Arf, signaling an anomaly. In this case, the source of the anomaly is clear: the theory has non-vanishing central charge, $c=1$, so it is not time-reversal invariant in the strict sense. We have to multiply by a suitable SPT in order to subtract off the central charge. In this case, $U(1)_{-1}$ does the trick, as this SPT has $c=-1$.

The Hilbert space of $U(1)_{-1}$ is straightforward: it suffices to take $k=1$ in the discussion above. Looking at the action of $\hat{\mathsf C}$ on the (one-dimensional) Hilbert space of $U(1)_{-1}$ we learn that $\hat{\mathsf C}=(-1)^{\Arf(s)}$. Therefore, multiplying a given theory by $U(1)_{\pm1}$ has the effect of redefining $\hat{\mathsf C}\to\hat{\mathsf C}\times(-1)^{\Arf(s)}$, which means that the theory $U(1)_k\times U(1)_{-1}$ has undeformed algebra, namely $\mathsf T^2=\hat{\mathsf C}$. The Arf deformation in the case of $U(1)_k$ was just signaling that we had not corrected the central charge down to zero; after doing so, the deformation disappears from the time-reversal algebra.

Finally, we make a few remarks concerning the $k$ even case. Now the theory is naturally bosonic, and can be made spin by tensoring with an invertible spin TQFT. If we are interested in time-reversal invariance, the natural choice is $U(1)_k\times U(1)_{-1}$, so as to have vanishing central charge. As the theory is a tensor product, one factor being bosonic, the total Hilbert space is straightforward:
\begin{equation}
\hat{\mathcal H}_s(U(1)_k\times U(1)_{-1})=\mathcal H(U(1)_k)\otimes \hat{\mathcal H}_s(U(1)_{-1})
\end{equation}
where $\mathcal H(U(1)_k)$ is the space of the bosonic theory $U(1)_k$, and $\hat{\mathcal H}_s(U(1)_{-1})$ is the space of the fermionic theory $U(1)_{-1}$. The Hilbert space of $U(1)_{-1}$ was discussed above, and that of $U(1)_k$ is well-known, being bosonic. In this sense, no new computation is required in the case of $U(1)_k$ with $k$ even. One can easily check through straightforward computation that the main conclusions are identical to those of the $k$ odd case, in particular, time-reversal satisfies $\mathsf T^2=\hat{\mathsf C}$, with no deformation. (If we reintroduce a non-zero value of the central charge, by multiplying by an extra factor of $U(1)_{\pm1}$, the deformation reappears, and we get $\mathsf T^2=(-1)^{\Arf(s)}\hat{\mathsf C}$, again signalling the anomaly due to $c$).

%\bibliography{references}
\clearpage
\printbibliography
\end{document}